%% file: CP3_Giant_Graviton.tex
\documentclass[a4paper,12pt]{JHEP3}
\pdfoutput=1

\usepackage{graphicx}
\usepackage{color}
\let\normalcolor\relax
\usepackage{amsmath}
\usepackage{amsfonts}
\usepackage{amsthm}
\usepackage{amssymb}
\usepackage{youngtab}
\usepackage{enumerate}

\title{The giant graviton on $AdS_{4}\times\mathbb{CP}^{3}$ - another step towards the emergence of geometry}
\author{Dino Giovannoni$^{1}$\footnote{dino.giovannoni@gmail.com}, Jeff Murugan$^{1,2}$\footnote{jeff@nassp.uct.ac.za} and
Andrea Prinsloo$^{1,2}$\footnote{andy.prinsloo@uct.ac.za} \\

$^{1}$Astrophysics, Cosmology \& Gravity Center and \\
Department of Mathematics and Applied Mathematics, \\
University of Cape Town, \\
Private Bag, Rondebosch, 7700, \\
South Africa. \\

$^{2}$National Institute for Theoretical Physics, \\
Private Bag X1, \\
Matieland, 7602, \\
South Africa.}

\abstract{We construct the giant graviton on $AdS_{4}\times\mathbb{CP}^{3}$ out of a four-brane embedded in and moving on the complex projective space. This configuration is dual to the totally anti-symmetric Schur polynomial operator $\chi_{R}(A_{1}B_{1})$ in the 2+1-dimensional, $\mathcal{N}$ = 6 super Chern-Simons ABJM theory. We demonstrate that this BPS solution of the D4-brane action is energetically degenerate with the point graviton solution and initiate a study of its spectrum of small fluctuations. Although the full computation of this spectrum proves to be analytically intractable, by perturbing around a ``small" giant graviton, we find good evidence for a dependence of the spectrum on the size, $\alpha_{0}$, of the giant. This is a direct result of the changing shape of the worldvolume as it grows in size.} 
\preprint{ACGC-150811}

\newcommand{\p}{\partial}

\begin{document}

\input{01-Intro}

\input{02-ABJM}

\input{03-PointGraviton}

\input{04-GiantGraviton}

\input{05-Fluct}

\input{06-Limits}

\input{07-Discussion}
\input{08-Acknow}

\appendix

\input{A-Background}
\input{B-Numerics}
\input{C-dAlembertian}

\input{Bib}

\end{document}

%% file: 01-Intro.tex
\section{Introduction} \label{section - intro}

{\it{\bf Lab$\cdot$o$\cdot$ra$\cdot$to$\cdot$ry} /'labre, t$\hat{o}$r$\bar{e}$ / noun: any place, situation, object, set of conditions, or the like, conducive to controlled experimentation, investigation, observation, etc.}

Since their inception in \cite{MST00} over a decade ago now, giant gravitons have matured into one of the best laboratories - if the above definition is anything to go by - that we have for studying the physics of D-branes and, by extension, the open strings that end on them. Indeed, directly or indirectly, giant gravitons have played a significant r$\hat{\mathrm o}$le in many of the biggest advances in string theory over these past ten years. These include (but are by no means limited to):
\vspace{-0.1cm}
\begin{enumerate}[i)]
 \item 
   the realization that D-branes are not described in the dual $SU(N)$ super-Yang-Mills 
   theory by {\it single-trace} operators but rather by {\it determinant-like} operators whose $\mathcal{R}$-charge 
   is $\sim \mathcal{O}(N)$. For the case of (excited) giant gravitons, these operators are known
   exactly. They are (restricted) Schur polynomials, 
   $\chi_{R}(\Phi) = \frac{1}{n!} \sum_{\sigma\in S_{n}}\chi_{R}(\sigma)\,tr\!
   \left(\sigma\Phi^{\otimes n}\right)$, built from fields in the Yang-Mills supermultiplet and labeled
   by Young diagrams with $n\sim\mathcal{O}(N)$ boxes, 
 \item 
   a complete classification all $\frac{1}{2}-$BPS geometries of type IIB supergravity in 
   \cite{Lin:2004nb} based on the free fermion description of giant graviton states given in 
   \cite{CJR, Berenstein1}, 
 \item 
   a detailed understanding of the structure of open string integrability in string theory as developed in 
   \cite{Hofman-Maldacena} and the corresponding statements about the integrablity of 
   $\mathcal{N}=4$ super Yang-Mills theory to be found in \cite{Berenstein:2006qk}. Indeed, 
   so powerful are the tools developed from giant graviton operators \cite{RdMK-et-al} that they 
   have recently even opened the door to the study of integrability beyond the planar level in the 
   gauge theory \cite{RdMK-non-planar} and, 
 \item 
   a concrete proposal for the realization of the idea that quantum gravity and spacetime 
   itself are emergent phenomena \cite{Emergent} (see also \cite{JM-RdMK} for a 
   summary of these ideas) encoded in the quantum interactions of a matrix model.
\end{enumerate} 
\vspace{-0.1cm}
It is this last research program that will be of most relevance to us in this article. The idea that spacetime, its local (geometrical) and global (topological) properties are not fundamental but emerge in some ``coarse-graining" limit of quantum gravity is not a new one and is certainly not unique to string theory. What string theory does bring to the table though is a concrete way to take such a limit via the AdS/CFT correspondence \cite{Maldacena97}. Broadly speaking, this gauge/gravity duality says that in the large $N$ limit, certain gauge theories (like 4-dimensional $\mathcal{N}=4$, $SU(N)$ super Yang-Mills theory) behave  more like gravity than gauge theories (and vice versa). It is in this sense that, in the AdS/CFT context, spacetime is {\it emergent}. This then begs the question:

\noindent
\framebox[\textwidth]{\it How is the geometry and topology of bulk physics encoded in the gauge theory?}

Recent advances in Schur operator technology, starting with \cite{CJR, Balasubramanian-et-al} and more recently developed in the series of articles \cite{RdMK-et-al}, have facilitated enormous strides toward answering these questions. For instance, it was convincingly argued in \cite{Balasubramanian:2004nb}, and later verified in great detail in \cite{RdMK-et-al}, that the fact that the giant graviton worldvolume is a compact space is encoded in the combinatorics of the Young diagrams that label the associated Schur operators. More precisely, any closed hypersurface (like the D3-brane worldvolume) must satisfy Gauss' 
law, thereby constraining how open strings may be attached to the D-brane. In the gauge theory, attaching open strings translates into adding a word of length $\sim\mathcal{O}(\sqrt{N})$ to the Schur polynomial corresponding to the giant or, equivalently, adding a box to a Young diagram. The Littlewood-Richardson rules that govern such additions precisely reproduce Gauss' law and consequently the topology of the spherical giant.

Geometry on the other hand is a {\it local} property of spacetime and if, as asserted by the 
gauge/gravity correspondence, the bulk spacetime and boundary gauge theory describe exactly the same physics with a different organization of degrees of freedom, this locality should also manifest on the boundary. In the first systematic study of this question, it was demonstrated - through a combinatorial tour de force - in \cite{Berenstein-shape} that the {\it shape} of a spherical D3-brane giant graviton can be read off from the spectrum of one loop anomalous dimensions of excitations of subdeterminant operators of the form $\displaystyle \mathcal{O}^{N-k}_{|D3\rangle} = 
\epsilon^{\mu_{1}\ldots\mu_{N}}\epsilon_{\rho_{1}\ldots\rho_{N}}\Phi^{\rho_{1}}_{\mu_{1}}
\cdots\Phi^{\rho_{N-k}}_{\mu_{N-k}}\delta^{\rho_{N-k+1}}_{\mu_{N-k+1}}\cdots
\delta^{\rho_{N}}_{\mu_{N}}$. Such excited operators are constructed by replacing one (or more) of 
the $\delta$'s with one (or more) words of the form $Z^{n}$. However, the combinatorics of these operators is, to say the least, formidable and the results obtained in \cite{Berenstein-shape} were restricted to {\it near maximal} sized giants. Here too, once it was realized that Schur polynomials (and their restrictions) furnish a more complete basis for giant graviton operators (and their excitations) \cite{CJR,RdMK-et-al,Balasubramanian-et-al}, rapid progress was made on many outstanding problems. 
These include:
\vspace{-0.1cm}
\begin{enumerate}[i)]
  \item
    Verification of the results reported in \cite{Berenstein-shape} and an extension ({\it a}) beyond the 
    near-maximal giant and ({\it b}) to multiple strings attached to the D-brane worldvolume 
    together with a dynamical mechanism for the emergence of the Chan-Paton factors for open 
    strings propagating on multiple membranes \cite{RdMK-et-al}.  
  \item
    A concrete construction of new $\frac{1}{2}-$BPS geometries \cite{RdMK-young-diagrams} 
    from coherent states of gravitons propagating on $AdS_{5}\times S^{5}$ through the study 
    of Schur polynomials with large $\mathcal{R}$-charge $\Delta\sim\mathcal{O}(N^{2})$ and even,
  \item
    A proposal for a mechanism of the emergence of the thermodynamic properties of gravity in the 
    presence of horizons, again through an analysis of heavy states with conformal dimension 
    $\sim\mathcal{O}(N^{2})$ in the dual gauge theory \cite{Babel, Minwalla-et-al}.
\end{enumerate}
\vspace{-0.1cm}
All in all, it is fair to say that the program to understand the emergence of spacetime in AdS/CFT has met with some success. Nevertheless, there remains much to do. Of the problems that remain, probably the most pressing is the question of how far beyond the 
$\frac{1}{2}-$BPS sector these results extend. This is, however, also one of the most difficult problems since, by definition, we would expect to lose much of the protection of supersymmetry and the powerful non-renormalization theorems that accompany it. 

On a more pragmatic level, one could well argue that the claim that spacetime geometry and topology are emergent properties of the gauge theory at large $N$ would be more convincing if said geometries and topologies were more, well, interesting than just the sphere\footnote{Although even a cursory glance at any of  \cite{RdMK-et-al,Balasubramanian-et-al} would be enough to convince the reader that there's nothing trivial about recovering the spherical geometry.}. For example, showing that Gauss' law is encoded in the combinatorics of the Young diagrams that label the Schur polynomials is a excellent step forward, but since it is a condition that must be satisfied by {\it any} compact worldvolume, by itself it is not a good characterization of topology. An obvious next step would be to understand how a
topological invariant such as {\it genus} is encoded in the gauge theory. The problem is that topologically and geometrically nontrivial giant graviton configurations are like the proverbial needle in the haystack: few and far between. More to the point, until very recently, there were no candidate dual operators to these giants in the literature.

The turnaround in this state of affairs came with the discovery of a new example of the AdS/CFT duality, this time between the type IIA superstring on $AdS_{4}\times \mathbb{CP}^{3}$ and an $\mathcal{N}=6$, super Chern-Simons theory on the 3-dimensional boundary of the AdS space - the so-called ABJM model \cite{ABJM,BLG}. While this new AdS$_{4}$/CFT$_{3}$ duality shares much in common with its more well-known and better understood higher dimensional counterpart 
(a well-defined perturbative expansion, integrability etc.), it is also sufficiently different that the hope that it will provide just as invaluable a testing ground as AdS$_{5}$/CFT$_{4}$ is not without justification. In particular, in a recent study of membranes in M-theory and their IIA decendants \cite{NT-giants}, a new class of giant gravitons with large angular momentum and a D0-brane charge was discovered with a toroidal worldvolume. More importantly, with the gauge theory in this case nearly as controlled as ${\mathcal N}=4$ super Yang-Mills theory, a class of $\frac{1}{2}-$BPS monopole operators has been mooted as the candidate duals to these giant torii in \cite{BP} by matching the energy of the quadratic fluctuations about the monopole configuration to that of the giant graviton. Of course, matching energies is a little like a ``3-sigma" event at a collider experiment: while nobody's booking tickets to Stockholm yet, it certainly points to something interesting going on. Much more work needs to be done to show how the full torus is recovered in the field theory. 

The situation is just as intriguing with respect to geometry. It is by now well known that giant gravitons on $AdS_{5}\times S^{5}$ come in two forms: both are spheres (one extended in the AdS space and one in the $S^{5}$), both are D3-branes and each is the Hodge dual of the other. Similarly, giant gravitons on $AdS_{4}\times \mathbb{CP}^{3}$ are expected to come in two forms also. The D2-brane ``AdS" giant graviton was constructed in \cite{NT-giants, HMPS}. This expands on the 2-sphere in $AdS_{4}$, is perturbatively stable and, apart from a non-vanishing coupling between the worldvolume gauge fields and the transverse fluctuations, exhibits a spectrum similar to that of the giant in $AdS_{5}$. The dual to this configuration - a D4-brane giant graviton wrapping some trivial cycle in the $\mathbb{CP}^{3}$ - has proven to be much more difficult to construct. This is due in no small part to its highly non-trivial geometry \cite{BT} and it is precisely this geometry, and the possibility of seeing it encoded in the ABJM gauge theory, that makes this giant so interesting. 

In this article we take the first steps toward extracting this geometry by constructing the D4-brane giant graviton in the type IIA string theory and studying its spectrum of small fluctuations. Our construction follows the methods developed in \cite{Mikhailov, HMP} for the giant graviton on $AdS_{5}\times T^{1,1}$ (and later extended to the maximal giant graviton\footnote{See also the recent works \cite{GLR,LPSS,HLP} for an independent analysis of the maximal giant graviton.} on $AdS_{4}\times\mathbb{CP}^{3}$ by two of us in \cite{MP}). By way of summary, guided by the structure of Schur polynomials in the ABJM model (see Section \ref{section - ABJM}), we formulate an ansatz for the D4-brane giant graviton and show that this solution minimizes the energy of the brane. We are also able to show how the giant grows with increasing momentum until, at maximal size, it ``factorizes" into two dibaryons, in excellent agreement with the factorization of the associated subdeterminant operator in the gauge theory.

%% file: 02-ABJM.tex
\section{Schurs and subdeterminants in ABJM theory} \label{section - ABJM}
\par{\emph{\textbf{Introduction to the ABJM model}}}
\smallskip

The ABJM model \cite{ABJM} is an $\mathcal{N}=6$ super Chern-Simons (SCS)-matter theory in 2+1-dimensions with a $U(N)_{k}\times U(N)_{-k}$ gauge group, and opposite level numbers $k$ and $-k$.  Aside from the gauge fields $A_{\mu}$ and $\hat{A}_{\mu}$, there are two sets of two chiral multiplets $(A_{i},\psi^{A_{i}}_{\alpha})$ and $(B_{i},\psi^{B_{i}}_{\alpha})$, corresponding to the chiral superfields $\mathcal{A}_{i}$ and $\mathcal{B}_{i}$ in $\mathcal{N}=2$ superspace, which transform in the $(N,\bar{N})$ and $(\bar{N},N)$ bifundamental representations respectively.

The ABJM superpotential takes the form
\begin{equation}
\mathcal{W} = \frac{2\pi}{k} \hspace{0.1cm} \epsilon^{ij}\epsilon^{kl} \hspace{0.15cm} \textrm{tr}
(\mathcal{A}_{i}\hspace{0.02cm}\mathcal{B}_{j}\hspace{0.02cm}\mathcal{A}_{k}\hspace{0.02cm}\mathcal{B}_{l}),
\end{equation}
which exhibits an explicit $SU(2)_{A}\times SU(2)_{B}$ $\mathcal{R}$-symmetry - the two $SU(2)$'s act on the doublets $(A_{1},A_{2})$ and $(B_{1},B_{2})$ respectively.  There is also an additional $SU(2)_{\mathcal{R}}$ symmetry, under which $(A_{1},B_{2}^{\dag})$ and $(A_{2},B_{1}^{\dag})$ transform as doublets, which enhances the symmetry group to $SU(4)_{\mathcal{R}}$ \cite{Klose-et-al}.

The scalar fields can be arranged into the multiplet $Y^{a}=(A_{1},A_{2},B_{1}^{\dag},B_{2}^{\dag})$, which transforms in the fundamental representation of $SU(4)_{\mathcal{R}}$, with hermitean conjugate $Y^{\dag}_{a}=(A_{1}^{\dag},A_{2}^{\dag},B_{1},B_{2})$.
The ABJM action can then be written as \cite{Klose-et-al,MZ-SCS}
\begin{eqnarray}
\nonumber & \!\! S = & \frac{k}{4\pi}\int{d^{3}x}~\textrm{tr}\left\{\epsilon^{\mu\nu\lambda}
\left(A_{\mu}\p_{\nu}A_{\lambda} + \frac{2i}{3}A_{\mu}A_{\nu}A_{\lambda} - \hat{A}_{\mu}\p_{\nu}\hat{A}_{\lambda} - \frac{2i}{3}\hat{A}_{\mu}\hat{A}_{\nu}\hat{A}_{\lambda}\right) \right. \\
\nonumber && \hspace{2.55cm} + \hspace{0.08cm} D_{\mu}^{\dag}Y^{\dag}_{a}D^{\mu}Y^{a}
+ \frac{1}{12}Y^{a}Y^{\dag}_{a}Y^{b}Y^{\dag}_{b}Y^{c}Y^{\dag}_{c}
+ \frac{1}{12}Y^{a}Y^{\dag}_{b}Y^{b}Y^{\dag}_{c}Y^{c}Y^{\dag}_{a} \\
&& \left. \hspace{2.5cm} - \hspace{0.08cm} \frac{1}{2}Y^{a}Y^{\dag}_{a}Y^{b}Y^{\dag}_{c}Y^{c}Y^{\dag}_{b} + \frac{1}{3}Y^{a}Y^{\dag}_{b}Y^{c}Y^{\dag}_{a}Y^{b}Y^{\dag}_{c}
+ \textrm{fermions}\right\}, \hspace{1.5cm}
\end{eqnarray}
where the covariant derivatives are defined as $D_{\mu}Y^{a} \equiv \p_{\mu}Y^{a} + iA_{\mu}Y^{a}  - iY^{a}\hat{A}_{\mu}$ and $D^{\dag}_{\mu}Y^{\dag}_{a} \equiv \p_{\mu}Y^{\dag}_{a} - iA_{\mu}Y^{\dag}_{a} + iY^{\dag}_{a}\hat{A}_{\mu}$. There are no kinetic terms associated with the gauge fields - they are dynamic degrees of freedom only by virtue of their coupling to matter.

The two-point correlation function for the free scalar fields in ABJM theory is\footnote{This two-point correlation function is the same as that quoted in \cite{NT-ppwaves} up to an overall $\tfrac{1}{4\pi}$ normalization.}
\begin{equation}
\langle \hspace{0.1cm} (Y^{a})^{\alpha}_{~~\gamma}(x_{1}) \hspace{0.15cm} (Y_{b}^{\dag})_{\beta}^{~~\epsilon}(x_{2}) \hspace{0.1cm} \rangle
= \frac{\delta^{\alpha}_{\beta} \hspace{0.15cm} \delta_{\gamma}^{\epsilon} \hspace{0.15cm} \delta^{a}_{b}}{|x_{1}-x_{2}|}.
\end{equation}
Note that the expression $|x_{1}-x_{2}|$ in the denominator is raised to the power of $2\Delta$ with $\Delta = \tfrac{1}{2}$ the conformal dimension of the ABJM scalar fields.

\par{\emph{\textbf{Schur polynomials and subdeterminants}}}
\smallskip

Schur polynomials and subdeterminant operators in the ABJM model cannot be constructed from individual scalar fields, as they are in the canonical case of $\mathcal{N} = 4$ super Yang-Mills (SYM) theory \cite{CJR,Balasubramanian-et-al}, since these fields are in the bifundamental representation of the gauge group and therefore carry indices in \emph{different} $U(N)$'s, which cannot be contracted.  However, it is possible, instead, to make use of composite scalar fields of the form\footnote{Operators constructed from the composite scalar fields $A_{1}A_{1}^{\dag}$, $A_{2}A_{2}^{\dag}$, $B^{\dag}_{1}B_{1}$ or $B^{\dag}_{2}B_{2}$ must be non-BPS as their conformal dimension cannot equal their $\mathcal{R}$-charge, which is zero.}
\begin{equation}
(Y^{a} \hspace{0.05cm} Y^{\dag}_{b})^{\alpha}_{\beta} = (Y^{a})^{\alpha}_{~~\gamma} \hspace{0.1cm}(Y^{\dag}_{b})_{\beta}^{~~\gamma},
\hspace{0.5cm} \textrm{with} \hspace{0.25cm} a \neq b,
\end{equation}
which carry indices in the \emph{same} $U(N)$.  We shall make use of the composite scalar field $Y^{1}Y^{\dag}_{3} = A_{1}B_{1}$ for definiteness.

Let us construct the Schur polynomial $\chi_{R}(A_{1}B_{1})$ of length $n$, with $R$ an irreducible representation of the permutation group $S_{n}$, which is labeled by a Young diagram with $n$ boxes:
\begin{eqnarray}
\nonumber && \hspace{-0.35cm} \chi_{R}\left(A_{1}B_{1}\right)
= \frac{1}{n!}\sum_{\sigma \hspace{0.05cm} \in \hspace{0.05cm} S_{n}} \hspace{0.1cm} \chi_{R}(\sigma) \hspace{0.1cm} \textrm{Tr}\left\{ \sigma (A_{1}B_{1} ) \right\}, \\
&& \hspace{-0.35cm} \textrm{with} \hspace{0.35cm}
 \textrm{Tr}\left\{ \sigma \left(A_{1}B_{1}\right) \right\} 
 \equiv 
\left(A_{1}B_{1}\right)^{\alpha_{1}}_{\alpha_{\sigma(1)}} \left(A_{1}B_{1}\right)^{\alpha_{2}}_{\alpha_{\sigma(2)}} \ldots \hspace{0.05cm} \left(A_{1}B_{1}\right)^{\alpha_{n}}_{\alpha_{\sigma(n)}}. \hspace{1.2cm}
\end{eqnarray}
This Schur polynomial is the character of $A_{1}B_{1}$ in the irreducible representation $R$ of the unitary group $U(N)$ associated with the same Young diagram via the Schur-Weyl duality. 

It was shown in \cite{Dey}, that by writing this Schur polynomials in terms of two separate permutations of the $A_{1}$'s and $B_{1}$'s:
\begin{eqnarray}
\nonumber && \hspace{-0.35cm}   \chi_{R}\left(A_{1}B_{1}\right)
= \frac{d_{R}}{(n!)^{2}}\sum_{\sigma, \rho \hspace{0.05cm} \in \hspace{0.05cm} S_{n}} \hspace{0.1cm} \chi_{R}(\sigma) \hspace{0.1cm} \chi_{R}(\rho) \hspace{0.1cm} 
\textrm{Tr}\left\{ \sigma \left(A_{1}\right) \rho \left(B_{1}\right) \right\}, \\
&& \hspace{-0.35cm} \textrm{with} \hspace{0.35cm} \textrm{Tr}\left\{ \sigma (A_{1}) \rho (B_{1} ) \right\} \equiv
\left(A_{1}\right)^{\alpha_{1}}_{~~~\beta_{\sigma(1)}} \ldots \hspace{0.05cm} \left(A_{1}\right)^{\alpha_{n}}_{~~~\beta_{\sigma(n)}}
\left(B_{1}\right)^{~~~\beta_{1}}_{\alpha_{\rho(1)}} \ldots \hspace{0.05cm} \left(B_{1}\right)^{~~~\beta_{n}}_{\alpha_{\rho(n)}},
\end{eqnarray}
the two point correlation function takes the form
\begin{equation}
\langle \hspace{0.1cm} \chi_{R}(A_{1}B_{1})(x_{1}) \hspace{0.20cm} \chi_{S}^{\dag}(A_{1}B_{1})(x_{2}) \hspace{0.1cm} \rangle
=  \frac{\left(f_{R}\right)^{2} \delta_{RS}}{(x_{1}-x_{2})^{2n}}
\hspace{0.6cm} \textrm{with} \hspace{0.4cm} f_{R} \equiv \frac{D_{R} \hspace{0.1cm} n! }{d_{R}}.
\end{equation}
Here $D_{R}$ and $d_{R}$ are the dimensions of the irreducible representations $R$ of the unitary group $U(N)$ and the permutation group $S_{n}$ respectively.  The two factors of $f_{R}$ are a result of the fact that two permutations are now necessary to treat the scalar fields $A_{1}$ and $B_{1}$ in the composite scalar field $A_{1}B_{1}$ separately\footnote{We would like to thank the anonymous referee for pointing out a flaw in our original argument.}.

These Schur polynomials are therefore orthogonal with respect to two-point correlation function in free ABJM theory \cite{Dey}.  They are also $\tfrac{1}{2}$-BPS and have conformal dimension $\Delta = n$ equal to their $\mathcal{R}$-charge.  Normalised Schurs $(f_{R})^{-1} \hspace{0.05cm} \chi_{R}(A_{1}B_{1})$ therefore form an orthonormal basis for this $\tfrac{1}{2}$-BPS sector of ABJM theory.

We shall focus on the special class of Schur polynomials associated with the totally anti-symmetric representation of the permutation group $S_{n}$ (labeled by a single column with $n$ boxes).   These are proportional to subdeterminant operators:
\begin{equation}
\chi_{^{^{\yng(1,1)}_{\vdots}}_{\yng(1)}}(A_{1}B_{1}) \hspace{0.05cm} \propto \hspace{0.05cm} \mathcal{O}^{\textrm{subdet}}_{n}\left(A_{1}B_{1}\right) = \frac{1}{n!} \hspace{0.1cm} \epsilon_{\alpha_{1}\ldots\alpha_{n}\alpha_{n+1}\ldots\alpha_{N}} \hspace{0.1cm} \epsilon^{\beta_{1}\ldots\beta_{n}\alpha_{n+1}\ldots\alpha_{N}} \hspace{0.1cm} \left(A_{1}B_{1}\right)^{\alpha_{1}}_{\beta_{1}} \ldots \hspace{0.05cm} \left(A_{1}B_{1}\right)^{\alpha_{n}}_{\beta_{n}}.
\end{equation}
As a result of the composite nature of the scalar fields from which they are constructed, these subdeterminants in ABJM theory factorize at maximum size $n=N$ into the product of two full determinant operators
\begin{equation}
\mathcal{O}^{\textrm{subdet}}_{N}\left(A_{1}B_{1}\right)
= \left(\det{A_{1}}\right)\left(\det{B_{1}}\right),
\end{equation}
with
\begin{eqnarray}
&& \det{A_{1}} \equiv \frac{1}{N!} \hspace{0.15cm} \epsilon_{\alpha_{1}\ldots\alpha_{N}} \hspace{0.15cm} \epsilon^{\beta_{1}\ldots\beta_{N}}
\hspace{0.1cm} (A_{1})^{\alpha_{1}}_{~~\beta_{1}} \ldots (A_{1})^{\alpha_{N}}_{~~\beta_{N}} \\
&& \det{B_{1}} \equiv \frac{1}{N!} \hspace{0.15cm} \epsilon^{\alpha_{1}\ldots\alpha_{N}} \hspace{0.15cm} \epsilon_{\beta_{1}\ldots\beta_{N}}
\hspace{0.1cm} (B_{1})_{\alpha_{1}}^{~~\beta_{1}} \ldots (B_{1})_{\alpha_{N}}^{~~\beta_{N}},
\end{eqnarray}
which are varieties of ABJM dibaryons.

This subdeterminant operator $\mathcal{O}^{\textrm{subdet}}_{n}(A_{1}B_{1})$ is dual to a D4-brane giant graviton, extended and moving on the complex projective space $\mathbb{CP}^{3}$.  The fact that it has a maximum size is merely a consequence of the compact nature of the space in which it lives.  We expect the worldvolume of the giant graviton to pinch off as its size increases, until it factorizes into two distinct D4-branes, each of which wraps a holomorphic cycle $\mathbb{CP}^{2}\subset\mathbb{CP}^{3}$ (they intersect on a $\mathbb{CP}^{1}$). These are dual to full determinant operators (see \cite{MP,GLR} for a description of dibaryons and the dual topologically stable D4-brane configurations.).

%% file: 03-PointGraviton.tex
\section{A point particle rotating on $\mathbb{CP}^{3}$} \label{section - point graviton}

The type IIA $AdS_{4}\times\mathbb{CP}^{3}$ background spacetime and our parametrization of the complex projective space are described in detail in Appendix \ref{appendix - background}.  Let us consider a point particle with mass $M$ moving along the  
$\chi(t) \equiv \tfrac{1}{2} \left( \psi + \phi_{1} + \phi_{2} \right)$ fibre direction in the complex projective space (a similar system was discussed in \cite{Pirrone}).  The induced metric on the worldline of the particle (situated at the centre of the $AdS_{4}$ space) can be obtained from the metric (\ref{metric-orig})
by setting $r=0$ and also $\psi(t) = 2 \hspace{0.05cm} \chi(t) + \phi_{1} + \phi_{2}$ with $\zeta$, $\theta_{i}$ and $\phi_{i}$ all constant. Hence the induced metric takes the form
\begin{equation}
ds^{2} = -R^{2}\left\{1 - \dot{\chi}^{2} \sin^{2}{(2\zeta)}\right\} dt^{2}.
\end{equation}

The action of the point particle is given by
\begin{equation}
S_{^{\textrm{point}}_{\textrm{particle}}} = -M \int{\sqrt{\left| ds^{2} \right| }} = \int{dt \hspace{0.2cm} L} \hspace{0.8cm} \textrm{with} \hspace{0.3cm} L = - MR\hspace{0.075cm} \sqrt{1 - \dot{\chi}^{2}\sin^{2}(2\zeta)},
\end{equation}
which is dependent on the constant value of $\zeta$ at which the particle is positioned.  The conserved momentum associated with the angular coordinate $\chi$ is 
\begin{equation}
P_{\chi} = \frac{MR \hspace{0.1cm} \dot{\chi}}{\sqrt{1 - \dot{\chi}^{2}\sin^{2}(2\zeta)}} \hspace{1.0cm} \Longrightarrow \hspace{0.5cm} \dot{\chi} = \frac{P_{\chi}}{\sin(2\zeta)\sqrt{P_{\chi}^{2} + M^{2}R^{2}\sin^{2}(2\zeta)}},
\end{equation}
from which it is possible to determine the energy $H = P_{\chi}\dot{\chi} - L$ of the point particle as a function of the momentum $P_{\chi}$:
\begin{equation}
H = \frac{1}{\sin(2\zeta)} \sqrt{P_{\chi}^{2} + M^{2}R^{2}\sin^{2}(2\zeta)}.
\end{equation}
This energy attains its minimum value $H = \sqrt{P_{\chi}^{2} + M^{2}R^{2}}$ when $\zeta = \tfrac{\pi}{4}$. The point graviton is associated with the massless limit $M \rightarrow 0$ in which the energy $H$ becomes equal to its angular momentum $P_{\chi}$, indicating a BPS state.

%% file: 04-GiantGraviton.tex
\section{The $\mathbb{CP}^{3}$ giant graviton} \label{section - giant graviton}

We may associate the four homogeneous coordinates $z^{a}$ of $\mathbb{CP}^{3}$ with the ABJM scalar fields in the multiplet $Y^{a} = (A_{1},A_{2},\bar{B}_{1},\bar{B}_{2})$.
Using the parameterization (\ref{cp3 parameterization}), the composite scalar fields $A_{i}B_{j}$ are therefore dual to
\begin{eqnarray}
& z^{1}\hspace{0.025cm}\bar{z}^{3} = \tfrac{1}{2}\sin\left(2\zeta\right)\sin{\tfrac{\theta_{1}}{2}}\sin{\tfrac{\theta_{2}}{2}} \hspace{0.075cm}
e^{\frac{1}{2}i(\psi - \phi_{1} - \phi_{2})} & \hspace{0.5cm} \longrightarrow \hspace{0.2cm} A_{1}B_{1} \hspace{0.5cm} \\
& z^{2}\hspace{0.025cm}\bar{z}^{4} \hspace{0.05cm} = \tfrac{1}{2}\sin\left(2\zeta\right)\cos{\tfrac{\theta_{1}}{2}}\cos{\tfrac{\theta_{2}}{2}} \hspace{0.075cm}
e^{\frac{1}{2}i(\psi + \phi_{1} + \phi_{2})} & \hspace{0.5cm} \longrightarrow \hspace{0.2cm} A_{2}B_{2} \hspace{0.5cm} \\
& z^{2}\hspace{0.025cm}\bar{z}^{3} = \tfrac{1}{2}\sin\left(2\zeta\right)\cos{\tfrac{\theta_{1}}{2}}\sin{\tfrac{\theta_{2}}{2}} \hspace{0.075cm}
e^{\frac{1}{2}i(\psi + \phi_{1} - \phi_{2})} & \hspace{0.5cm} \longrightarrow \hspace{0.2cm} A_{2}B_{1} \hspace{0.5cm} \\
& z^{1}\hspace{0.025cm}\bar{z}^{4} = \tfrac{1}{2}\sin\left(2\zeta\right)\sin{\tfrac{\theta_{1}}{2}}\cos{\tfrac{\theta_{2}}{2}} \hspace{0.075cm}
e^{\frac{1}{2}i(\psi - \phi_{1} + \phi_{2})} & \hspace{0.5cm} \longrightarrow \hspace{0.2cm} A_{1}B_{2}. \hspace{0.5cm}
\end{eqnarray}
Aside from the additional factor of $\tfrac{1}{2}\sin{\left(2\zeta\right)}$, these combinations bear an obvious resemblance to the parameterization \cite{Candelas} of the base manifold $T^{1,1}$ of a cone $\mathcal{C}$ in $\mathbb{C}^{4}$.
We may therefore adapt the ansatz of \cite{Mikhailov,HMP}, which describes a D3-brane giant graviton on $AdS_{5}\times T^{1,1}$, to construct a D4-brane giant graviton on $AdS_{4}\times\mathbb{CP}^{3}$.

\subsection{Giant graviton ansatz}

Our ansatz for a D4-brane giant graviton on $AdS_{4}\times\mathbb{CP}^{3}$, which is positioned at the centre of the anti-de Sitter space, takes the form
\begin{equation} \label{gg-constraint}
\sin{\left(2\zeta\right)}\sin{\tfrac{\theta_{1}}{2}}\sin{\tfrac{\theta_{2}}{2}} = \sqrt{1-\alpha^{2}},
\end{equation}
where the constant $\alpha \in \left[0,1\right]$ describes the size of the giant.
Motion is along the angular direction $\chi \equiv \tfrac{1}{2}\left(\psi - \phi_{1} - \phi_{2}\right)$, as in the case of the D2-brane dual giant graviton on $AdS_{4}\times\mathbb{CP}^{3}$ studied in \cite{HMPS}.
This is also analogous to the direction of motion of the giant graviton \cite{Mikhailov,HMP} on $AdS_{5}\times T^{1,1}$, up to a constant multiple, which we have included to account for the difference between the conformal dimensions of the scalar fields in Klebanov-Witten and ABJM theory.

Since this giant graviton is extended and moving on the complex projective space, it is confined to the background $\mathbb{R}\times\mathbb{CP}^{3}$ with metric
\begin{equation}
ds^{2} = R^{2}\left\{-dt^{2} + ds_{\textrm{radial}}^{2} + ds_{\textrm{angular}}^{2}\right\},
\end{equation}
where the radial and angular parts of the metric are given by
\begin{eqnarray}
&& \!\! ds_{\textrm{radial}}^{2}
  = 4 \, d\zeta^{2} + \cos^{2}{\zeta} \; d\theta_{1}^{2} + \sin^{2}{\zeta} \hspace{0.075cm} d\theta_{2}^{2} \\
\nonumber && \!\! ds_{\textrm{angular}}^{2}
 = \cos^{2}{\zeta}\sin^{2}{\zeta}
  \left[d\psi + \cos{\theta_{1}} \; d\phi_{1} + \cos{\theta_{2}} \; d\phi_{2}\right]^{2} \\
&& \hspace{1.75cm} + \cos^{2}{\zeta}\sin^{2}{\theta_{1}} \; d\phi_{1}^{2} + \sin^{2}{\zeta}\sin^{2}{\theta_{2}} \; d\phi_{2}^{2}.
\end{eqnarray}
Only the 2-form and 6-form field strengths (\ref{F2-orig}) and (\ref{F6-orig}) remain non-trivial.

Let us now define new sets of radial coordinates $z_{i} \equiv \cos^{2}{\tfrac{\theta_{i}}{2}}$ and $y \equiv \cos{\left(2\zeta\right)}$, and angular coordinates
$\chi \equiv \tfrac{1}{2}\left(\psi - \phi_{1} - \phi_{2}\right)$ and $\varphi_{i} \equiv \phi_{i}$ in terms of which the radial and angular metrics can be written as follows:
\begin{eqnarray}
&& ds_{\textrm{radial}}^{2} = \frac{dy^{2}}{\left(1-y^{2}\right)}
+ \frac{1}{2}\left(1+y\right)\frac{dz_{1}^{2}}{z_{1}\left(1-z_{1}\right)}
+ \frac{1}{2}\left(1-y\right)\frac{dz_{2}^{2}}{z_{2}\left(1-z_{2}\right)} \hspace{1.5cm} \\
\nonumber && \\
\nonumber && ds_{\textrm{angular}}^{2} = \left(1-y^{2}\right)
\left[d\chi + z_{1} \hspace{0.075cm} d\varphi_{1} + z_{2} \hspace{0.075cm} d\varphi_{2}\right]^{2} \\
&& \hspace{2.0cm} + \hspace{0.05cm} 2\left(1+y\right)z_{1}\left(1-z_{1}\right) d\varphi_{1}^{2}
+ 2\left(1-y\right)z_{2}\left(1-z_{2}\right) d\varphi_{2}^{2}.
\end{eqnarray}
The constant dilaton still satisfies $e^{2\Phi} = \tfrac{4R^{2}}{k^{2}}$, while the non-trivial field strength forms on $\mathbb{R}\times \mathbb{CP}^{3}$ are given by
\begin{eqnarray}
\nonumber && F_{2} = \tfrac{1}{2}k\left\{dy \wedge
\left[d\chi + z_{1} \hspace{0.075cm} d\varphi_{1} + z_{2} \hspace{0.075cm} d\varphi_{2}\right]
        + \left(1+y\right) dz_{1}\wedge d\varphi_{1} - \left(1-y\right) dz_{2}\wedge d\varphi_{2} \right\} \\
&& \\
&& F_{6} = \tfrac{3}{2}kR^{4} \left(1-y^{2}\right) dy \wedge dz_{1} \wedge dz_{2} \wedge d\chi \wedge d\varphi_{1} \wedge d\varphi_{2}. \label{F6}
\end{eqnarray}
Our giant graviton ansatz
\begin{equation} \label{gg-constraint-yz}
\left(1-y^{2}\right)\left(1-z_{1}\right)\left(1-z_{2}\right) = 1-\alpha^{2}
\end{equation}
describes a surface in 3D radial $(y,z_{1},z_{2})$ space. (Horizontal slices parallel to the $z_{1}z_{2}$-plane, at fixed $y \in [-\alpha,\alpha]$, are shifted hyperbolae.)  The maximal giant graviton $\alpha=1$ can be viewed as part of a rectangular box with sides $z_{1}=1$, $z_{2}=1$ and $y = \pm 1$.
Note that the top and bottom sides $y=\pm 1$ result in coordinate singularities\footnote{When $y=1$ (or $y = -1$) all dependence on the second 2-sphere (or first 2-sphere) disappears.} and therefore yield no contribution to the worldvolume of the maximal giant graviton.


\begin{figure}[htb!]
\begin{center}
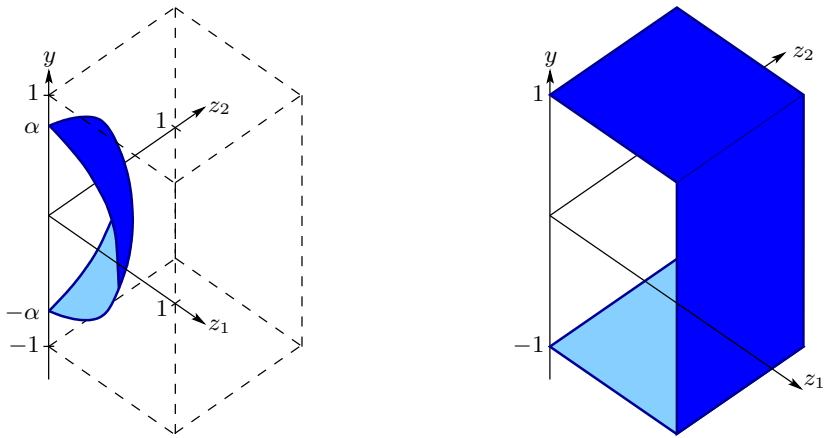
\caption{A sketch of the submaximal and maximal $\mathbb{CP}^{3}$ giants in radial $(y,z_{1},z_{2})$ space.}
\label{giant-radial}
\end{center}
\end{figure}


This ansatz for the giant graviton on $AdS_{4}\times\mathbb{CP}^{3}$ is similar to the ansatz \cite{HMP} for the giant graviton on $AdS_{5}\times T^{1,1}$.  The 5-dimensional compact space $T^{1,1}$, in which this D3-brane is embedded, consists of two 2-spheres and a non-trivial $U(1)$ fibre - motion is along the fibre direction. In the case of our D4-brane giant embedded in the 6-dimensional compact space $\mathbb{CP}^{3}$, there is an additional radial coordinate $y$, which controls the (now variable) related sizes of the two 2-spheres in the complex projective space.  In both cases, the giant graviton splits up into two pieces at maximal size.

To obtain both halves of the maximal giant graviton as a limiting case $\alpha \rightarrow 1$ of the submaximal giant graviton, we could parameterize the two regions $z_{1} \leq z_{2}$ and $z_{1} \geq z_{2}$ separately.  Note that, as result of the symmetry of the problem, these would yield identical contributions to the D4-brane action.  However, for convenience, we shall simply parameterize the full worldvolume of the submaximal giant graviton using the coordinates $\sigma^{a} = (t,y,z_{1},\varphi_{1},\varphi_{2})$
 with ranges
 \begin{equation}
 y \in [-\alpha,\alpha], \hspace{0.8cm} z_{1}  \in \left[ 0, \hspace{0.05cm} \frac{\alpha^{2}-y^{2}}{1-y^{2}} \right]  \hspace{0.5cm} \textrm{and}  \hspace{0.5cm} \varphi_{i} \in [0,2\pi].
\end{equation}

\subsection{D4-brane action}

The D4-brane action $S_{\textrm{D4}} = S_{\textrm{DBI}} + S_{\textrm{WZ}}$, which describes the dynamics of our giant graviton, consists of Dirac-Born-Infeld (DBI) and Wess-Zumino (WZ) terms:
\begin{equation} \label{DBI-general}
S_{\textrm{DBI}} = -T_{4}\int_{\Sigma} ~ d^{5}\sigma ~ e^{-\Phi}\sqrt{-\det{ \left( \mathcal{P}\left[g\right] + 2\pi F \right) }},
\end{equation}
and
\begin{equation} \label{WZ-general}
S _{\textrm{WZ}}  
= \pm T_{4} \int_{\Sigma} ~ \left\{\mathcal{P}\left[C_{5}\right] + \mathcal{P}\left[C_{3}\right] \wedge \left(2 \pi F \right) 
+ \frac{1}{2} \hspace{0.05cm} \mathcal{P}\left[C_{1}\right] \wedge \left(2\pi F\right) \wedge \left( 2\pi F\right) \right\},
\end{equation}
with $T_{4} \equiv \frac{1}{\left(2\pi\right)^{4}}$ the tension.  Here we have included the possibility of a non-trivial worldvolume gauge field $F$. Since the form field $C_{3}$ has components only in $AdS_{4}$, the corresponding term in the WZ action vanishes when pulled-back to the worldvolume $\Sigma$ of the giant graviton - an object extended only in $\mathbb{CP}^{3}$.

Now, it is consistent (as an additional specification in our giant graviton ansatz) to turn off all worldvolume fluctuations.  Note that these should be included when we turn our attention to the spectrum of small fluctuations.  Hence the D4-brane action becomes
\begin{equation}
S_{\textrm{D4}} 
= -T_{4}\int_{\Sigma} ~ d^{5}\sigma ~ e^{-\Phi}\sqrt{-\det{ \left( \mathcal{P}\left[g\right] \right) }} 
~ \pm T_{4} \int_{\Sigma} ~ \mathcal{P}\left[C_{5}\right].
\end{equation}

\medskip
\par{\emph{\textbf{Dirac-Born-Infeld action}}}
\smallskip

The induced radial metric on the worldvolume of the giant graviton can be obtained by setting $z_{2}\left(z_{1}\right)$ from the constraint (\ref{gg-constraint}). Hence
\begin{eqnarray}
\nonumber && \!\! ds_{^{\textrm{ind}}_{\textrm{rad}}}^{2}
= \frac{\left[ \left(1-y^{2}\right)z_{2} + 2y^{2}\left(1-y\right) \left(1-z_{2}\right) \right]}{2\left(1-y^{2}\right)^{2} z_{2}} \hspace{0.075cm} dy^{2}
+ \frac{2y\left(1-y\right)\left(1-z_{2}\right)}{\left(1-y^{2}\right)\left(1-z_{1}\right)^{2} z_{2}} \hspace{0.075cm} dy \hspace{0.05cm} dz_{1} \\
&& \hspace{1.40cm} + \hspace{0.075cm} \frac{\left[\left(1+y\right)\left(1-z_{1}\right)z_{2} + \left(1-y\right)z_{1}\left(1-z_{2}\right)\right]}
{2z_{1}\left(1-z_{1}\right)^{2}z_{2}} \hspace{0.075cm} dz_{1}^{2}.
\end{eqnarray}
The determinant in the coordinates $(y,z_{1})$ is then given by
\begin{equation}
\det g_{^{\textrm{ind}}_{\textrm{rad}}} = 
\frac{ \left[\tfrac{1}{2}\left(1+y\right)\left(1-z_{1}\right) + \tfrac{1}{2}\left(1+y\right)\left(1-z_{2}\right) - \left(1-\alpha^{2}\right)\right]}{z_{1} \left(1-z_{1}\right)^{2} z_{2}}.
\end{equation}

The temporal and angular part of the induced metric on the worldvolume of the giant graviton takes the form
\begin{eqnarray}
\nonumber && \!\! ds_{^{\textrm{ ind}} _{t,\hspace{0.05cm}\textrm{ang}}}^{2}
= - \hspace{0.075cm} dt^{2} + \left(1-y^{2}\right)
\left[\dot{\chi} \hspace{0.075cm} dt + z_{1} \hspace{0.075cm}d\varphi_{1} + z_{2}\hspace{0.075cm}d\varphi_{2}\right]^{2} 
+ \hspace{0.075cm} 2 z_{1} \left(1-z_{1}\right) d\varphi_{1}^{2} + 2 z_{2} \left(1-z_{2}\right) d\varphi_{2}^{2}, \hspace{1.0cm}
\end{eqnarray}
which has the following determinant
\begin{eqnarray}
\nonumber && \hspace{-0.35cm} \det g_{^{\textrm{ ind}} _{t,\hspace{0.05cm}\textrm{ang}}} = - \left\{ \left(C_{\textrm{ang}}\right)_{11}  - \dot{\chi}^{2} \left[\det{g}_{\textrm{ang}}\right] \right\} \\
&& \hspace{-0.35cm} \hspace{1.625cm} = - \hspace{0.075cm} 4 \left(1-y^{2}\right)z_{1}z_{2} \\
\nonumber && \hspace{-0.35cm} \hspace{2.1cm} \times \left\{ \left[\tfrac{1}{2}\left(1+y\right)\left(1-z_{1}\right) + \tfrac{1}{2}\left(1+y\right)\left(1-z_{2}\right) - \left(1-\alpha^{2}\right)\right] + \left(1-\dot{\chi}^{2}\right) \left(1-\alpha^{2}\right) \right\}.
\end{eqnarray}

The determinant of the pullback of the metric to the worldvolume of the giant graviton in the coordinates $(t,y,z_{1},\varphi_{1},\varphi_{2})$ is therefore given by
\begin{eqnarray}
&& \!\! \det{\mathcal{P}\left[g\right]} = -\frac{4R^{10}}{\left(1-z_{1}\right)^{2}}
\left[\tfrac{1}{2}\left(1+y\right)\left(1-z_{1}\right) + \tfrac{1}{2}\left(1-y \right)\left(1-z_{2}\right) - \left(1-\alpha^{2}\right)\right]^{2} \\
\nonumber && \hspace{4.0cm} \times \left\{1 + \frac{\left(1-\dot{\chi}^{2}\right)\left(1-\alpha^{2}\right)}
{\left[\tfrac{1}{2}\left(1+y\right)\left(1-z_{1}\right) + \tfrac{1}{2}\left(1-y \right)\left(1-z_{2}\right) - \left(1-\alpha^{2}\right)\right]}\right\},
\end{eqnarray}
while $e^{-\Phi} = \tfrac{k}{2R}$.  Integrating over $\varphi_{1}$ and $\varphi_{2}$, we obtain the DBI action
\begin{equation}
S_{\textrm{DBI}} = \int{dt} \hspace{0.15cm} L_{\textrm{DBI}} \hspace{0.8cm} \textrm{with} \hspace{0.4cm} L_{\textrm{DBI}} = \int_{-\alpha}^{\alpha} dy \int_{0}^{\frac{\alpha^{2}-y^{2}}{1-y^{2}}} dz_{1} \hspace{0.2cm} \mathcal{L}_{\textrm{DBI}}(y,z_{1})
\end{equation}
associated with the radial DBI Lagrangian density
\begin{eqnarray}
\nonumber && \mathcal{L}_{\textrm{DBI}}(y,z_{1}) = -\frac{N}{2} \frac{1}{\left(1-z_{1}\right)}
\left[\tfrac{1}{2}\left(1+y\right)\left(1-z_{1}\right) + \tfrac{1}{2}\left(1-y \right)\left(1-z_{2}\right) - \left(1-\alpha^{2}\right)\right] \hspace{0.6cm} \\
&& \hspace{-0.25cm} \hspace{2.725cm} \times \sqrt{1 + \frac{\left(1-\dot{\chi}^{2}\right)\left(1-\alpha^{2}\right)}
{\left[\tfrac{1}{2}\left(1+y\right)\left(1-z_{1}\right) + \tfrac{1}{2}\left(1-y \right)\left(1-z_{2}\right) - \left(1-\alpha^{2}\right)\right]}}, \hspace{1.5cm}
\end{eqnarray}
where $z_{2}(z_{1}) = 1 - \frac{(1-\alpha^{2})}{(1-y^{2})(1-z_{1})}$
and the ABJM duality associates the rank $N$ of the product gauge group with the flux $N \equiv \frac{kR^{4}}{2\pi^{2}}$ of the 6-form field strength through the complex projective space.

\medskip
\par{\emph{\textbf{Wess-Zumino action}}}
\smallskip

In order to calculate the WZ action, we need to determine the 5-form field $C_{5}$ associated with the 6-form field strength $F_{6}=dC_{5}$. (The former is only defined up to an exact form of integration.)  Usually we would change to orthogonal radial worldvolume coordinates $(\alpha,u,v)$ and then integrate $F_{6}$ on $\alpha$ subject to the condition $C_{5}(\alpha=0) = 0$.  However, in this case, it is not immediately obvious how to determine $u$ and $v$, so we must proceed via an alternative route.

Consider the 5-form field
\begin{eqnarray} \label{5-form-yzcoords}
&& \hspace{-0.25cm} C_{5} = \tfrac{1}{2}kR^{4}
\left\{ y \left(1-y^{2}\right) dz_{1} \wedge dz_{2} - \left(1-y\right) z_{1} \hspace{0.075cm} dy \wedge dz_{2} + \left(1+y\right) z_{2} \hspace{0.075cm} dy \wedge dz_{1}\right\} \hspace{1.5cm} \\
\nonumber && \hspace{-0.25cm} \hspace{11.0cm} \wedge \hspace{0.075cm} d\chi  \wedge d\varphi_{1}  \wedge d\varphi_{2},
\end{eqnarray}
which satisfies both $F_{6}=dC_{5}$ and $C_{5}(y = z_{1} = z_{2}=0) = 0$. When pulled back to the worldvolume of the giant graviton, this becomes
\begin{eqnarray}
&& \hspace{-0.25cm} \mathcal{P}\left[C_{5}\right] = \frac{kR^{4} \hspace{0.025cm} \dot{\chi}}{\left(1-z_{1}\right)} \hspace{0.075cm}
\left[\tfrac{1}{2}\left(1+y\right)\left(1-z_{1}\right) + \tfrac{1}{2}\left(1-y\right)\left(1-z_{2}\right) - \left(1-\alpha^{2}\right)\right] \\
\nonumber && \hspace{-0.25cm} \hspace{10.0cm} dt \hspace{0.025cm} \wedge dy \wedge dz_{1} \wedge d\varphi_{1} \wedge d\varphi_{2}, 
\end{eqnarray}
where $z_{2}(z_{1})$ follows directly from the giant graviton constraint (\ref{gg-constraint}). The WZ action is therefore given by
\begin{eqnarray}
\nonumber && \hspace{-0.25cm}S_{\textrm{WZ}} = \int{dt} \hspace{0.15cm} L_{\textrm{WZ}} \hspace{0.8cm} \textrm{with} \hspace{0.4cm} L_{\textrm{WZ}} = \int_{-\alpha}^{\alpha} dy \int_{0}^{\frac{\alpha^{2}-y^{2}}{1-y^{2}}} dz_{1} \hspace{0.2cm}
\mathcal{L}_{\textrm{WZ}}(y,z_{1}), \\
\end{eqnarray}
with radial WZ Lagrangian density
\begin{equation}
\mathcal{L}_{\textrm{WZ}}(y,z_{1}) = \pm \frac{N}{2}\frac{\dot{\chi}}{\left(1-z_{1}\right)}
\left[\tfrac{1}{2}\left(1+y\right)\left(1-z_{1}\right) + \tfrac{1}{2}\left(1-y \right)\left(1-z_{2}\right) - \left(1-\alpha^{2}\right)\right],
\end{equation}
where, again, $z_{2}(z_{1}) = 1 - \frac{(1-\alpha^{2})}{(1-y^{2})(1-z_{1})}$.  The $\pm$ distinguishes between branes and anti-branes.  We shall confine our attention to the positive sign, indicative of a D4-brane.

\medskip
\par{\emph{\textbf{Full D4-brane action}}}
\smallskip

We can combine the DBI and WZ terms in the action to obtain the D4-brane action
\begin{eqnarray}  \label{eq:Lagrangian}
&& \hspace{-0.25cm}S_{\textrm{D4}} =  \int{dt} \hspace{0.15cm} L_{\textrm{D4}} \hspace{0.8cm} \textrm{with} \hspace{0.4cm} 
L_{\textrm{D4}} =  \int_{-\alpha}^{\alpha} dy \int_{0}^{\frac{\alpha^{2}-y^{2}}{1-y^{2}}} dz_{1} \hspace{0.2cm} \mathcal{L}_{\textrm{D4}}(y,z_{1})
\end{eqnarray}
associated with the radial Lagrangian density
\begin{eqnarray} \label{eq:LagrangianDensity}
&& \mathcal{L}_{\textrm{D4}}(y,z_{1}) = -\frac{N}{2} \frac{1}{\left(1-z_{1}\right)}
\left[\tfrac{1}{2}\left(1+y\right)\left(1-z_{1}\right) + \tfrac{1}{2}\left(1-y \right)\left(1-z_{2}\right) - \left(1-\alpha^{2}\right)\right] \hspace{0.6cm} \\
\nonumber && \hspace{-0.25cm} \hspace{2.55cm} \times \left\{ \sqrt{1 + \frac{\left(1-\dot{\chi}^{2}\right)\left(1-\alpha^{2}\right)}
{\left[\tfrac{1}{2}\left(1+y\right)\left(1-z_{1}\right) + \tfrac{1}{2}\left(1-y \right)\left(1-z_{2}\right) - \left(1-\alpha^{2}\right)\right]}} - \dot{\chi} \right\}, \hspace{1.0cm}
\end{eqnarray}
where $z_{2}(z_{1}) = 1 - \frac{(1-\alpha^{2})}{(1-y^{2})(1-z_{1})}$ and $N \equiv \frac{kR^{4}}{2\pi^{2}}$ denotes the flux of the 6-form field strength through the complex projective space.

\subsection{Energy and momentum}

The conserved momentum conjugate to the coordinate $\chi$ takes the form
\begin{equation}
\label{eq:Momentum}
P_{\chi} = \int_{-\alpha}^{\alpha} dy \int_{0}^{\frac{\alpha^{2}-y^{2}}{1-y^{2}}} dz_{1} \hspace{0.2cm} \mathcal{P}_{\chi}(y,z_{1}),
\end{equation}
written in terms of the momentum density
\begin{eqnarray} \label{eq:MomentumDensity}
\nonumber && \mathcal{P}_{\chi}(y,z_{1}) = \frac{N}{2} \frac{1}{\left(1-z_{1}\right)}
\left[\tfrac{1}{2}\left(1+y\right)\left(1-z_{1}\right) + \tfrac{1}{2}\left(1-y \right)\left(1-z_{2}\right) - \left(1-\alpha^{2}\right)\right] \hspace{0.6cm} \\
&& \hspace{-0.25cm} \hspace{2.3cm} \times \left\{\frac{\frac{\left(1-\alpha^{2}\right)\dot{\chi}}{\left[\tfrac{1}{2}\left(1+y\right)\left(1-z_{1}\right) + \tfrac{1}{2}\left(1-y \right)\left(1-z_{2}\right) - \left(1-\alpha^{2}\right)\right] }}{\sqrt{1 + \frac{\left(1-\dot{\chi}^{2}\right)\left(1-\alpha^{2}\right)}
{\left[\tfrac{1}{2}\left(1+y\right)\left(1-z_{1}\right) + \tfrac{1}{2}\left(1-y \right)\left(1-z_{2}\right) - \left(1-\alpha^{2}\right)\right] }}} + 1 \right\}.
\end{eqnarray}

The energy $H = P_{\chi}\dot{\chi} - L$ of this D4-brane configuration can hence be determined as a function of its size $\alpha$ and the angular velocity $\dot{\chi}$ as follows:
\begin{equation}
\label{eq:Energy}
H =  \int_{-\alpha}^{\alpha} dy \int_{0}^{\frac{\alpha^{2}-y^{2}}{1-y^{2}}} dz_{1} \hspace{0.2cm} \mathcal{H}(y,z_{1})
\end{equation}
with 
\begin{eqnarray} \label{eq:EnergyDensity}
\mathcal{H}(y,z_{1}) = \frac{N}{2}\frac{1}{\left(1-z_{1}\right)}
\frac{\left[\tfrac{1}{2}\left(1+y\right)\left(1-z_{1}\right) + \tfrac{1}{2}\left(1-y\right)\left(1-z_{2}\right)\right]}
{\sqrt{1 + \frac{\left(1-\dot{\chi}^{2}\right)\left(1-\alpha^{2}\right)}
{\left[\tfrac{1}{2}\left(1+y\right)\left(1-z_{1}\right) + \tfrac{1}{2}\left(1-y \right)\left(1-z_{2}\right) - \left(1-\alpha^{2}\right)\right]}}}
\end{eqnarray}
the Hamiltonian density.  Here $z_{2}(z_{1}) = 1 - \frac{(1-\alpha^{2})}{(1-y^{2})(1-z_{1})}$ is an implicit function of the worldvolume coordinates $y$ and $z_{1}$, so that we can write
$\left(1-y\right)\left(1-z_{2}\right) = \tfrac{1 - \alpha^{2}}{(1 + y )(1 - z_{1})}$, a combination ubiquitous in the above expressions.

Note that the first contribution to the momentum is due to angular motion along the $\chi$ direction. At maximal size $\alpha = 1$, the D4-brane is no longer moving and this term vanishes. 
The momentum is then determined entirely by the second contribution, resulting from the extension of the D4-brane in the complex projective space.  

\subsection{Giant graviton solution}
\begin{figure}[htb!]
\begin{center}
 \setlength{\unitlength}{1cm}
    \begin{picture}(14,11)(0,0)
      \put(0,0){ \includegraphics[scale=1,trim=3.4cm 9.3cm 3.4cm 9cm]{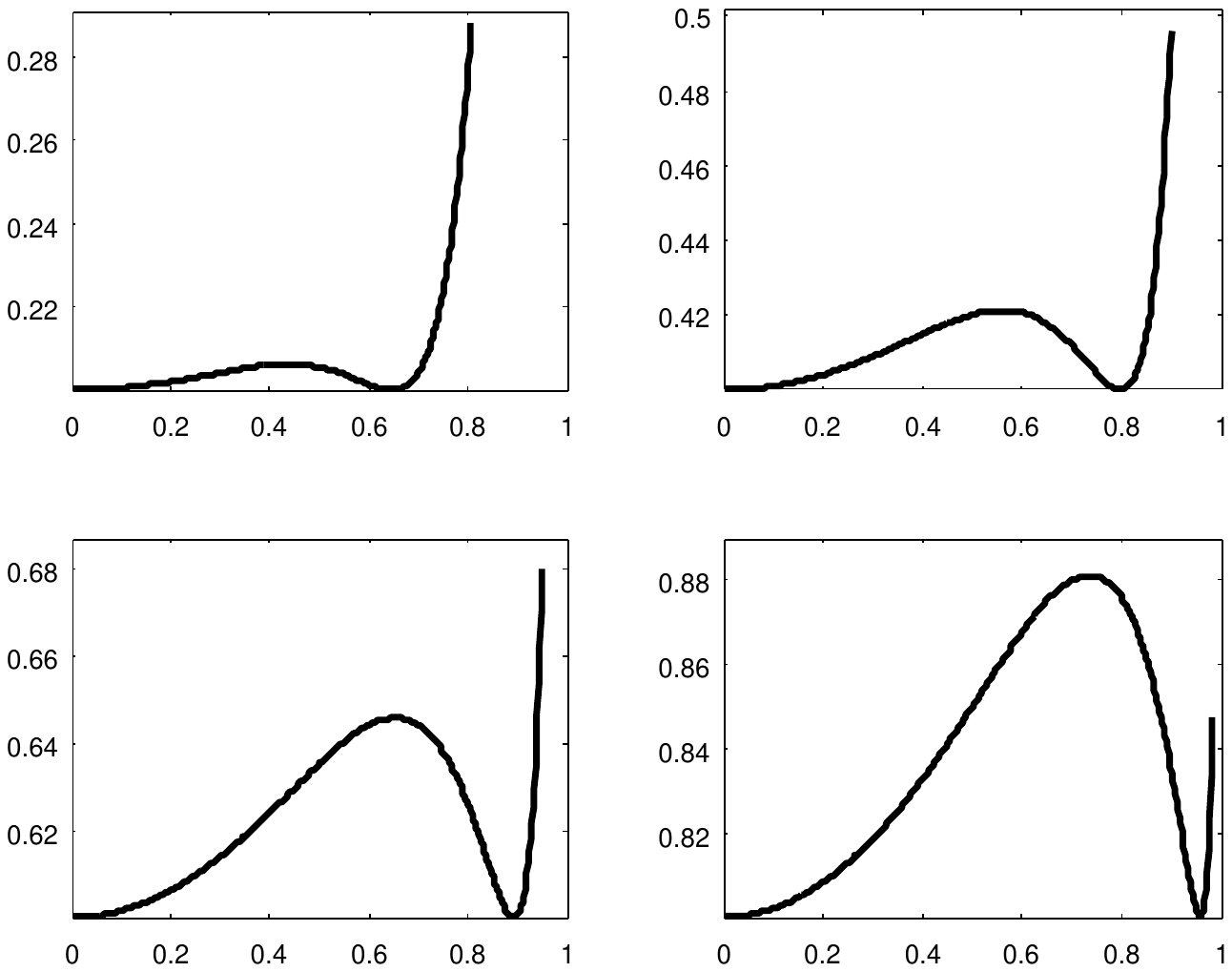}}

      \put(4,  0.1){$\alpha$}
      \put(4,  5.7){$\alpha$}
      \put(11, 0.1){$\alpha$}
      \put(11, 5.7){$\alpha$}
      \put(0.2,  2.8){$E$}
      \put(0.2,  8.4){$E$}
      \put(7.2,  2.8){$E$}
      \put(7.2,  8.4){$E$}
      \put(1.4,  10.8){(a) $P_{\chi} = 0.2$}
      \put(8.4,  10.8){(b) $P_{\chi} = 0.4$}
      \put(1.4,  5.2){(c) $P_{\chi} = 0.6$}
      \put(8.4,  5.2){(d) $P_{\chi} = 0.8$}
 \end{picture}
\end{center}
\vspace{-7mm}
\caption{The energy of the D4-brane configuration, plotted as a function of the size $\alpha$ at fixed momentum $P_{\chi}$, in units of the flux $N$.} 
\label{fig:EnergyCurves}
\end{figure}

The task now is to solve for the finite size $\alpha_{0}$ giant graviton configuration, which is associated with a minimum in the energy $H(\alpha,P_{\chi})$, plotted as a function of $\alpha$ at some fixed momentum $P_{\chi}$.  
Unfortunately, inverting $P_{\chi}(\dot{\chi})$ for $\dot{\chi}(P_{\chi})$ analytically and then substituting the result into the energy $H(\alpha,\dot{\chi})$ to obtain $H(\alpha,P_{\chi})$ is problematic.  
We hence resort to the numerical integration of the momentum (\ref{eq:Momentum}) and energy (\ref{eq:Energy}), as described in Appendix \ref{appendix - numerics}, to produce the standard energy plots for this D4-brane configuration, which are shown in Figure \ref{fig:EnergyCurves}.

\begin{figure}[htb!]
\begin{center}
 \setlength{\unitlength}{1cm}
  \begin{picture}(8.7,7.2)(0,0)
   \put(0,0){\includegraphics[scale=0.6,trim=3.4cm 9cm 3.4cm 9cm]{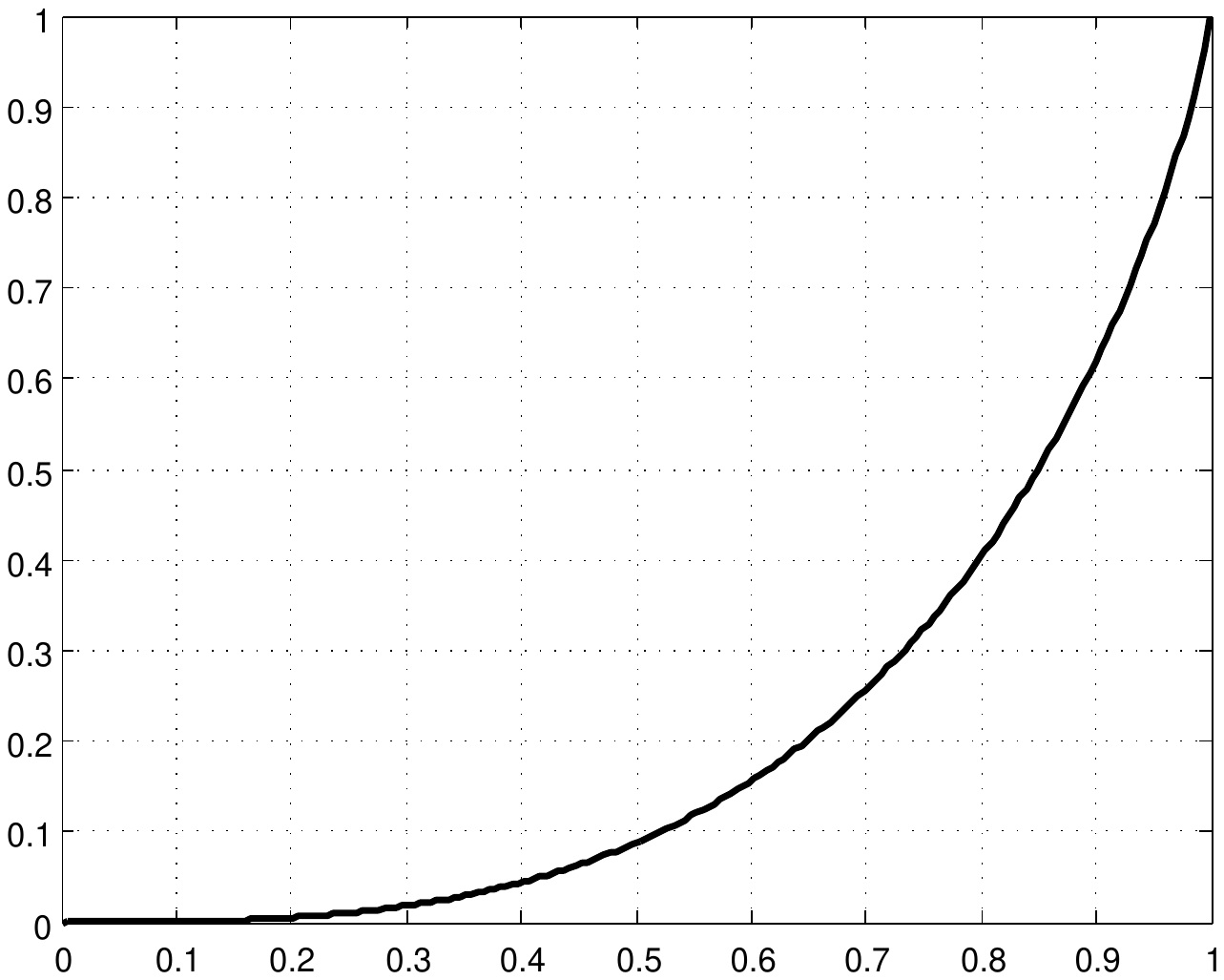}}
    \put(4, 0.1){$\alpha_{0}$}
    \put(-0.1, 3.5){$E$}
 \end{picture}
\end{center}
\vspace{-0.75cm}
\caption{The energy of the giant graviton as a function of its size $\alpha_{0}$ (in units of $N$).} 
\label{fig:EnergyGiant}
\end{figure}

The giant graviton solution, visible as the finite size $\alpha = \alpha_{0}$ minimum in the energy, always occurs when $\dot{\chi} = 1$ and is energetically degenerate with the point graviton solution at $\alpha = 0$ (previously described in Section \ref{section - point graviton}).  Now, substituting $\dot{\chi} = 1$ into the momentum and energy integrals (\ref{eq:Momentum}) and (\ref{eq:Energy}) respectively, we obtain
\begin{equation} \label{gg-energy-momentum}
H = P_{\chi} = \frac{N}{4} \int_{-\alpha_{0}}^{\alpha_{0}} dy \int_{0}^{\frac{\alpha_{0}^{2}-y^{2}}{1-y^{2}}} dz_{1}
\left[\left(1+y\right) + \frac{\left(1-\alpha_{0}^{2}\right)}{\left(1+y\right)\left(1-z_{1}\right)^{2}}\right].
\end{equation}
This integral is perfectly tractable!  
The energy and momentum of the submaximal giant graviton solution (plotted in Figure \ref{fig:EnergyGiant}) can hence be determined as follows:
\begin{equation}
H = P_{\chi} = N \left\{ \alpha_{0}  + \frac{1}{2} \left(1-\alpha_{0}^{2}\right) \ln{\left(\frac{1-\alpha_{0}}{1+\alpha_{0}}\right)}\right\},
\end{equation}
which is defined for all $\alpha_{0} \in (0,1)$.  Note that the maximal giant graviton limit, in which $\alpha \rightarrow 1$, is well-defined and yields $H = P_{\chi} = N$ as expected
(being twice the energy of a $\mathbb{CP}^{2}$ dibaryon \cite{MP,GLR}).  

We have therefore completed our construction of the submaximal giant graviton in type IIA string theory on $AdS_{4}\times \mathbb{CP}^{3}$ - which we refer to as the $\mathbb{CP}^{3}$ giant graviton (indicating the space in which the D4-brane is extended, rather than the shape of the object, which changes as the size $\alpha_{0}$ increases).  
We expect this to be a BPS configuration, although we have not yet computed the number of supersymmetries preserved by the Killing-Spinor equations.  
This is dual to the subdeterminant operator $\mathcal{O}_{n}(A_{1}B_{1})$ in ABJM theory.  
The equality between the energy and momentum $P_{\chi}$ agrees with the fact that the conformal dimension of the subdeterminant $\Delta = n$ is the same as its $\mathcal{R}$-charge.

\vspace{-0.25cm}
\begin{figure}[htb!] \label{GiantEvolution}
\begin{center}
\vspace{0.65cm}
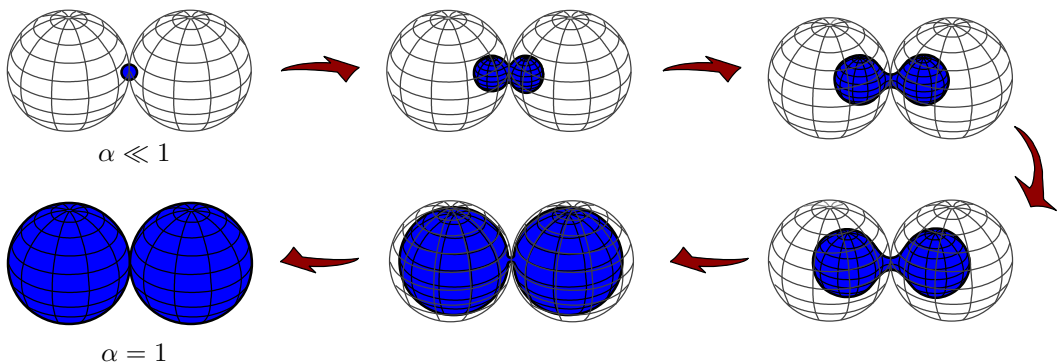
\caption{A cartoon representation of the growth of the $\mathbb{CP}^{3}$ giant graviton.} \label{fig:GravitonEvolution}
\label{fig:GiantGrow3Da}
\end{center}
\end{figure}
\vspace{-0.25cm}
In Figure~\ref{fig:GiantGrow3Da} we show a heuristic picture of the growth of the giant graviton in the complex projective space.  
The small submaximal giant is a nearly spherical configuration, similar in nature to the canonical case. 
As the size increases, however, its worldvolume pinches off, until it factorizes into two D4-branes, wrapped on different $\mathbb{CP}^{2}$ subspaces and intersecting on a $\mathbb{CP}^{1}$
(these are the $\mathbb{CP}^{2}$ dibaryons of \cite{MP,GLR}).  
We thereby observe the factorization of the subdeterminant operators in ABJM theory into two full determinants from the gravitational point of view, which is a direct result of the product nature of the 
SCS-matter gauge group.

%% file: CP3Submaximal.pdf_t
\begin{picture}(0,0)%
\includegraphics{CP3Submaximal.pdf}%
\end{picture}%
\setlength{\unitlength}{4144sp}%
\begingroup\makeatletter\ifx\SetFigFont\undefined%
\gdef\SetFigFont#1#2#3#4#5{%
  \reset@font\fontsize{#1}{#2pt}%
  \fontfamily{#3}\fontseries{#4}\fontshape{#5}%
  \selectfont}%
\fi\endgroup%
\begin{picture}(4978,2920)(-4276,-3194)
\put(-3082,-2137){\makebox(0,0)[rb]{\smash{{\SetFigFont{9}{10.8}{\familydefault}{\mddefault}{\updefault}{\color[rgb]{0,0,0}$1$}%
}}}}
\put(-1181,-3134){\makebox(0,0)[lb]{\smash{{\SetFigFont{9}{10.8}{\familydefault}{\mddefault}{\updefault}{\color[rgb]{0,0,0}(b) Maximal giant graviton $\alpha = 1$}%
}}}}
\put(-855,-624){\makebox(0,0)[lb]{\smash{{\SetFigFont{9}{10.8}{\familydefault}{\mddefault}{\updefault}{\color[rgb]{0,0,0}$y$}%
}}}}
\put(613,-591){\makebox(0,0)[lb]{\smash{{\SetFigFont{9}{10.8}{\familydefault}{\mddefault}{\updefault}{\color[rgb]{0,0,0}$z_2$}%
}}}}
\put(679,-2548){\makebox(0,0)[lb]{\smash{{\SetFigFont{9}{10.8}{\familydefault}{\mddefault}{\updefault}{\color[rgb]{0,0,0}$z_1$}%
}}}}
\put(-2845,-2221){\makebox(0,0)[lb]{\smash{{\SetFigFont{9}{10.8}{\familydefault}{\mddefault}{\updefault}{\color[rgb]{0,0,0}$z_1$}%
}}}}
\put(-3823,-624){\makebox(0,0)[lb]{\smash{{\SetFigFont{9}{10.8}{\familydefault}{\mddefault}{\updefault}{\color[rgb]{0,0,0}$y$}%
}}}}
\put(-3845,-851){\makebox(0,0)[rb]{\smash{{\SetFigFont{9}{10.8}{\familydefault}{\mddefault}{\updefault}{\color[rgb]{0,0,0}$1$}%
}}}}
\put(-3842,-1048){\makebox(0,0)[rb]{\smash{{\SetFigFont{9}{10.8}{\familydefault}{\mddefault}{\updefault}{\color[rgb]{0,0,0}$\alpha$}%
}}}}
\put(-3842,-2157){\makebox(0,0)[rb]{\smash{{\SetFigFont{9}{10.8}{\familydefault}{\mddefault}{\updefault}{\color[rgb]{0,0,0}$-\alpha$}%
}}}}
\put(-3842,-2352){\makebox(0,0)[rb]{\smash{{\SetFigFont{9}{10.8}{\familydefault}{\mddefault}{\updefault}{\color[rgb]{0,0,0}$-1$}%
}}}}
\put(-855,-2352){\makebox(0,0)[rb]{\smash{{\SetFigFont{9}{10.8}{\familydefault}{\mddefault}{\updefault}{\color[rgb]{0,0,0}$-1$}%
}}}}
\put(-855,-852){\makebox(0,0)[rb]{\smash{{\SetFigFont{9}{10.8}{\familydefault}{\mddefault}{\updefault}{\color[rgb]{0,0,0}$1$}%
}}}}
\put(-3086,-1016){\makebox(0,0)[rb]{\smash{{\SetFigFont{9}{10.8}{\familydefault}{\mddefault}{\updefault}{\color[rgb]{0,0,0}$1$}%
}}}}
\put(-4261,-3134){\makebox(0,0)[lb]{\smash{{\SetFigFont{9}{10.8}{\familydefault}{\mddefault}{\updefault}{\color[rgb]{0,0,0}(a) Submaximal giant graviton $0<\alpha <1$}%
}}}}
\put(-2845,-917){\makebox(0,0)[lb]{\smash{{\SetFigFont{9}{10.8}{\familydefault}{\mddefault}{\updefault}{\color[rgb]{0,0,0}$z_2$}%
}}}}
\end{picture}%

%% file: GiantGrow3Da.pdf_t
\begin{picture}(0,0)%
\includegraphics{GiantGrow3Da.pdf}%
\end{picture}%
\setlength{\unitlength}{4144sp}%
\begingroup\makeatletter\ifx\SetFigFont\undefined%
\gdef\SetFigFont#1#2#3#4#5{%
  \reset@font\fontsize{#1}{#2pt}%
  \fontfamily{#3}\fontseries{#4}\fontshape{#5}%
  \selectfont}%
\fi\endgroup%
\begin{picture}(6235,2163)(614,-2124)
\put(1377,-2060){\makebox(0,0)[b]{\smash{{\SetFigFont{10}{12.0}{\familydefault}{\mddefault}{\updefault}{\color[rgb]{0,0,0}$\alpha = 1$}%
}}}}
\put(1377,-871){\makebox(0,0)[b]{\smash{{\SetFigFont{10}{12.0}{\familydefault}{\mddefault}{\updefault}{\color[rgb]{0,0,0}$\alpha \ll 1$}%
}}}}
\end{picture}%

%% file: 05-Fluct.tex
\section{Fluctuation analysis} \label{fluctuation analysis} \label{section - fluctuations}

This section contains a general analysis of small fluctuations about the giant graviton on $AdS_{4}\times\mathbb{CP}^{3}$.  We obtain the D4-brane action and equations of motion describing this perturbed configuration.  Included are both scalar and worldvolume fluctuations - we cannot initially rule out the possibility that these may couple, as in case of the spherical dual giant graviton \cite{NT-giants,HMPS}.  Our ultimate goal is to determine whether any dependence on the $\alpha_{0}$, which parameterizes the changing shape and size of the giant, is manifest in the fluctuation spectrum.

\subsection{Coordinates of $AdS_{4}\times\mathbb{CP}^{3}$ best suited to the fluctuation analysis}

\emph{\textbf{Anti-de Sitter spacetime}}
\smallskip

The metric (\ref{metric-orig}) of the anti-de Sitter spacetime AdS$_{4}$ can be rewritten in terms of an alternative set of cartesian coordinates $v_{k}$, which are are more convenient for the purposes of the fluctuation analysis \cite{DJM}\footnote{The original coordinates $r$, $\tilde{\theta}$ and $\tilde{\varphi}$ have a coordinate singularity at $r=0$, which is precisely the position of the D4-brane giant graviton.}.  Here we define
\begin{equation}
v_{1} = r\cos{\tilde{\theta}}, \hspace{1.0cm}
v_{2} = r\sin{\tilde{\theta}}\cos{\tilde{\varphi}} \hspace{1.0cm} \textrm{and} \hspace{1.0cm}
v_{3} = r\sin{\tilde{\theta}}\sin{\tilde{\varphi}},
\end{equation}
in terms of which the $AdS_{4}$ metric can be written as
\begin{equation}
ds_{AdS_{4}}^{2} = -\left(1 + \sum_{k}v_{k}^{2}\right)dt^{2}
+ \sum_{i,j}\left(\delta_{ij} - \frac{v_{i}v_{j}}{\left(1 + \sum_{k}v_{k}^{2}\right)}\right)dv_{i}dv_{j}.
\end{equation}
The 4-form field strength (\ref{F4-orig}) becomes
\begin{equation}
F_{4} = -\tfrac{3}{2} \hspace{0.05cm} kR^{2} \hspace{0.1cm} dt \wedge dv_{1} \wedge dv_{2} \wedge dv_{3},
\end{equation}
which is associated with the 3-form potential
\begin{equation}
C_{3} = \tfrac{1}{2} \hspace{0.05cm} kR^{2} \hspace{0.1cm} dt \wedge
\left( v_{1} dv_{2} \wedge dv_{3} + v_{2} dv_{3} \wedge dv_{1} + v_{3} dv_{1} \wedge dv_{2}\right).
\end{equation}

\medskip
\emph{\textbf{Complex projective space}}
\smallskip

The metric of the complex projective space is given by
\begin{equation}
ds_{\mathbb{CP}^{3}}^{2} = \tfrac{1}{4}\left(ds_{\textrm{radial}}^{2} + ds_{\textrm{angular}}^{2} \right),
\end{equation}
where we shall assume the following generic forms for the radial and angular metrics:
\begin{eqnarray}
&& \hspace{-0.25cm} ds_{\textrm{radial}}^{2} = g_{\alpha\alpha} \hspace{0.05cm} d\alpha^{2} + g_{x_{1}x_{1}} \hspace{0.05cm} dx_{1}^{2} + g_{x_{2}x_{2}} \hspace{0.05cm} dx_{2}^{2} 
+ 2 g_{\alpha x_{1}} \hspace{0.05cm} d\alpha \hspace{0.05cm} dx_{1} + 2 g_{\alpha x_{2}} \hspace{0.05cm} d\alpha \hspace{0.05cm} dx_{2} + 2 g_{x_{1}x_{2}} \hspace{0.05cm} dx_{1} \hspace{0.05cm} dx_{2} \\
&& \hspace{-0.25cm} ds_{\textrm{angular}}^{2} = g_{\chi\chi} \hspace{0.05cm} d\chi^{2} + g_{\varphi_{1}\varphi_{1}} \hspace{0.05cm} d\varphi_{1}^{2} + g_{\varphi_{2}\varphi_{2}} d\varphi_{2}^{2} 
+ 2 g_{\chi\varphi_{1}} \hspace{0.05cm} d\chi d\varphi_{1} + 2 g_{\chi\varphi_{2}} \hspace{0.05cm} d\chi d\varphi_{2} 
+ 2 g_{\varphi_{1}\varphi_{2}} \hspace{0.05cm} d\varphi_{1} \hspace{0.05cm} d\varphi_{2} \hspace{1.0cm}
\end{eqnarray}
in the radial coordinates $\alpha$, $x_{1}$ and $x_{2}$, and the angular coordinates $\chi$, $\varphi_{1}$ and $\varphi_{2}$.

The 6-form field strength is given by
\begin{equation}
F_{6} = \frac{3}{2} \hspace{0.05cm} kR^{4} \sqrt{ \left[\det{g_{\textrm{rad}}} \right] \left[ \det{g_{\textrm{ang}}} \right] } \hspace{0.2cm} d\alpha \wedge dx_{1} \wedge dx_{2} \wedge d\chi \wedge d\varphi_{1} \wedge d\varphi_{2},
\end{equation}
which is associated with the 5-form potential\footnote{It is not immediately obvious that $F_{6} = dC_{5}$.  However, it is possible to check this expression for $C_{5}$ in one particular set of radial coordinates 
(for example, $\alpha$, $y$ and $z_{1}$) and then note that it is invariant under any radial coordinate transformation $(\alpha,x_{1},x_{2}) \rightarrow (\alpha, \tilde{x}_{1}(\alpha,x_{1},x_{2}), \tilde{x}_{2}(\alpha,x_{1},x_{2}))$ which keeps $\alpha$ fixed.}
\begin{eqnarray}
&& C_{5} = \frac{1}{2} \sqrt{\left(C_{\textrm{rad}}\right)_{11} \left[\left(C_{\textrm{rad}}\right)_{11} - \det{g_{\textrm{ang}}}\right]} \\
\nonumber &&\hspace{1.0cm} \times \left\{ dx_{1}\wedge dx_{2} - \frac{\left(C_{\textrm{rad}}\right)_{12}}{\left(C_{\textrm{rad}}\right)_{11}} \hspace{0.1cm} d\alpha \wedge dx_{1} 
+ \frac{\left(C_{\textrm{rad}}\right)_{13}}{\left(C_{\textrm{rad}}\right)_{11}} \hspace{0.1cm} d\alpha \wedge dx_{2} \right\} d\chi \wedge d\varphi_{1} \wedge d\varphi_{2},
\end{eqnarray}
while the 2-form field strength can generically be written in terms of the K\"{a}hler form on the complex projective space (\ref{F2-orig}).

Note that, throughout this section, we studiously avoid any reference to a particular choice of radial worldvolume coordinates\footnote{Except that we assume $x_{1}$ and $x_{2}$ have fixed coordinate ranges which are independent of $\alpha$.} $x_{1}$ and $x_{2}$.  We leave the metric components and their cofactors (as well as their derivatives) unspecified.  We anticipate that, in the subsequent section, it may be convenient to make use of several different sets of coordinates $x_{i}$, each of which is best suited to describe a certain limiting case.

\subsection{Fluctuation ansatz}

Our ansatz for the scalar fluctuations about the worldvolume of the submaximal $\mathbb{CP}^{3}$ giant graviton takes the form
\begin{equation}
v_{k}(\sigma^{a}) = \varepsilon \hspace{0.075cm} \delta v_{k}(\sigma^{a}), \hspace{0.6cm}
\alpha(\sigma^{a}) = \alpha_{0} + \varepsilon \hspace{0.075cm} \delta\alpha(\sigma^{a})
\hspace{0.6cm} \textrm{and} \hspace{0.6cm}
\chi(\sigma^{a}) = t + \varepsilon \hspace{0.075cm} \delta\chi(\sigma^{a}),
\end{equation}
whereas the worldvolume fluctuations can be taken into account by setting
\begin{equation}\
F(\sigma^{a}) = \varepsilon \hspace{0.1cm} \tfrac{R^{2}}{2\pi}\hspace{0.1cm}  \delta F(\sigma^{a}),
\end{equation}
with $\varepsilon$ a small parameter. The dependence of the fluctuations on the worldvolume coordinates $\sigma^{a} = (t,x_{1},x_{2},\varphi_{1},\varphi_{2})$ has been shown here explicitly. 

\subsection{D4-brane action to second order}

We shall now determine the D4-brane action associated with this perturbed $\mathbb{CP}^{3}$ giant graviton configuration, keeping terms quadratic in $\varepsilon$. 

\medskip
\emph{\textbf{Dirac-Born Infeld action}}
\smallskip

The DBI action (\ref{DBI-general}) can be simplified to the form
\begin{equation}
S_{\textrm{DBI}} = - \frac{kR^{4}}{2(2\pi)^{4}} \hspace{0.05cm}  \int d^{5}\sigma
\sqrt{- \det \left(h + \varepsilon \hspace{0.075cm} \delta \mathcal{F} \right)}, \hspace{0.5cm}
\textrm{with} \hspace{0.25cm} \delta \mathcal{F} \equiv \left(2\pi R^{-2} \right) \delta F,
\end{equation} 
where the components of the (scaled) pullback of the metric 
$h_{ab} \equiv R^{-2} \left( \mathcal{P}[g]\right)_{ab}$
to the worldvolume of the perturbed D4-brane can be expanded in orders of $\varepsilon$ as follows:
\begin{eqnarray}
\nonumber & \hspace{-0.15cm} h_{ab} = & \left\{ -\left( 1 - g_{\chi\chi} \right) \p_{a}t \hspace{0.05cm} \p_{b}t + g_{x_{1}x_{1}} \hspace{0.05cm} \p_{a}x_{1} \hspace{0.05cm} \p_{b}x_{1} 
+ g_{x_{2}x_{2}} \hspace{0.05cm} \p_{a}x_{2} \hspace{0.05cm}  \p_{b}x_{2} + g_{x_{1}x_{2}} \left(\p_{a}x_{1} \hspace{0.05cm} \p_{b}x_{2} + \p_{a}x_{2} \hspace{0.05cm} \p_{b}x_{1} \right) \right. \\
\nonumber && \hspace{0.2cm} + \hspace{0.075cm} g_{\varphi_{1}\varphi_{1}} \hspace{0.05cm} \p_{a}\varphi_{1} \hspace{0.05cm} \p_{b}\varphi_{1} 
+ g_{\varphi_{2}\varphi_{2}} \hspace{0.05cm} \p_{a}\varphi_{2} \hspace{0.05cm} \p_{b}\varphi_{2} 
+ g_{\chi\varphi_{1}}\left( \p_{a}t \hspace{0.05cm} \p_{b}\varphi_{1} + \p_{a}\varphi_{1} \hspace{0.05cm} \p_{b}t \right)  \\
\nonumber && \left. \hspace{0.155cm} + \hspace{0.075cm} g_{\chi\varphi_{2}}\left( \p_{a}t \hspace{0.05cm} \p_{b}\varphi_{2} + \p_{a}\varphi_{2} \hspace{0.05cm} \p_{b}t \right) 
+ g_{\varphi_{1}\varphi_{2}}\left( \p_{a}\varphi_{1} \hspace{0.05cm} \p_{b}\varphi_{2} + \p_{a}\varphi_{2} \hspace{0.05cm} \p_{b}\varphi_{1} \right) \right\} \\
\nonumber && \\
\nonumber && + \hspace{0.075cm} \varepsilon \left\{ g_{\alpha x_{1}} \left[ \left( \p_{a}\delta\alpha \right) \p_{b}x_{1} + \p_{a}x_{1} \left( \p_{b}\delta\alpha \right) \right]  
+ g_{\alpha x_{2}} \left[ \left( \p_{a}\delta\alpha \right) \p_{b}x_{2} + \p_{a}x_{2} \left( \p_{b}\delta\alpha \right) \right] \right. \\
\nonumber && \hspace{0.75cm} + \hspace{0.075cm} g_{\chi\chi} \left[ \p_{a}t \left(\p_{b}\delta\chi \right) + \left( \p_{a}\delta\chi \right) \p_{b}t \right] 
+ \hspace{0.075cm} g_{\chi\varphi_{1}} \left[ \left(\p_{a}\delta\chi \right) \p_{b}\varphi_{1} + \p_{a}\varphi_{1} \left( \p_{b}\delta\chi \right) \right] \\
&& \left. \hspace{0.705cm} + \hspace{0.075cm} g_{\chi\varphi_{2}} \left[ \left(\p_{a}\delta\chi \right) \p_{b}\varphi_{2} + \p_{a}\varphi_{2} \left( \p_{b}\delta\chi \right) \right]  \right\} \\ 
\nonumber && \\
\nonumber && + \hspace{0.075cm} \varepsilon^{2} \left\{ - \left( {\textstyle \sum\limits_{k} } \hspace{0.05cm} \delta v_{k}^{2} \right) \delta_{a}t \hspace{0.05cm} \delta_{b}t 
+ { \textstyle \sum\limits_{k} } \left( \p_{a} \delta v_{k} \right) \left( \p_{b} \delta v_{k} \right) + g_{\alpha\alpha} \left( \p_{a}\delta\alpha \right)\left( \p_{b}\delta\alpha \right) 
+ g_{\chi\chi} \left( \p_{a}\delta\chi \right) \left( \p_{b}\delta\chi \right)  \right\}.
\end{eqnarray}
Note that the metric components $g_{\mu\nu}(\alpha,x_{1},x_{2})$ can also be expanded in orders of $\varepsilon$ using $\alpha = \alpha_{0} + \varepsilon \hspace{0.05cm} \delta\alpha$.  We shall not write
out any of these expansions of the metric or its cofactors until the end - it will then turn out that only certain specific combinations need be determined beyond leading order.

It can be shown that, in the DBI action, the scalar fluctuations $\delta v_{k}$, $\delta\alpha$ and $\delta\chi$, and worldvolume fluctuations $\delta \mathcal{F}_{ab}$ decouple:
\begin{equation}
S_{\textrm{DBI}} = - \frac{kR^{4} }{2(2\pi)^{2}}
\left( \int {d^{5}\sigma} \left\{  \sqrt{-a_{0}} 
\left[1 - \varepsilon \hspace{0.05cm} a_{1} 
+ \frac{1}{2} \hspace{0.05cm} \varepsilon^{2}\left( a_{2} - a_{1}^{2} \right) \right] \right\} 
+ \frac{1}{2} \hspace{0.05cm} \varepsilon^{2} \int_{\Sigma}{\delta \mathcal{F} \wedge \ast \delta \mathcal{F}} \right),
\end{equation}
where we have expanded the determinant of the induced metric on the pullback of the perturbed D4-brane worldvolume 
\begin{equation}
\det{h} \approx -a_{0}\left( 1 - 2 \hspace{0.05cm} \varepsilon \hspace{0.05cm} a_{1} + \varepsilon^{2} \hspace{0.05cm} a_{2} \right)
\end{equation}
in orders of $\varepsilon$.  Note that the Hodge dual $\ast \delta \mathcal{F}$ of the fluctuation $\delta \mathcal{F}$ of the worldvolume field strength form is constructed using the rescaled induced metric $h_{ab}$ on the worldvolume of the original $\mathbb{CP}^{3}$ giant graviton.

It now remains for us to find explicit expressions for $a_{0}$, $a_{1}$ and $a_{2}$:
\begin{eqnarray}
&& \hspace{-0.35cm} a_{0} = (C_{\textrm{rad}})_{11} \left[(C_{\textrm{ang}})_{11} - \det{g_{\textrm{ang}}} \right] \\
\nonumber && \\
\nonumber && \hspace{-0.35cm}  a_{1} = \frac{(C_{\textrm{rad}})_{12}}{(C_{\textrm{rad}})_{11}} \left( \p_{x_{1}}\delta\alpha \right) + \frac{(C_{\textrm{rad}})_{13}}{(C_{\textrm{rad}})_{11}} \left( \p_{x_{2}}\delta\alpha \right) 
+ \frac{\det{g_{\textrm{ang}}}}{\left[(C_{\textrm{ang}})_{11} - \det{g_{\textrm{ang}}} \right] } \hspace{0.1cm} \dot{\delta\chi} \\
&& \hspace{-0.25cm} \hspace{0.8cm} + \hspace{0.075cm} \frac{(C_{\textrm{ang}})_{12}}{\left[(C_{\textrm{ang}})_{11} - \det{g_{\textrm{ang}}} \right] } \left( \p_{\varphi_{1}} \delta\chi \right) 
+ \frac{(C_{\textrm{ang}})_{13}}{\left[(C_{\textrm{ang}})_{11} - \det{g_{\textrm{ang}}} \right] }  \left( \p_{\varphi_{2}} \delta\chi \right) \hspace{1.2cm} \\
\nonumber && \\
\nonumber &&  \hspace{-0.35cm} a_{2} - a_{1}^{2} = \sum_{k} \left\{ \left( \p \hspace{0.1cm} \delta v_{k} \right)^{2} 
+ \frac{(C_{\textrm{ang}})_{11}}{ \left[(C_{\textrm{ang}})_{11} - \det{g_{\textrm{ang}}} \right] } \hspace{0.1cm} \delta v_{k}^{2}  \right\} \\
\nonumber && \hspace{-0.35cm} \hspace{1.75cm} + \hspace{0.05cm} \frac{\det{g}_{\textrm{rad}}}{\left(C_{\textrm{rad}}\right)_{11}}  \left(\partial \hspace{0.1cm} \delta\alpha \right)^{2} 
+ \frac{\det{g}_{\textrm{ang}}}{\left[\left(C_{\textrm{ang}}\right)_{11} - \det{g_{\textrm{ang}}}\right]}  \left(\partial \hspace{0.1cm} \delta\chi \right)^{2} \\
\nonumber && \hspace{-0.35cm} \hspace{1.75cm} + \hspace{0.075cm} 2 \hspace{0.075cm} \frac{\left(C_{\textrm{rad}} \right)_{12}}{(C_{\textrm{rad}})_{11}} \frac{\det{g_{\textrm{ang}}}}{\left[(C_{\textrm{ang}})_{11} - \det{g_{\textrm{ang}}} \right]} 
\left[ \dot{\delta\chi} \left(\p_{x_{1}} \delta\alpha \right) - \dot{\delta\alpha} \left(\p_{x_{1}} \delta\chi \right) \right]  \\
&& \hspace{-0.35cm} \hspace{1.75cm} + \hspace{0.075cm} 2 \hspace{0.075cm} \frac{\left(C_{\textrm{rad}} \right)_{13}}{(C_{\textrm{rad}})_{11}} \frac{\det{g_{\textrm{ang}}}}{\left[(C_{\textrm{ang}})_{11} - \det{g_{\textrm{ang}}} \right]} 
\left[ \dot{\delta\chi} \left(\p_{x_{2}} \delta\alpha \right) - \dot{\delta\alpha} \left(\p_{x_{2}} \delta\chi \right) \right]  \\
\nonumber && \hspace{-0.35cm} \hspace{1.75cm} + \hspace{0.075cm} 2 \hspace{0.075cm} \frac{\left(C_{\textrm{rad}} \right)_{12}}{(C_{\textrm{rad}})_{11}} \frac{\left(C_{\textrm{ang}} \right)_{12}}{\left[(C_{\textrm{ang}})_{11} - \det{g_{\textrm{ang}}} \right]} 
\left[ \left(\p_{\varphi_{1}} \delta\chi \right) \left(\p_{x_{1}} \delta\alpha \right) - \left(\p_{\varphi_{1}} \delta\alpha \right) \left(\p_{x_{1}} \delta\chi \right) \right] \\
\nonumber && \hspace{-0.35cm} \hspace{1.75cm} + \hspace{0.075cm} 2 \hspace{0.075cm}\frac{\left(C_{\textrm{rad}} \right)_{13}}{(C_{\textrm{rad}})_{11}} \frac{\left(C_{\textrm{ang}} \right)_{12}}{\left[(C_{\textrm{ang}})_{11} - \det{g_{\textrm{ang}}} \right]} 
\left[\left(\p_{\varphi_{1}} \delta\chi \right) \left(\p_{x_{2}} \delta\alpha \right) - \left(\p_{\varphi_{1}} \delta\alpha \right) \left(\p_{x_{2}} \delta\chi \right) \right] \\
 \nonumber && \hspace{-0.35cm} \hspace{1.75cm} + \hspace{0.075cm} 2 \hspace{0.075cm} \frac{\left(C_{\textrm{rad}} \right)_{12}}{(C_{\textrm{rad}})_{11}} \frac{\left(C_{\textrm{ang}} \right)_{13}}{\left[(C_{\textrm{ang}})_{11} - \det{g_{\textrm{ang}}} \right]} 
 \left[ \left(\p_{\varphi_{2}} \delta\chi \right) \left(\p_{x_{1}} \delta\alpha \right) - \left(\p_{\varphi_{2}} \delta\alpha \right) \left(\p_{x_{1}} \delta\chi \right) \right] \\
\nonumber && \hspace{-0.35cm} \hspace{1.70cm} + \hspace{0.075cm} 2 \hspace{0.075cm} \frac{\left(C_{\textrm{rad}} \right)_{13}}{(C_{\textrm{rad}})_{11}} \frac{\left(C_{\textrm{ang}} \right)_{13}}{\left[(C_{\textrm{ang}})_{11} - \det{g_{\textrm{ang}}} \right]} 
 \left[ \left(\p_{\varphi_{2}} \delta\chi \right) \left(\p_{x_{2}} \delta\alpha \right) - \left(\p_{\varphi_{2}} \delta\alpha \right) \left(\p_{x_{2}} \delta\chi \right) \right] \hspace{1.2cm}
\end{eqnarray}
in terms of the determinants and cofactors of the radial and angular metrics, with $\left(\p \hspace{0.075cm} f\right)^{2}$ the gradiant squared of a function $f$ on the worldvolume of the $\mathbb{CP}^{3}$ giant graviton 
(see Appendix \ref{appendix - dAlembertian}). 

\medskip
\emph{\textbf{Wess-Zumino action}}
\smallskip

The WZ action (\ref{WZ-general}) can be written as
\begin{equation}
S_{\textrm{WZ}} = \frac{kR^{4}}{(2\pi)^{4}} \int_{\Sigma} 
\left\{ \left( k^{-1}R^{-4} \right) \mathcal{P}\left[C_{5}\right]  
+ \frac{1}{2} \hspace{0.05cm} \varepsilon^{2} \hspace{0.1cm} k^{-1} \hspace{0.05cm}  \mathcal{P}\left[ C_{1} \right] \wedge \delta F \wedge \delta F \right\},
\end{equation}
since $\mathcal{P}\left[ C_{3} \right]$ is cubic in $\varepsilon$ and hence negligible.  Note that, while the pullback of the 5-form potential $\mathcal{P}\left[ C_{5} \right]$ must be expanded to 
quadratic order in $\varepsilon$:
\begin{eqnarray}
&& \hspace{-0.35cm} \left( k^{-1}R^{-4} \right) \mathcal{P}\left[C_{5}\right] = \frac{1}{2} \hspace{0.05cm} b_{0}\left( 1 + \varepsilon \hspace{0.05cm} b_{1} + \varepsilon^{2} \hspace{0.05cm} b_{2}\right) 
dt \wedge dx_{1} \wedge dx_{2} \wedge d\varphi_{1} \wedge d\varphi_{2}, 
\end{eqnarray}
it is only necessary to keep the leading order terms in $\mathcal{P}\left[ C_{1} \right]$, which involve no scalar fluctuations.  The WZ action then simplifies as follows:
\begin{eqnarray}
&& S_{\textrm{WZ}} = \frac{kR^{4}}{2(2\pi)^{4}} \left( \int d^{5}\sigma \left\{ b_{0} \left[ 1 + \varepsilon \hspace{0.05cm} b_{1} + \varepsilon^{2} \hspace{0.05cm} b_{2} \right] \right\}
+  \varepsilon^{2} \int_{\Sigma} B \wedge \delta \mathcal{F} \wedge \delta \mathcal{F}  \right), \hspace{1.2cm}
\end{eqnarray}
where $B =   k^{-1} \hspace{0.05cm}  \mathcal{P}\left[ C_{1} \right]$, and the coefficients $b_{0}$, $b_{1}$ and $b_{2}$ are given  by
\begin{eqnarray}
&& \hspace{-0.35cm} b_{0} = \sqrt{\left(C_{\textrm{rad}}\right)_{11}\left[ \left(C_{\textrm{ang}}\right)_{11} - \det{g_{\textrm{ang}}} \right]}  = \sqrt{a_{0}} \hspace{1.2cm}  \\
\nonumber && \\
&& \hspace{-0.35cm} b_{1} = \dot{\delta\chi} - \frac{\left(C_{\textrm{rad}}\right)_{12}}{\left(C_{\textrm{rad}}\right)_{11}}  \hspace{0.05cm} \left(\p_{x_{1}}\delta \alpha\right) 
- \frac{\left(C_{\textrm{rad}}\right)_{13}}{\left(C_{\textrm{rad}}\right)_{11}}  \hspace{0.05cm}  \left(\p_{x_{2}}\delta \alpha\right) \\
\nonumber && \\
&& \hspace{-0.35cm} b_{2} =  - \hspace{0.075cm} \frac{\left(C_{\textrm{rad}}\right)_{12}}{\left(C_{\textrm{rad}}\right)_{11}}  \hspace{0.05cm} 
\left[\dot{\delta\chi} \left(\p_{x_{1}}\delta\alpha\right) - \dot{\delta\alpha} \left(\p_{x_{1}}\delta\chi\right) \right] 
-  \frac{\left(C_{\textrm{rad}}\right)_{13}}{\left(C_{\textrm{rad}}\right)_{11}}  \hspace{0.05cm} \left[\dot{\delta\chi} \left(\p_{x_{2}}\delta\alpha\right) - \dot{\delta\alpha} \left(\p_{x_{2}}\delta\chi\right) \right]. \hspace{1.2cm}
\end{eqnarray}

\medskip
\emph{\textbf{D4-brane action}}
\smallskip

We can combine the DBI and WZ actions to obtain the D4-brane action describing small fluctuations around the D4-brane giant graviton on AdS$_{4}\times\mathbb{CP}^{3}$.  Contrary to our initial 
expectations, based on the result of a similar fluctuation analysis for the D2-brane dual giant graviton \cite{HMPS}, the scalar fluctuations $\delta\alpha$ and $\delta \chi$ do decouple from the 
worldvolume fluctuations 
$\delta \mathcal{F}$.  The D4-brane action $S_{\textrm{D4}} = S_{\textrm{scalar}} + S_{\textrm{worldvolume}}$ splits into two parts:
\begin{eqnarray}
&& \!\!\!\! S_{\textrm{scalar}} =  - \frac{kR^{4}}{2(2\pi)^{4}}  \int d^{5}\sigma \left\{ \sqrt{a_{0}}  \left[ - \varepsilon\left(  a_{1} + b_{1} \right) 
+ \varepsilon^{2}\left( \frac{1}{2} \left( a_{2} -  a_{1}^{2}\right) - b_{2} \right)\right] \right\}  \\
\nonumber && \\
&& \!\!\!\! S_{\textrm{worldvolume}} = - \frac{kR^{4}}{2(2\pi)^{4}} \hspace{0.15cm} \varepsilon^{2} 
\int_{\Sigma} \left\{ \frac{1}{4} \hspace{0.1cm} \delta \mathcal{F} \wedge \ast \delta \mathcal{F} - B \wedge \delta \mathcal{F} \wedge \delta \mathcal{F} \right\},
\hspace{0.4cm} \textrm{with} \hspace{0.2cm} \delta \mathcal{F} = d \hspace{0.05cm} \delta\mathcal{A}, \hspace{1.2cm}
\end{eqnarray}
which will separately yield the equations of motion for the scalar and worldvolume fluctuations respectively.  

Let us focus for the moment on the scalar fluctuations. Note that only $\delta\chi$ derivative terms 
\begin{eqnarray} \label{action - first order}
\nonumber && \hspace{-0.10cm} -\varepsilon\sqrt{a_{0}} \left(  a_{1} + b_{1} \right) =  - \varepsilon\sqrt{\left(C_{\textrm{rad}}\right)_{11}\left[ \left(C_{\textrm{ang}}\right)_{11} - \det{g_{\textrm{ang}}} \right]} \\
\nonumber && 
\hspace{3.58cm} \times \left\{ \frac{(C_{\textrm{ang}})_{11}}{\left[(C_{\textrm{ang}})_{11} - \det{g_{\textrm{ang}}} \right]} \hspace{0.1cm} \dot{\delta\chi} 
+ \frac{(C_{\textrm{ang}})_{12}}{\left[(C_{\textrm{ang}})_{11} - \det{g_{\textrm{ang}}} \right]} \left(\delta_{\varphi_{1}}\delta\chi \right) \right. \\
&& \hspace{4.1cm} \left. + \hspace{0.075cm} \frac{(C_{\textrm{ang}})_{13}}{\left[(C_{\textrm{ang}})_{11} - \det{g_{\textrm{ang}}} \right]} \left(\delta_{\varphi_{2}}\delta\chi \right) \right\}  
\end{eqnarray}
appear in the first order scalar action.  The contributions to the $\delta\alpha$ derivative terms from the DBI and WZ actions cancel out  - they are actually only there in these individual actions because we are making 
use of non-orthogonal radial coordinates.  

The above expression still needs to be evaluated at $\alpha = \alpha_{0} + \varepsilon \hspace{0.05cm} \delta\alpha$ and expanded in orders of $\varepsilon$.  This expansion will yield both first order terms in the action (which are clearly total derivatives) and additional second order contributions:
\begin{eqnarray}
- \varepsilon \hspace{0.075cm} \sqrt{a_{0}} \left(  a_{1} + b_{1} \right) \hspace{0.1cm} \approx \hspace{0.1cm} \varepsilon \left\{ \textrm{total derivatives}\right\} \hspace{0.1cm} 
+ \hspace{0.075cm} \varepsilon^{2} \left\{ \cdots 2 \cdots \right\}  \hspace{1.1cm}
\end{eqnarray}
with
\begin{eqnarray}
\nonumber  &&  \hspace{-0.45cm} \left\{ \cdots 2 \cdots \right\} 
= - \hspace{0.075cm} \p_{\alpha}\left\{ \sqrt{(C_{\textrm{rad}})_{11}\left[(C_{\textrm{ang}})_{11} - \det{g_{\textrm{ang}}} \right] } \hspace{0.2cm}
\frac{(C_{\textrm{ang}})_{11}}{\left[(C_{\textrm{ang}})_{11} - \det{g_{\textrm{ang}}} \right] } \right\} 
\delta\alpha \hspace{0.1cm} \dot{\delta\chi} \hspace{1.2cm} \\
\nonumber && \hspace{-0.45cm} \hspace{2.375cm} - \hspace{0.075cm} \p_{\alpha}
\left\{ \sqrt{(C_{\textrm{rad}})_{11}\left[(C_{\textrm{ang}})_{11} - \det{g_{\textrm{ang}}} \right] } \hspace{0.2cm}
\frac{(C_{\textrm{ang}})_{12}}{\left[(C_{\textrm{ang}})_{11} - \det{g_{\textrm{ang}}} \right] } \right\} 
\delta\alpha \left( \p_{\varphi_{1}} \delta\chi \right) \\
&& \hspace{-0.45cm} \hspace{2.375cm} - \hspace{0.075cm} \p_{\alpha}\left\{ \sqrt{(C_{\textrm{rad}})_{11}\left[(C_{\textrm{ang}})_{11} - \det{g_{\textrm{ang}}} \right] } \hspace{0.2cm}
\frac{(C_{\textrm{ang}})_{13}}{\left[(C_{\textrm{ang}})_{11} - \det{g_{\textrm{ang}}} \right] } \right\}
\delta\alpha \left( \p_{\varphi_{2}} \delta\chi \right) \hspace{1.5cm}
\end{eqnarray} 
where the coefficients are now evaluated at $\alpha = \alpha_{0}$, the fixed size of the giant.

The manifestly second order term in the scalar action can also be simplified.  We shall neglect surface terms and hence obtain
\begin{equation}
\varepsilon^{2} \sqrt{a_{0}} \left[ \frac{1}{2}\left( a_{2} -  a_{1}^{2}\right) - b_{2} \right] = \varepsilon^{2}  \left\{ \cdots 1 \cdots \right\} 
\end{equation}
with
\begin{eqnarray}
&& \hspace{-0.45cm} \left\{ \cdots 1 \cdots \right\} 
 =  \sqrt{(C_{\textrm{rad}})_{11} \left[(C_{\textrm{ang}})_{11} - \det{g_{\textrm{ang}}} \right]}  
  \left\{ \frac{1}{2} \sum_{k} \left[ \left( \p \hspace{0.1cm} \delta v_{k} \right)^{2} 
 + \frac{(C_{\textrm{ang}})_{11}}{ \left[(C_{\textrm{ang}})_{11} - \det{g_{\textrm{ang}}} \right] } \hspace{0.1cm} \delta v_{k}^{2} \right] \right. \hspace{1.0cm} \\
 \nonumber && \left. \hspace{-0.45cm} \hspace{7.7cm} + \hspace{0.075cm} \frac{1}{2} \hspace{0.05cm} 
 \frac{ \left[ \det{g}_{\textrm{rad}} \right] }{ (C_{\textrm{rad}})_{11} } \left(\p \hspace{0.1cm} \delta\alpha \right)^{2}  
 +  \frac{1}{2} \hspace{0.05cm} 
 \frac{\left[ \det{g}_{\textrm{ang}} \right]}{\left[(C_{\textrm{ang}})_{11} - \det{g_{\textrm{ang}}} \right]} \left(\p \hspace{0.1cm} \delta\chi \right)^{2} \right\} \\
 \nonumber && \\
  \nonumber && \\
\nonumber && \hspace{-0.45cm} \hspace{1.9cm} - \hspace{0.075cm} \p_{x_{1}} \left\{ \sqrt{(C_{\textrm{rad}})_{11} \left[(C_{\textrm{ang}})_{11} - \det{g_{\textrm{ang}}} \right]} \hspace{0.2cm}
\frac{\left(C_{\textrm{rad}} \right)_{12}}{(C_{\textrm{rad}})_{11}} \hspace{0.1cm} \frac{\left(C_{\textrm{ang}} \right)_{11}}{\left[(C_{\textrm{ang}})_{11} - \det{g_{\textrm{ang}}} \right]}  \right\}
 \delta\alpha \hspace{0.1cm} \dot{\delta\chi} \\
\nonumber && \hspace{-0.45cm} \hspace{1.9cm} - \hspace{0.075cm} \p_{x_{2}} \left\{ \sqrt{(C_{\textrm{rad}})_{11} \left[(C_{\textrm{ang}})_{11} - \det{g_{\textrm{ang}}} \right]} \hspace{0.2cm}
\frac{\left(C_{\textrm{rad}} \right)_{13}}{(C_{\textrm{rad}})_{11}} \hspace{0.1cm} \frac{\left(C_{\textrm{ang}} \right)_{11}}{\left[(C_{\textrm{ang}})_{11} - \det{g_{\textrm{ang}}} \right]}  \right\}
 \delta\alpha \hspace{0.1cm} \dot{\delta\chi} \\
 \nonumber && \hspace{-0.45cm} \hspace{1.9cm} - \hspace{0.075cm} \p_{x_{1}} \left\{ \sqrt{(C_{\textrm{rad}})_{11} \left[(C_{\textrm{ang}})_{11} - \det{g_{\textrm{ang}}} \right]} \hspace{0.2cm}
\frac{\left(C_{\textrm{rad}} \right)_{12}}{(C_{\textrm{rad}})_{11}} \hspace{0.1cm} \frac{\left(C_{\textrm{ang}} \right)_{12}}{\left[(C_{\textrm{ang}})_{11} - \det{g_{\textrm{ang}}} \right]}  \right\}
 \delta\alpha \left(\p_{\varphi_{1}} \chi\right) \\
  \nonumber && \hspace{-0.45cm} \hspace{1.9cm} - \hspace{0.075cm} \p_{x_{2}} \left\{ \sqrt{(C_{\textrm{rad}})_{11} \left[(C_{\textrm{ang}})_{11} - \det{g_{\textrm{ang}}} \right]} \hspace{0.2cm}
\frac{\left(C_{\textrm{rad}} \right)_{13}}{(C_{\textrm{rad}})_{11}} \hspace{0.1cm} \frac{\left(C_{\textrm{ang}} \right)_{12}}{\left[(C_{\textrm{ang}})_{11} - \det{g_{\textrm{ang}}} \right]}  \right\}
 \delta\alpha \left(\p_{\varphi_{1}} \chi\right) \\
 \nonumber && \hspace{-0.45cm} \hspace{1.9cm} - \hspace{0.075cm} \p_{x_{1}} \left\{ \sqrt{(C_{\textrm{rad}})_{11} \left[(C_{\textrm{ang}})_{11} - \det{g_{\textrm{ang}}} \right]} \hspace{0.2cm}
\frac{\left(C_{\textrm{rad}} \right)_{12}}{(C_{\textrm{rad}})_{11}} \hspace{0.1cm} \frac{\left(C_{\textrm{ang}} \right)_{13}}{\left[(C_{\textrm{ang}})_{11} - \det{g_{\textrm{ang}}} \right]}  \right\}
 \delta\alpha \left(\p_{\varphi_{2}} \chi\right) \\
  \nonumber && \hspace{-0.45cm} \hspace{1.9cm} - \hspace{0.075cm} \p_{x_{2}} \left\{ \sqrt{(C_{\textrm{rad}})_{11} \left[(C_{\textrm{ang}})_{11} - \det{g_{\textrm{ang}}} \right]} \hspace{0.2cm}
\frac{\left(C_{\textrm{rad}} \right)_{13}}{(C_{\textrm{rad}})_{11}} \hspace{0.1cm} \frac{\left(C_{\textrm{ang}} \right)_{13}}{\left[(C_{\textrm{ang}})_{11} - \det{g_{\textrm{ang}}} \right]}  \right\}
 \delta\alpha \left(\p_{\varphi_{2}} \chi\right) 
 \end{eqnarray} 
 
 The scalar action to second order in $\varepsilon$ therefore takes the form
\begin{equation}
 S_{\textrm{scalar}} =  - \frac{\varepsilon^{2}kR^{4}}{2(2\pi)^{4}}  \int d^{5}\sigma \hspace{0.2cm} \mathcal{L}_{\textrm{scalar}},  \hspace{1.2cm}
 \end{equation}
with $\mathcal{L}_{\textrm{scalar}} = \left\{ \cdots 1 \cdots \right\} + \left\{ \cdots 2 \cdots \right\}$
the combination of the two previously defined expressions.  We now observe that this scalar Lagrangian density can now be written in the more convenient form
\begin{eqnarray}
\nonumber && \mathcal{L}_{\textrm{scalar}} 
= \sqrt{-h} \left\{ \frac{1}{2}\sum_{k} \left[ \left( \p \hspace{0.1cm} \delta v_{k} \right)^{2}  - h^{tt} \hspace{0.1cm} \delta v_{k}^{2}\right] 
+ \frac{1}{2}\frac{1}{g_{\textrm{rad}}^{\alpha\alpha}} \left(\p \hspace{0.1cm} \delta\alpha \right)^{2}
+ \frac{1}{2} \frac{1}{\left(g_{\textrm{ang}}^{\chi\chi} - 1\right)} \left(\p \hspace{0.1cm} \delta\chi \right)^{2} \right\} \\
&& \hspace{1.2cm} + \hspace{0.075cm} \frac{1}{2} \hspace{0.075cm} \p_{i}\left[\sqrt{-h} \hspace{0.15cm} \frac{g_{\textrm{rad}}^{\alpha i}}{g_{\textrm{rad}}^{\alpha\alpha}} \hspace{0.15cm} h^{tb} \right] 
\left[\delta\alpha \left(\p_{b}\delta\chi\right) - \delta\chi \left(\p_{b}\delta\alpha \right)\right]
\end{eqnarray}
and, integrating by parts,
\begin{equation}
\mathcal{L}_{\textrm{scalar}}  = -\frac{1}{2} \sqrt{-h} \left\{\cdots\right\},
\end{equation}
with
\begin{eqnarray}
&& \hspace{-0.35cm} \left\{\cdots\right\} = 
\sum_{k} \left[ \left( \Box \hspace{0.075cm} \delta v_{k} \right)  + h^{tt} \hspace{0.1cm} \delta v_{k} \right] \delta v_{k}  \\
\nonumber && \hspace{-0.35cm} \hspace{1.5cm} + \hspace{0.075cm} \frac{1}{g_{\textrm{rad}}^{\alpha\alpha}} \hspace{0.075cm}\left[ \left(\Box \hspace{0.075cm} \delta\alpha \right) 
+ g_{\textrm{rad}}^{\alpha\alpha}\hspace{0.15cm} \p_{a} \left( \frac{1}{g_{\textrm{rad}}^{\alpha\alpha}}\right) h^{ab} \left( \p_{b}\delta\alpha \right)
- \frac{g_{\textrm{rad}}^{\alpha\alpha}}{\sqrt{-h}} \hspace{0.15cm} \p_{i} \left(\sqrt{-h} \hspace{0.15cm} \frac{g_{\textrm{rad}}^{\alpha i}}{g_{\textrm{rad}}^{\alpha\alpha}} \hspace{0.15cm} h^{tb}\right) \left( \p_{b}\delta\chi \right) \right] \delta\alpha \\
\nonumber && \hspace{-0.35cm} \hspace{1.5cm} + \hspace{0.075cm} \frac{1}{\left(g_{\textrm{ang}}^{\chi\chi} - 1\right)} 
\left[ \left(\Box \hspace{0.075cm} \delta\chi \right)
+ \left(g_{\textrm{ang}}^{\chi\chi}-1\right) \hspace{0.075cm} \p_{a} \left( \frac{1}{g_{\textrm{ang}}^{\chi\chi}-1}\right) h^{ab} \left( \p_{b}\delta\chi \right) \right. \\
\nonumber && \hspace{-0.35cm} \hspace{8.275cm}  \left. - \hspace{0.075cm} \frac{\left(g_{\textrm{ang}}^{\chi\chi} - 1\right)}{\sqrt{-h}} \hspace{0.15cm} \p_{i} \left(\sqrt{-h} \hspace{0.15cm} \frac{g_{\textrm{rad}}^{\alpha i}}{g_{\textrm{rad}}^{\alpha\alpha}} \hspace{0.15cm} h^{tb}\right) \left( \p_{b}\delta\alpha \right)  \right] \delta\chi,
\end{eqnarray}
where $i$ and $j$ run over the radial coordinates $\alpha$, $x_{1}$ and $x_{2}$. We make use of the volume element $\sqrt{-h}$, the inverse metric components $h^{ab}$ and the d'Alembertian $\Box$ on the worldvolume of the giant graviton, which are defined in Appendix \ref{appendix - dAlembertian}.
We also need several components of the inverse radial metric
\begin{equation}
g_{\textrm{rad}}^{\alpha\alpha} = \frac{\left(C_{\textrm{rad}}\right)_{11}}{ \det{g_{\textrm{rad}}} }, \hspace{0.8cm} g_{\textrm{rad}}^{\alpha x_{1}} = \frac{\left(C_{\textrm{rad}}\right)_{12}}{ \det{g_{\textrm{rad}}} } \hspace{0.8cm} \textrm{and} \hspace{0.8cm}
g_{\textrm{rad}}^{\alpha x_{2}} = \frac{\left(C_{\textrm{rad}}\right)_{13}}{ \det{g_{\textrm{rad}}} },
\end{equation}
and the first component of the inverse angular metric
\begin{eqnarray}
g_{\textrm{ang}}^{\chi\chi} = \frac{\left(C_{\textrm{ang}}\right)_{11}}{ \det{g_{\textrm{ang}}} }.
\end{eqnarray}
Once the derivatives with respect to $\alpha$ have been taken, all the above expressions are evaluated at $\alpha = \alpha_{0}$, the fixed size of the giant graviton.
 
The equations of motion for the scalar fluctuations are therefore given by
\begin{eqnarray} 
&& \hspace{-0.25cm}  \left( \Box \hspace{0.075cm} \delta v_{k} \right)  + h^{tt} \hspace{0.1cm} \delta v_{k} = 0 \label{eom-vk}\\
&& \hspace{-0.25cm}  \left(\Box \hspace{0.075cm} \delta\alpha \right) 
+ g_{\textrm{rad}}^{\alpha\alpha}\hspace{0.15cm} \p_{a} \left( \frac{1}{g_{\textrm{rad}}^{\alpha\alpha}}\right) h^{ab} \left( \p_{b}\delta\alpha \right)
- \frac{g_{\textrm{rad}}^{\alpha\alpha}}{\sqrt{-h}} \hspace{0.15cm} \p_{i} \left(\sqrt{-h} \hspace{0.15cm} \frac{g_{\textrm{rad}}^{\alpha i}}{g_{\textrm{rad}}^{\alpha\alpha}} \hspace{0.15cm} h^{tb}\right) \left( \p_{b}\delta\chi \right) = 0 
\hspace{1.2cm} 
\label{eom-alpha} \\
\nonumber && \hspace{-0.25cm}  \left(\Box \hspace{0.075cm} \delta\chi \right) 
+ \left(g_{\textrm{ang}}^{\chi\chi}-1\right) \hspace{0.075cm} \p_{a} \left( \frac{1}{g_{\textrm{ang}}^{\chi\chi}-1}\right) h^{ab} \left( \p_{b}\delta\chi \right)
+ \frac{\left(g_{\textrm{ang}}^{\chi\chi} - 1\right)}{\sqrt{-h}} \hspace{0.15cm} \p_{i} \left(\sqrt{-h} \hspace{0.15cm} \frac{g_{\textrm{rad}}^{\alpha i}}{g_{\textrm{rad}}^{\alpha\alpha}} \hspace{0.15cm} 
h^{tb}\right) \left( \p_{b}\delta\alpha \right) = 0. \\
&& \label{eom-chi}
\end{eqnarray}
The $\mathbb{CP}^{3}$ fluctuations $\delta\alpha$ and $\delta\chi$ are clearly coupled. It is not immediately obvious, without making a specific choice for the radial worldvolume coordinates $x_{1}$ and $x_{2}$,  how to define new 
$\mathbb{CP}^{3}$ fluctuations  $\delta \beta_{\pm}$, in terms of a linear combination of $\delta\alpha$ and $\delta\chi$, such that the equations of motion for $\delta \beta_{+}$ and $\delta \beta_{-}$ decouple.
However, once these equations of motion have been decoupled, the obvious ans\"{a}tze
\begin{eqnarray}
&& \hspace{-0.35cm} \delta v_{k}(t,x_{1},x_{2},\varphi_{1},\varphi_{2}) = e^{i\omega_{k} t}  \hspace{0.075cm}  e^{i m_{k} \varphi_{1}}  \hspace{0.075cm}  e^{i n_{k} \varphi_{2}} \hspace{0.075cm} f_{k}(x_{1},x_{2}) \\
&& \hspace{-0.35cm} \delta \beta_{\pm}(t,x_{1},x_{2},\varphi_{1},\varphi_{2}) = e^{i\omega_{\pm} t}  \hspace{0.075cm}  e^{i m_{\pm} \varphi_{1}}  \hspace{0.075cm}  e^{i n_{\pm} \varphi_{2}} \hspace{0.075cm}  f_{\pm}(x_{1},x_{2})
\end{eqnarray} 
should reduce these problems to second order decoupled partial differential equations for $f_{k}(x_{1},x_{2})$ and $f_{\pm}(x_{1},x_{2})$.  We are interested in solving for the spectrum of eigenfrequencies
$\omega_{k}$ and $\omega_{\pm}$ in terms of the two pairs of integers $m_{k}$ and $n_{k}$, and $m_{\pm}$ and $n_{\pm}$ respectively.

%% file: 06-Limits.tex
\section{Some instructive limits} \label{section - limits}

In this section, we make a specific choice of the generic radial worldvolume coordinates $x_{1}$ and $x_{2}$ of Section \ref{section - fluctuations}.  
Our parameterization describes the full radial worldvolume of a submaximal giant graviton of size $\alpha_{0}$.  
Although it should, theoretically, be possible to write down the equations of motion (\ref{eom-vk})-(\ref{eom-chi}) explicitly, it appears that these are too complex to obtain in full generality, even assisted by a numerical
package such as Maple.  
We therefore confine our attention to the limiting case of the small giant graviton: the equations of motion are found to leading order and next-to-leading order in $\alpha_{0}$.
Although we anticipate no dependence on the size $\alpha_{0}$ at leading order, we hope to observe an $\alpha_{0}$-dependence in the spectrum at next-to-leading order, indicating that we are starting to probe the non-trivial geometry of the giant's worldvolume.  The spectrum of the maximal giant graviton - being simply that of two dibaryons - is already known \cite{MP}.

\subsection{Radial worldvolume coordinates}

The radial worldvolume of a submaximal giant graviton of size $\alpha_{0}$ shall now be described using two sets of nested polar coordinates\footnote{Note that this parameterization breaks the $y^{2}$-$z_{i}$ symmetry of the giant graviton constraint.  This is perfectly reasonable, however, given the different coordinate ranges of $y$ and $z_{i}$.} $(r_{1}(\alpha_{0},\theta),\theta)$ and $(r_{2}(\alpha_{0},\theta,\phi),\phi)$:

The giant graviton constraint equation (\ref{gg-constraint-yz}) describes a surface in the radial space $(y,z_{1},z_{2})$.  Let us first turn off one of the $z_{i}$ coordinates, say $z_{2}$, and parameterize the intersection of this surface with the 
$yz_{1}$-plane.  Setting $z_{1} \equiv z$ and $z_{2} = 0$ yields
\begin{equation}
\left(1-y^{2}\right)\left(1-z\right) = 1 - \alpha_{0}^{2},
\end{equation}
which is described by the polar ansatz $y \equiv r_{1} \cos{\theta}$ and $\sqrt{z} \equiv r_{1} \sin{\theta}$, if the polar radius $r_{1}(\alpha_{0},\theta)$ satisfies
\begin{equation} \label{constraint-r1}
\sin^{2}(2\theta) \hspace{0.05cm} r_{1}^{4} - 4r_{1}^{2} + 4\alpha_{0}^{2} = 0.
\end{equation}

To obtain the full surface, we need to extend this curve into the 3-dimensional radial space by requiring that the $z_{i}$ coordinates now satisfy
\begin{equation}
\left( 1-z_{1} \right) \left( 1-z_{2} \right) = 1-z = 1- r_{1}^{2}\sin^{2}{\theta}.
\end{equation}
Another polar ansatz $\sqrt{z_{1}} \equiv r_{2}\cos{\phi}$ and $\sqrt{z_{2}} \equiv r_{2}\sin{\phi}$ then yields the complete parameterization, if $r_{2}(\alpha_{0},\theta,\phi)$ obeys
\begin{equation} \label{constraint-r2}
\sin^{2}(2\phi) \hspace{0.05cm} r_{2}^{4} - 4r_{2}^{2} + 4 r_{1}^{2} \sin^{2}{\theta} = 0.
\end{equation}

Promoting $\alpha$ to a radial coordinate and defining
\begin{eqnarray}
&& y = r_{1}(\alpha,\theta) \cos{\theta} \\
&& z_{1} = r_{2}^{2}(\alpha,\theta,\phi) \cos^{2}{\phi} \\
&& z_{2} = r_{2}^{2}(\alpha,\theta,\phi) \sin^{2}{\phi},
\end{eqnarray}
with the polar radii $r_{1}$ and $r_{2}$ the positive roots of\footnote{We have chosen the solution to each of the quadratic constraint equations (\ref{constraint-r1}) and (\ref{constraint-r2}) which avoids the singularities at 
$\theta=0$ and $\theta = \pi$, and $\phi=0$ respectively.}
\begin{eqnarray}
&& r_{1}^{2}(\alpha,\theta) = \frac{2}{\sin^{2}(2\theta)} \left\{ 1 - \sqrt{1-\alpha^{2}\sin^{2}(2\theta)} \right\} \\
&& r_{2}^{2}(\alpha,\theta,\phi) = \frac{2}{\sin^{2}(2\phi)} \left\{ 1 - \sqrt{1-  r_{1}^{2}(\alpha,\theta) \sin^{2}{\theta} \sin^{2}(2\phi)} \right\},
\end{eqnarray}
we observe that $\alpha = \alpha_{0}$ describes the radial worldvolume of the submaximal giant graviton.
Here the radial worldvolume coordinates $x_{1} \equiv \theta \in [0,\pi]$ and $x_{2} \equiv \phi \in [0,\tfrac{\pi}{2}]$ have fixed ranges (which is required by our general fluctuation analysis in Section \ref{section - fluctuations}).

\subsection{Small giant graviton}

\emph{\textbf{Leading order in $\alpha_{0}^{2}$}}
\smallskip

Let us now focus on the small giant graviton, for which $0 < \alpha_{0} \ll 1$.  We can expand the square roots in $r_{1}$ and $r_{2}$ to leading order in $\alpha$ to obtain $r_{1}(\theta) \approx \alpha$ 
and $r_{2}(\theta,\phi) \approx \alpha \sin{\theta}$.  Our radial coordinates then become
\begin{eqnarray} \label{radial-LO-sphere}
&& y \approx \alpha\cos{\theta} \\
&& z_{1} \approx \alpha^{2} \sin^{2}{\theta} \cos^{2}{\phi} \\
&& z_{2} \approx \alpha^{2} \sin^{2}{\theta} \sin^{2}{\phi}
\end{eqnarray}
in the vicinity of the $\alpha = \alpha_{0}$ surface.  This approximate radial projection of the giant graviton is simply a 2-sphere in $(y,\sqrt{z_{1}},\sqrt{z_{2}})$-space.

The equations of motion were obtained from (\ref{eom-vk})-(\ref{eom-chi}) to leading order in $\alpha_{0}$.  Rescaling 
$\delta\tilde{\alpha} \equiv \alpha_{0}\hspace{0.05cm} \delta\alpha$, our results can be summarized as follows:
\begin{eqnarray}
   \left[M^{ab}\, \p_{a}\p_{b} +   \hat{k}^{a} \p_{a} + 1 \right] \delta v_{k} \label{eom-numeric-vk} &=& 0\\
   \left[M^{ab}\, \p_{a}\p_{b} +       {k}^{a} \p_{a} \right] \delta\tilde{\alpha}   + \left[ {\ell}^{a} \p_{a} \right] \delta\chi &=&0\label{eom-numeric-alpha}  \\
   \left[M^{ab}\, \p_{a}\p_{b} + \tilde{k}^{a} \p_{a} \right] \delta\chi   - \left[ \tilde{\ell}^{a} \p_{a} \right]\delta \tilde{\alpha} &=&0,  \label{eom-numeric-chi}
\end{eqnarray}
where the inverse metric on the worldvolume of the giant graviton, rescaled by a factor of $(h^{tt})^{-1}$ for convenience, is approximated to leading order as follows:
\begin{equation}
M^{ab} \approx M^{ab}_{(1)} =
\begin{pmatrix}
  1  		& 		0  	& 0 		& -\frac{1}{2}	& -\frac{1}{2} \\
  0  		& -\frac{1}{2}	& 0 	& 0		& 0 \\
  0  		& 0  		& F_{1} 	& 0 		& 0 \\
  -\frac{1}{2}  & 0  		& 0 	& F_{1}\sec^{2}{\phi} + \frac{1}{4} 		& \frac{1}{4} \\
  -\frac{1}{2}  & 0  		& 0 	& \frac{1}{4}	& F_{1}\csc^{2}{\phi} + \frac{1}{4}
\end{pmatrix}, 
\end{equation}
while
\begin{eqnarray}
&& \hspace{-0.35cm} \hat{k}^{a} \approx \hat{k}^{a}_{(1)} \equiv  \begin{pmatrix}
    0 & F_{2} & F_{4} & 0 & 0 
  \end{pmatrix} \\
&& \hspace{-0.35cm} k^{a} \approx k^{a}_{(1)} \hspace{0.5cm} \textrm{and} \hspace{0.5cm} \tilde{k}^{a} \approx \tilde{k}^{a}_{(1)},  \hspace{0.5cm} \textrm{with} \hspace{0.3cm} k^{a}_{(1)}  = \tilde{k}^{a}_{(1)} 
\equiv
  \begin{pmatrix}
  0 & F_{3} & F_{4} & 0 & 0
  \end{pmatrix} \\
&& \hspace{-0.35cm} \ell^{a} \approx \ell^{a}_{(1)}  \hspace{0.59cm} \textrm{and} \hspace{0.5cm}  \tilde{\ell}^{a}\approx \tilde{\ell}^{a}_{(1)},   \hspace{0.6cm} \textrm{with} \hspace{0.3cm} \ell^{a}_{(1)} = \tilde{\ell^{a}}_{(1)} 
\equiv F_{5}
  \begin{pmatrix}
    -2 & 0 & 0 &  1 & 1 
  \end{pmatrix},
\end{eqnarray}
in terms of the following functions of the radial worldvolume coordinates $\theta$ and $\phi$:
\begin{eqnarray}
&& F_{1} = -\frac{(2-\sin^{2}{\theta})}{4 \sin^{2}{\theta}}  \\
&& F_{2} = -\frac{3}{2}\cot{\theta} \\
&& F_{3} = -\frac{1}{2}\left[\frac{4}{(2-\sin^{2}{\theta})} + 1 \right]\cot{\theta} 
\end{eqnarray}
\begin{eqnarray}
&& F_{4} = F_{1}\left(\cot{\phi} - \tan{\phi}\right) \\
&& F_{5} = \frac{1}{(2-\sin^{2}{\theta})} .
\end{eqnarray}
We are now able to decouple the leading order equations of motion (\ref{eom-numeric-alpha})-(\ref{eom-numeric-chi}) for the $\mathbb{CP}^{3}$ scalar fluctuations by defining 
$\delta\beta_{\pm} \equiv \tilde{\delta\alpha} \pm i\delta\chi$ to obtain
\begin{equation} \label{eom-numeric-beta}
 \left[M^{ab} \hspace{0.05cm}  \p_{a}\p_{b} + {k}^{a} \p_{a} \mp i \hspace{0.05cm} {\ell}^{a} \p_{a} \right] \delta\beta_{\pm} \approx 0. 
\end{equation}

Let us now make the ans\"{a}tze
\begin{eqnarray}
&& \hspace{-0.35cm} \delta v_{k}(t,\theta,\phi,\varphi_{1},\varphi_{2}) = e^{i\omega_{k} t}  \hspace{0.075cm}  e^{i m_{k} \varphi_{1}}  \hspace{0.075cm}  e^{i n_{k} \varphi_{2}} \hspace{0.075cm} f_{k}(\theta,\phi) \label{exp-ansatz-vk}\\
&& \hspace{-0.35cm} \delta \beta_{\pm}(t,\theta,\phi,\varphi_{1},\varphi_{2}) = e^{i\omega_{\pm} t}  \hspace{0.075cm}  e^{i m_{\pm} \varphi_{1}}  \hspace{0.075cm}  e^{i n_{\pm} \varphi_{2}} \hspace{0.075cm}  f_{\pm}(\theta,\phi), 
\label{exp-ansatz-beta}
\end{eqnarray} 
with $m_{k}$ and $n_{k}$, and $m_{\pm}$ and $n_{\pm}$ integers.  The leading order decoupled equations of motion (\ref{eom-numeric-vk}) and (\ref{eom-numeric-beta}) become
\begin{eqnarray}
\nonumber && \hspace{-0.35cm} \left\{ \tfrac{1}{2} \hspace{0.075cm} \p_{\theta}^{2} - F_{1} \hspace{0.05cm}\p_{\phi}^{2}  - F_{2} \hspace{0.05cm} \p_{\theta} - F_{4} \hspace{0.075cm} \p_{\phi} \right. \\
&& \hspace{-0.35cm} \hspace{0.25cm} \left.  +  \left[\tilde{\omega}_{k}^{2} + \left(F_{1}\sec^{2}{\phi}\right) m_{k}^{2} + \left(F_{1}\csc^{2}{\phi}\right) n_{k}^{2}  - 1 \right] \right\} f_{k}(\theta,\phi)= 0 \\
\nonumber && \\
\nonumber && \hspace{-0.35cm} \left\{ \tfrac{1}{2} \hspace{0.075cm} \p_{\theta}^{2} - F_{1} \hspace{0.05cm}\p_{\phi}^{2}  - F_{3} \hspace{0.05cm} \p_{\theta} - F_{4} \hspace{0.075cm} \p_{\phi} \right. \\
&& \hspace{-0.35cm} \hspace{0.25cm} \left. + \left[\tilde{\omega}_{\pm}^{2} \pm 2F_{5} \hspace{0.075cm} \tilde{\omega}_{\pm} + \left(F_{1}\sec^{2}{\phi}\right) m_{\pm}^{2} + \left(F_{1}\csc^{2}{\phi}\right) n_{\pm}^{2} \right] \right\} f_{\pm}(\theta,\phi) = 0, 
\end{eqnarray}
where we have shifted the eigenfrequencies as follows: 
\begin{equation} \label{shifted-freq}
\tilde{\omega}_{k} = \omega_{k} - \tfrac{1}{2}\left(m_{k} + n_{k}\right) \hspace{1.0cm} \textrm{and} \hspace{1.0cm} \tilde{\omega}_{\pm} = \omega_{\pm} - \tfrac{1}{2}\left(m_{\pm} + n_{\pm}\right).
\end{equation}

These second order partial differential equations admit separable ans\"{a}tze
\begin{equation} 
f_{k}(\theta,\phi) \equiv \Theta_{k}(\theta) \hspace{0.075cm} \Phi_{k}(\phi) \hspace{1.0cm} \textrm{and} \hspace{1.0cm}
f_{\pm}(\theta,\phi) \equiv \Theta_{\pm}(\theta) \hspace{0.075cm} \Phi_{\pm}(\phi),
\end{equation}
which reduce the problems to
\begin{eqnarray}
&& \hspace{-0.25cm} \frac{d^{2}\Theta_{k}}{d \hspace{0.025cm} \theta^{2}} + 3\cot{\theta} \hspace{0.15cm} \frac{d \hspace{0.025cm} \Theta_{k}}{d \hspace{0.025cm} \theta} 
+ \left[ 2\left(\tilde{\omega}_{k}^{2} - 1 \right) - \frac{\lambda_{k}(2-\sin^{2}{\theta})}{2\sin^{2}{\theta}}\right] \Theta_{k} = 0 \hspace{1.2cm} \label{LO-separable-Theta-k} \\
&& \hspace{-0.25cm} \frac{d^{2}\Phi_{k}}{d \hspace{0.025cm} \phi^{2}} + \left( \cot{\phi} - \tan{\phi} \right) \frac{d \hspace{0.025cm} \Phi_{k}}{d \hspace{0.025cm} \phi} + \left[ \lambda_{k} - m_{k}^{2}\sec^{2}{\phi} - n_{k}^{2}\csc^{2}{\phi} \right] \Phi_{k} = 0  \hspace{1.2cm} \label{LO-separable-Phi-k}
\end{eqnarray}
and
\begin{eqnarray}
\nonumber && \hspace{-0.25cm} \frac{d^{2}\Theta_{\pm}}{d \hspace{0.025cm} \theta^{2}} + \left[ \frac{4}{(2 - \sin^{2}{\theta})} + 1 \right]\cot{\theta} \hspace{0.15cm} \frac{d \hspace{0.025cm} \Theta_{\pm}}{d \hspace{0.025cm} \theta} 
+ \left[ 2 \hspace{0.025cm} \tilde{\omega}_{\pm}^{2} \pm \frac{4 \hspace{0.025cm} \tilde{\omega}_{\pm}}{(2 - \sin^{2}{\theta})} -  \frac{\lambda_{\pm}(2-\sin^{2}{\theta})}{2\sin^{2}{\theta}}\right] \Theta_{\pm} = 0 
\label{LO-separable-Theta-pm} \\
&& \\
&& \hspace{-0.25cm} \frac{d^{2}\Phi_{\pm}}{d\hspace{0.025cm}\phi^{2}} + \left( \cot{\phi} - \tan{\phi} \right) \frac{d\hspace{0.025cm}\Phi_{\pm}}{d \hspace{0.025cm} \phi} 
+ \left[\lambda_{\pm} - m_{\pm}^{2}\sec^{2}{\phi} - n_{\pm}^{2}\csc^{2}{\phi} \right] \Phi_{\pm} = 0, \hspace{1.2cm} \label{LO-separable-Phi-pm}
\end{eqnarray}
with $\lambda_{k}$ and $\lambda_{\pm}$ constant.  The solutions of these second order ordinary differential equations, on the intervals $\theta \in [0,\pi]$ and $\phi \in [0,\tfrac{\pi}{2}]$ respectively, can be obtained in terms of hypergeometric and Heun functions, as we shall now briefly describe.  It is clear, however, even without the solutions, that the spectrum of energy eigenvalues $\omega_{k}$ and $\omega_{\pm}$ is independent of the size 
$\alpha_{0}$ of the giant graviton to leading order.

The differential equations (\ref{LO-separable-Phi-k}) and (\ref{LO-separable-Phi-pm}), which describe the $\phi$ dependence of the scalar fluctuations of the AdS and $\mathbb{CP}^{3}$ coordinates respectively, take the same generic form. Taking the ansatz $\Phi(z) = z^{\frac{1}{2}|m|} \left(1-z\right)^{\frac{1}{2}|n|} g(z)$, with $z \equiv \cos^{2}{\phi} \in [0,1]$, these can be written in the standard hypergeometric form\footnote{Here we drop the $k$ and $\pm$ subscripts temporarily, since the same differential equation for $\Phi(\phi)$ applies in both cases.}
\begin{equation}
z\left(1-z\right) \frac{d^{2}g}{d \hspace{0.025cm} z^{2}} \hspace{0.05cm} +  \hspace{0.05cm} \left[ \left(|m|+1\right) - \left(|m|+|n|+2\right) \right] \frac{d \hspace{0.025cm} g}{d \hspace{0.025cm} z}  \hspace{0.05cm} 
-  \hspace{0.05cm} \frac{1}{4} \left[ \left( |m| + |n| + 1 \right)^{2} - \left(\lambda + 1\right) \right] g = 0.
\end{equation}
Similar problems were studied in \cite{HMP,MP,Gubser}.  The solutions $g(z) = F(a,b,c;z)$ are hypergeometric functions, dependent on the usual constants
\begin{equation}
a, \hspace{0.05cm} b \equiv \frac{1}{2} \left\{ |m| + |n| + 1 \pm \sqrt{\lambda + 1} \right\} \hspace{1.0cm} \textrm{and} \hspace{1.0cm} c \equiv |m| + 1, \hspace{1.0cm}
\end{equation}
which are regular on the interval $[0,1]$ when $a$ or $b$ is a non-positive integer.  Hence
\begin{equation}
 |m| + |n| + 1 \pm \sqrt{\lambda + 1}   = -2s_{1}, \hspace{1.0cm} \textrm{with} \hspace{0.5cm} s_{1} \in \{0,1,2,\ldots\},
\end{equation}
from which it follows that $\lambda = l\left(l+2\right)$, with $l \equiv 2s_{1} + |m| + |n|$.  Notice that these constants $\lambda$ are just the usual eigenvalues of the Laplacian \cite{MZ-SCS} on the 
complex projective space $\mathbb{CP}^{2}$. 

Let us first consider the second order differential equation  (\ref{LO-separable-Theta-k}), which describes the $\theta$ dependence of the scalar fluctuations of the AdS directions.
If we now set $\Theta_{k}(x) \equiv  x^{\frac{l_{k}}{2}} (1-x)^{\frac{l_{k}}{2}}  h_{k}(x)$, with $x \equiv \sin^{2}{\tfrac{\theta}{2}} \in [0,1]$, this can be written in the standard hypergeometric form
\begin{equation}
x\left(1-x\right) \frac{d^{2}h_{k}}{dx^{2}} + \left[ \left( l_{k} + 2 \right) - 2(l_{k} + 2) x \right] \frac{dh_{k}}{dx} 
- \left[ \tfrac{1}{2} \hspace{0.05cm} l_{k}^{2} + 2l_{k} - 2\left( \tilde{\omega}_{k}^{2} - 1 \right) \right] h_{k} = 0,
\end{equation}
where $\lambda_{k} = l_{k}\left(l_{k} + 2\right)$, with $l_{k} \equiv 2s_{k,1} + |m_{k}| + |n_{k}|$, are the eigenvalues of the $\Phi_{k}$ differential equation (\ref{LO-separable-Phi-k}).
The solutions $h_{k}(x) = F(a_{k},b_{k},c_{k};x)$ are associated with the usual hypergeometric parameters
\begin{equation}
a_{k},b_{k} =  \left( l_{k} + \tfrac{3}{2} \right) \pm \sqrt{\tfrac{1}{2} \hspace{0.05cm} l_{k}^{2} + l_{k} + \tfrac{9}{4} + 2\left(\tilde{\omega}_{k}^{2} - 1 \right) } 
\hspace{0.5cm} \textrm{and} \hspace{0.5cm}
c_{k} = l_{k} + 2.
\end{equation}
For regularity on $[0,1]$, we require that either $a_{k}$ or $b_{k}$ be a non-positive integer:
\begin{equation}
\left( l_{k} + \tfrac{3}{2} \right)- \sqrt{\tfrac{1}{2} \hspace{0.05cm} l_{k}^{2} + l_{k} + \tfrac{9}{4} + 2\left(\tilde{\omega}_{k}^{2} - 1 \right) }  = -s_{k,2}, 
\hspace{0.6cm} \textrm{with} \hspace{0.3cm} s_{k,2} \in \{0,1,2,\ldots\}.
\end{equation}
We can hence determine an equation for the shifted frequencies squared of the AdS fluctuations about the small giant graviton to leading order in $\alpha_{0}$:
\begin{eqnarray}
\nonumber & \hspace{-0.25cm}\tilde{\omega}_{k}^{2} & = \left[\omega_{k} - \tfrac{1}{2} \left( m_{k} + n_{k} \right) \right]^{2}  \\
&& = \tfrac{1}{2} \left[ 2s_{k,1} + s_{k,2} + |m_{k}| + |n_{k}|  + \tfrac{3}{2} \right]^{2} - \tfrac{1}{4} \left[2s_{k,1} + |m_{k}| + |n_{k}| + 1 \right]^{2} + \tfrac{1}{8} \hspace{1.2cm}
\end{eqnarray}
in terms of the non-negative integers $s_{k,1}$ and $s_{k,2}$.  Notice that there are no complex energy eigenvalues, indicating stability.  As expected, there are also no zero modes associated with the fluctuations in the AdS spacetime.

Let us focus momentarily on the s-modes, obtained by setting $s_{k,1} = s_{k,2} = 0$.  We can express these lowest frequencies as follows:
\begin{equation}
\omega_{k} = \tfrac{1}{2}\left(m_{k} + n_{k} \right) \pm \left[\tfrac{1}{2} \left( |m_{k}| + |n_{k}| \right) + 1 \right],
\end{equation}
which can be divided into two cases, depending on the relative signs of $m_{k}$ and $n_{k}$.  More specifically, we find that 
\begin{eqnarray} 
\nonumber & \omega_{k} = \textrm{sign}(m_{k})\left[ |m_{k}| + |n_{k}| + 1 \right] \hspace{0.3cm} \textrm{or} \hspace{0.3cm} \omega_{k} = -\textrm{sign}(m_{k}) 1, \hspace{0.7cm}
& \textrm{when $m_{k}n_{k} \geq 0$} \hspace{1.35cm} \\
& \hspace{-0.5cm} \omega_{k} = \textrm{sign}(m_{k}) \left[|m_{k}| + 1 \right] \hspace{0.3cm} \textrm{or} \hspace{0.3cm} \omega_{k} = \textrm{sign}(n_{k}) \left[|n_{k}| + 1 \right], \hspace{0.275cm}
& \textrm{when $m_{k}n_{k} < 0$}. \hspace{1.35cm}
\end{eqnarray}

We shall now consider the second order differential equation  (\ref{LO-separable-Theta-pm}), which describes the $\theta$ dependence of the scalar fluctuations of the 
transverse $\mathbb{CP}^{3}$ coordinates.  Setting 
$\Theta_{\pm}(\tilde{x}) \equiv   x^{\frac{l_{\pm}}{2}} (1-x)^{\frac{l_{\pm}}{2}} h_{\pm}(\tilde{x})$, where $\tilde{x} = 4x(1-x)$ and $x \equiv \sin^{2}{\tfrac{\theta}{2}}$ as before\footnote{Note that $\theta$ runs over the interval $[0,\pi]$,
so that $\tilde{x} = \sin^{2}{\theta}$ double covers the interval $[0,1]$, while $x \equiv \sin^{2}{\tfrac{\theta}{2}}$ covers it only once.}, we obtain a Heun differential equation
\begin{eqnarray}
\nonumber && \hspace{-0.35cm} \frac{d^{2}h_{\pm}}{d\tilde{x}^{2}} + \left[\frac{\left(l_{\pm}+2\right)}{\tilde{x}} + \frac{\tfrac{1}{2}}{\left(\tilde{x}-1\right)} + \frac{(-1)}{\left( \tilde{x}-2 \right)} \right] \frac{dh_{\pm}}{d\tilde{x}}  \\
&& \hspace{-0.35cm} + \hspace{0.075cm} \frac{ \left[\tfrac{1}{8} \hspace{0.025cm} l_{\pm}^{2} - \tfrac{1}{2} \hspace{0.025cm} \tilde{\omega}_{\pm}^{2}\right]\tilde{x} - \left[\tfrac{1}{4}\left(l_{\pm} + 1\right)^{2} - (\tilde{\omega}_{\pm} + \tfrac{1}{2} )^{2}\right]  }{\tilde{x}\left(\tilde{x}-1\right)\left(\tilde{x}-2\right)} 
\hspace{0.1cm} h_{\pm} = 0.
\end{eqnarray}
Again $\lambda_{\pm} = l_{\pm}\left(l_{\pm} + 2\right)$, with $l_{\pm} \equiv 2s_{\pm,1} + |m_{\pm}| + |n_{\pm}|$, are the eigenvalues of the $\Phi_{\pm}$ differential equation  (\ref{LO-separable-Phi-pm}).  The Heun solutions
$h_{\pm}(\tilde{x}) = F(2,q_{\pm};a_{\pm},b_{\pm},c_{\pm},d_{\pm};\tilde{x})$ depend on the parameters
\begin{eqnarray}
\nonumber \hspace{0.0cm} a_{\pm}, \hspace{0.05cm} b_{\pm} = \frac{1}{2} \left\{ l_{\pm} + \tfrac{1}{2} \pm \sqrt{\tfrac{1}{2} \hspace{0.05cm} l_{\pm}^{2} + l_{\pm} + \tfrac{1}{4} + 2\hspace{0.05cm} \tilde{\omega}_{\pm}^{2} } \right\},
\hspace{0.45cm} c_{\pm} = l_{\pm} + 2, \hspace{0.45cm} d_{\pm} = \tfrac{1}{2}, \hspace{0.45cm} e_{\pm} = -1  \\
\end{eqnarray}
and the accessory parameter
\begin{equation}
q_{\pm} = \tfrac{1}{4}\left(l_{\pm} + 1\right)^{2} - \left( \tilde{\omega}_{\pm} \pm \tfrac{1}{2} \right)^{2}.
\end{equation}
There are  several different regular classes of Heun functions \cite{Heun}.  All possible regular solutions are obtained, in this case, by requiring that either $a_{\pm}$ or $b_{\pm}$ be a non-positive integer or half-integer:
\begin{equation}
\left( l_{\pm} + \tfrac{1}{2} \right) - \sqrt{\tfrac{1}{2} \hspace{0.05cm} l_{\pm}^{2} + l_{\pm} + \tfrac{1}{4} + 2 \hspace{0.05cm} \tilde{\omega}_{k}^{2} }  = -s_{\pm,2}, 
\hspace{0.6cm} \textrm{with} \hspace{0.3cm} s_{\pm,2} \in \{0,1,2,\ldots\}.
\end{equation}
It is hence possible to find an equation for the shifted frequencies squared of the $\mathbb{CP}^{3}$ fluctuations about the small giant graviton to leading order in $\alpha_{0}$:
\begin{eqnarray}
\nonumber & \hspace{-0.25cm} \tilde{\omega}_{\pm}^{2} & = \left[\omega_{\pm} - \tfrac{1}{2} \left( m_{\pm} + n_{\pm} \right) \right]^{2}  \\
& \hspace{-0.25cm} & = \tfrac{1}{2} \left[ 2s_{\pm,1} + s_{\pm,2} + |m_{\pm}| + |n_{\pm}|  + \tfrac{1}{2} \right]^{2} - \tfrac{1}{4} \left[2s_{\pm,1} + |m_{\pm}| + |n_{\pm}| + 1 \right]^{2} + \tfrac{1}{8} \hspace{1.2cm}
\end{eqnarray}
in terms of the non-negative integers $s_{k,1}$ and $s_{k,2}$.  Notice that, again, there are no complex energy eigenvalues, indicating stability.

The s-modes are associated with the lowest frequencies, obtained by setting $s_{\pm,1} = s_{\pm,2} = 0$,  which are given by
\begin{equation}
\omega_{\pm} = \tfrac{1}{2}\left(m_{\pm} + n_{\pm} \right) \pm \tfrac{1}{2} \left( |m_{\pm}| + |n_{\pm}| \right).
\end{equation}
If the integers $m_{\pm}$ and $n_{\pm}$ have the same sign, this yields simply $\omega_{\pm} = (m_{\pm} + n_{\pm})$ or $\omega_{\pm} = 0$, whereas, if $m_{\pm}$ and $n_{\pm}$ have different signs, we obtain
$\omega_{\pm} = m_{\pm}$ or $\omega_{\pm} = n_{\pm}$.  Notice that there are zero modes associated with these $\mathbb{CP}^{3}$ fluctuations.  This is to be expected, since changing the size $\alpha_{0}$ of the giant does not cost any extra energy. We anticipate that these lowest frequencies should match the conformal dimensions of BPS excitations of the dual ABJM subdeterminant operator.

\medskip
\emph{\textbf{Next-to-leading order in $\alpha_{0}$}}
\smallskip

The equations of motion to next-to-leading order in $\alpha_{0}$ can again be written in the form (\ref{eom-numeric-vk})-(\ref{eom-numeric-chi}), where we now include an additional higher order term in the rescaled inverse
worldvolume metric 
$ M^{ab} \approx M^{ab}_{(1)} + \alpha_{0} \, M^{ab}_{(2)}$,  with
\begin{eqnarray}
 M^{ab}_{(2)} \equiv \cos{\theta}
\begin{pmatrix}
 0		& 0				&	0	& \tfrac{1}{2}    & -\tfrac{1}{2}  \\ 
 0			&  \tfrac{1}{2} \cos{(2\phi)}	& -\cot{\theta} \sin{(2\phi)}	& 0						& 0				\\ 
 0			&  -\cot{\theta}\sin{(2\phi)}			& -\tfrac{1}{2}\cot^{2}{\theta}\cos{(2\phi)}	& 0		         & 0				\\ 
 \tfrac{1}{2} & 0				& 0		& \tfrac{1}{2}\cot^{2}{\theta}\sec^{2}{\phi} & 0	\\ 
 -\tfrac{1}{2} & 0				& 0		& 0	& \tfrac{1}{2}\cot^{2}{\theta}\csc^{2}{\phi} 
\end{pmatrix}. \hspace{1.0cm}
\end{eqnarray}
The other next-to-leading order coefficients are
\begin{eqnarray}
&& \hspace{-0.35cm} \hat{k}^{a} \approx \hat{k}^{a}_{(1)} + \alpha_{0} \hspace{0.05cm}  \hat{k}^{a}_{(2)}, \hspace{0.49cm} \textrm{with} \hspace{0.3cm} 
   \hat{k}^{a}_{(2)} \equiv  \begin{pmatrix} 0 &  S_{2} & \hat{S}_{4} & 0 & 0\end{pmatrix} \\
&& \hspace{-0.35cm} k^{a} \approx k^{a}_{(1)} + \alpha_{0} \hspace{0.05cm}  k^{a}_{(2)}, \hspace{0.5cm} \textrm{with} \hspace{0.3cm} 
    k^{a}_{(2)} \equiv \begin{pmatrix}    0 & S_{3} & S_{4} & 0 & 0   \end{pmatrix} \\
&& \hspace{-0.35cm}\tilde{k}^{a} \approx \tilde{k}^{a}_{(1)} + \alpha_{0} \hspace{0.05cm}  \tilde{k}^{a}_{(2)}, \hspace{0.5cm} \textrm{with} \hspace{0.3cm} 
    \tilde{k}^{a}_{(2)} = k^{a}_{(2)} \\
&& \hspace{-0.35cm} \ell^{a} \approx \ell^{a}_{(1)} + \alpha_{0} \hspace{0.05cm}  \ell^{a}_{(2)} ,\hspace{0.65cm} \textrm{with} \hspace{0.3cm} 
    \ell^{a}_{(2)} \equiv S_{5} \begin{pmatrix}  -2 & 0 & 0 &  1 & 1  \end{pmatrix} - S_{6} \begin{pmatrix}   0 & 0 & 0 &  1 & -1   \end{pmatrix}  \\
&& \hspace{-0.35cm}\tilde{\ell}^{a} \approx \tilde{\ell}^{a}_{(1)} + \alpha_{0} \hspace{0.05cm}  \tilde{\ell}^{a}_{(2)},  \hspace{0.65cm} \textrm{with} \hspace{0.3cm} 
    \tilde{\ell}^{a}_{(2)} = \ell^{a}_{(2)},
\end{eqnarray}
where
\begin{eqnarray}
&& S_{2}(\theta,\phi) \equiv -\frac{1}{2} \csc{\theta} \cos{(2\phi)} \\
&& S_{3}(\theta,\phi) \equiv -\frac{\cos^{2}{\theta}(4+\sin^4\theta)}{2\sin{\theta}\left( 2 - \sin^{2}{\theta} \right)^{2}} \hspace{0.05cm} \cos{(2\phi)} \\
&& \hat{S}_{4}(\theta,\phi) \equiv -\frac{\cos{\theta}}{\sin^{2}{\theta}}\left[\cos^{2}{\theta}\csc{(2\phi)} - \sin{(2\phi)}\right] \\
&& S_{4}(\theta,\phi) \equiv -\cot^{2}{\theta}\cos{\theta} \csc{(2\phi)} + \frac{\cos{\theta}\left(3+\cos^{4}{\theta}\right)}{2\sin^{2}{\theta}\left(2 - \sin^{2}{\theta}\right)} \hspace{0.05cm} \sin{(2\phi)}  \\
&& S_{5}(\theta,\phi) \equiv \frac{\sin^{2}{\theta}\cos{\theta}}{(2-\sin^{2}{\theta})^{2}} \hspace{0.05cm} \cos{(2\phi)} \\
&& S_{6}(\theta) \equiv -\frac{\cos{\theta}\left(4-\sin^{2}{\theta}\right)}{2(2-\sin^{2}{\theta})}.
\end{eqnarray}
These are simply the next-to-leading order functions to be associated with the leading order functions $F_{2}$, $F_{3}$ and $F_{4}$ (which appears in both $\hat{k}^{a}$ and $k^{a}$).

Notice that, since $k^{a} \approx \tilde{k}^{a}$ and $\ell^{a} \approx \tilde{\ell}^{a}$ to next-to-leading order in $\alpha_{0}$, the equations of motion for the $\mathbb{CP}^{3}$ fluctuations $\delta\tilde{\alpha}$ and $\delta\chi$ can still be decoupled by setting $\delta\beta_{\pm} \equiv \tilde{\delta\alpha} \pm i\delta\chi$.  These equations of motion now become (\ref{eom-numeric-beta}), as before, except in that $M^{ab}$, $k^{a}$ and $\ell^{a}$ now include next-to-leading order terms.

Again taking ans\"{a}tze of the form (\ref{exp-ansatz-vk})-(\ref{exp-ansatz-beta}), describing the oscillatory behaviour of the temporal and angular worldvolume coordinates, the next-to-leading order equations of motion can be written as
\begin{eqnarray}
\nonumber && \hspace{-0.35cm} \left\{ \left[ \tfrac{1}{2} - \tfrac{1}{2}\alpha_{0} \cos{\theta}\cos{(2\phi)}\right] \p^{2}_{\theta} - \left[ F_{1} - \tfrac{1}{2} \alpha_{0} \cos{\theta}\cot^{2}{\theta}\cos{(2\phi)} \right] \p^{2}_{\phi} \right. \\
\nonumber && \hspace{-0.25cm} + \left[ 2\alpha_{0} \cos{\theta} \cot{\theta} \sin{(2\phi)} \right] \p_{\theta}\p_{\phi}  - \left[F_{2} + \alpha_{0} S_{2} \right] \p_{\theta} - \left[ F_{4} + \alpha_{0}\hat{S}_{4} \right] \p_{\phi} \\
\nonumber && \hspace{-0.25cm} + \left[\tilde{\omega}_{k}^{2} + \alpha_{0} \cos{\theta} \hspace{0.15cm} \tilde{\omega}_{k} \left(m_{k} - n_{k}\right) 
+ \left( F_{1}\sec^{2}{\phi} + \tfrac{1}{2}\alpha_{0} \cos{\theta} \left(\cot^{2}{\theta}\sec^{2}{\phi} + 1\right) \right)m_{k}^{2} \right. \\
&& \hspace{-0.25cm} \left. \left. \hspace{0.5cm} + \left( F_{1}\csc^{2}{\phi} + \tfrac{1}{2}\alpha_{0} \cos{\theta} \left(\cot^{2}{\theta}\csc^{2}{\phi} - 1\right) \right)n_{k}^{2} - 1 \right] \right\} f_{k}(\theta,\phi) = 0 \\
&& \nonumber \\
\nonumber && \hspace{-0.35cm} \left\{ \left[ \tfrac{1}{2} - \tfrac{1}{2}\alpha_{0} \cos{\theta}\cos{(2\phi)}\right] \p^{2}_{\theta} - \left[ F_{1} - \tfrac{1}{2} \alpha_{0} \cos{\theta}\cot^{2}{\theta}\cos{(2\phi)} \right] \p^{2}_{\phi}  \right. \\
\nonumber && \hspace{-0.25cm} + \left[2\alpha_{0} \cos{\theta} \cot{\theta} \sin{(2\phi)} \right] \p_{\theta}\p_{\phi} - \left[F_{3} + \alpha_{0} S_{3} \right] \p_{\theta} - \left[ F_{4} + \alpha_{0} S_{4} \right] \p_{\phi} \\
\nonumber && \hspace{-0.25cm} + \left[\tilde{\omega}_{\pm}^{2} + \alpha_{0} \cos{\theta} \hspace{0.15cm} \tilde{\omega}_{\pm} \left(m_{\pm} - n_{\pm}\right) \pm 2\left(F_{5} + \alpha_{0}S_{5}\right)\tilde{\omega}_{\pm}
 \pm \alpha_{0} S_{6}\left(m_{\pm}-n_{\pm}\right) \right. \\
\nonumber && \hspace{-0.25cm} \hspace{0.57cm} + \left( F_{1}\sec^{2}{\phi} + \tfrac{1}{2}\alpha_{0} \cos{\theta} \left(\cot^{2}{\theta}\sec^{2}{\phi} + 1\right) \right)m_{\pm}^{2} \\
&& \hspace{-0.25cm} \left. \left. \hspace{0.5cm} 
 + \left( F_{1}\csc^{2}{\phi} + \tfrac{1}{2}\alpha_{0} \cos{\theta} \left(\cot^{2}{\theta}\csc^{2}{\phi} - 1\right) \right)n_{\pm}^{2} 
 \right]  \right\} f_{\pm}(\theta,\phi) = 0, 
\end{eqnarray}
in terms of the shifted eigenfrequencies (\ref{shifted-freq}).  These second order partial differential equations no longer admit separable ans\"{a}tze. Note that  $m_{k}$ and $n_{k}$, as well as $m_{\pm}$ and $n_{\pm}$, must be independent of the size $\alpha_{0}$ - these are integers and hence cannot be continuously varied as we change $\alpha_{0}$.  However, we expect the frequencies  $\omega_{k}(\alpha_{0})$ and $\omega_{\pm}(\alpha_{0})$ associated with each pair of integers to pick up an $\alpha_{0}$ dependence, together with the eigenfunctions $f_{k}(\alpha_{0},\theta,\phi)$ and 
$f_{\pm}(\alpha_{0},\theta,\phi)$, since there is now an explicit dependence on $\alpha_{0}$ in the next-to-leading order equations of motion.

\medskip
\emph{\textbf{Next-to-next-to-leading order in $\alpha_{0}$}}
\smallskip

To obtain the leading and next-to-leading order equations of motion, it was sufficient to make use of the spherical parameterization (\ref{radial-LO-sphere}) of the radial worldvolume.  The next-to-leading order $\alpha_{0}$ terms came from including additional $\alpha_{0}$ terms in the metric and not from changing our parameterization of the radial surface.
At higher orders, however, we need to include additional 
$O(\alpha^{3})$ terms in the functions $r_{1}$ and $r_{2}$, which describe the deviation of the radial worldvolume away from the spherical:
\begin{eqnarray} \label{radial-NNLO}
&& r_{1}(\theta) \approx \alpha \left\{ 1 + \tfrac{1}{2}\alpha^{2}\sin^{2}{\theta} \cos^{2}{\theta} \right\}  \\
&& r_{2}(\theta,\phi) \approx  \alpha \sin{\theta} \left\{ 1 + \tfrac{1}{2} \alpha^{2} \sin^{4}{\theta}\left( \cos^{2}{\theta} + \sin^{2}{\theta}\cos^{2}{\phi} \sin^{2}{\phi} \right) \right\}.
\end{eqnarray}
We should hence make use of the radial coordinates
\begin{eqnarray}
&& y \approx \alpha  \left\{ 1 + \tfrac{1}{2}\alpha^{2}\sin^{2}{\theta} \cos^{2}{\theta} \right\} \cos{\theta} \\
&& z_{1} \approx \alpha^{2} \sin^{2}{\theta} \left\{ 1 + \alpha^{2} \sin^{4}{\theta}\left( \cos^{2}{\theta} + \sin^{2}{\theta}\cos^{2}{\phi} \sin^{2}{\phi} \right) \right\} \cos^{2}{\phi} \\
&& z_{2} \approx \alpha^{2} \sin^{2}{\theta} \left\{ 1 + \alpha^{2} \sin^{4}{\theta}\left( \cos^{2}{\theta} + \sin^{2}{\theta}\cos^{2}{\phi} \sin^{2}{\phi} \right) \right\} \sin^{2}{\phi}.
\end{eqnarray}
We have not written down the next-to-next-to-leading order equations of motion for the scalar fluctuations $\delta v_{k}$, $\delta\tilde{\alpha}$ and $\delta\chi$, since an $\alpha_{0}$-dependence (at least at the level of the decoupled equations of motion) has already been observed at next-to-leading order.  However, in this case, we anticipate that the  equations of motion for the $\mathbb{CP}^{3}$ scalar fluctuations $\delta\tilde{\alpha}$ and $\delta\chi$ will no longer trivially decouple.

%% file: 07-Discussion.tex
\section{Discussion and an outlook to the future} \label{section - discussion}
Showing that all of spacetime and its various properties, size, shape, geometry, topology, locality and causality, are phenomena that are not fundamental but emergent through a vast number of quantum interactions is as ambitious a goal as any in the history of physics. While it is not usually understood as one of the goals of string theory {\it per s\'e}\footnote{Indeed, over the past decade, it has been a fertile pursuit for a number of research programs in quantum gravity \cite{Emergent-gravity}.}, string theory does bring a formidable set of tools to bear on the problem via the AdS/CFT correspondence. 

This article aims to draw attention to the question of how the {\it nontrivial} geometry of a D4-brane giant graviton in type IIA string theory on $AdS_{4} \times \mathbb{CP}^{3}$ is encoded in the dual ABJM super Chern-Simons theory. To this end, we have focused on the gravity side of the correspondence and, in particular, on the construction of the giant graviton solution. In this sense, this work can be seen as a natural extension of the research program initiated in \cite{HMP} and continued in \cite{MP}. In the former we showed how to implement Mikhailov's holomorphic curve prescription \cite{Mikhailov} to construct giant gravitons on $AdS_{5}\times T^{1,1}$. Guided by that construction and the similarities between the ABJM and Klebanov-Witten models, we formulate an ansatz for the D4-brane  giant graviton extended and moving in $\mathbb{CP}^{3}$ and show that it is energetically degenerate 
with the point graviton. We show also that as the giant grows to maximal size it pinches off into two D4-branes, each wrapping a $\mathbb{CP}^{2}\subset\mathbb{CP}^{3}$ with opposite orientation (preserving the D4-brane charge neutrality of the configuration). This is in excellent agreement with the expectation from the gauge theory in which the operators dual to the giant graviton are ({\it i}) determinant-like and ({\it ii}) built from composite fields of the form $AB$, which factorize at maximal size into dibaryon operators as $\mathrm{det}(AB) = \mathrm{det}(A)\,\mathrm{det}(B)$.   

The spectrum of small fluctuations about this solution, however, has proven to be a much more technically challenging problem. Encouraged by our success in computing the fluctuation spectrum of the giant graviton on $AdS_{5}\times T^{1,1}$, we pursued an analogous line of computation here only to find the resulting system of fluctuation equations not analytically tractable in general. We were, however, able to make some progress in the case of a {\it small} giant graviton (parameterized by $0< \alpha_{0}\ll 1$). Here we were able to solve the decoupled fluctuation equations exactly in terms of hypergeometric and Heun functions. We found that, for both the scalar fluctuations of the $AdS_{4}$ and $\mathbb{CP}^{3}$ transverse coordinates, all eigenvalues are real indicating that the D4-brane giant is, at least to this order in the approximation, perturbatively stable. The zero-mode structure of the spectrum is also in keeping with our expectations: there are no zero modes in the ${AdS_{4}}$ part of the spectrum and a zero mode in the spectrum of $\mathbb{CP}^{3}$ fluctuations corresponding to the fact that it costs no extra energy to increase the size of the giant. More generally though, we were unable to find a global parameterization of the D-brane worldvolume for which the entire spectrum could be read off. Still, there are several interesting observations that can be made:
\vspace{-0.025cm}
\begin{enumerate}[i)]
  \item
    Unlike the spherical dual D2-brane giant graviton \cite{NT-giants,HMPS} for which mixing between
    longitudinal (worldvolume) and transverse (scalar) fluctuations gives rise to a massless Goldstone mode
    that hints towards a solution carrying both momentum and D0-brane charge, no such coupling between 
     gauge field and scalar fluctuations occurs for the D4-brane giant.
  \item
    While our parameterization does not allow us to solve the fluctuation equations in full 
    generality, by expanding in $\alpha_{0}$, we see hints of a dependance on the size of  
    the giant in the spectrum at subleading order in the perturbation series. Should this prove a 
    robust feature of 
    the spectrum, as we expect from our study of the $T^{1,1}$ giant, it will furnish one of the 
    most novel tests of the Giant Graviton/Schur Polynomial correspondence to date. This in 
    itself is, in our opinion, sufficient reason to continue the study of this solution. 
\end{enumerate} 
Evidently then, our study of the D4-brane giant graviton presents just as many (if not more) questions than answers. These include:
\begin{enumerate}[i)]
  \item {\it How much supersymmetry does the D4-brane giant preserve?} To answer this, a detailed
  analysis of the Killing spinor equations along the lines of \cite{NT-giants,GMT}, needs to be undertaken. 
  \item {\it Are these configurations perturbatively stable?} Even though, as we have demonstrated, 
  the D4-brane giant is energetically degenerate with the point graviton, it remains to be shown that
  the fluctuation spectrum is entirely real {\it i.e. there are no tachyonic modes present}.
  \item {\it What are the precise operators dual to the giant and its excitations?} 
  Based on the lessons learnt from 
  $\mathcal{N}=4$ SYM theory, it seems clear that the operators in the ABJM model dual to giant gravitons
  are Schur polynomials constructed from composite scalars in the supermultiplet 
  (see Section \ref{section - ABJM} and the related work in \cite{Dey}). What is not clear is whether the associated 
  {\it restricted} Schur polynomials, which correspond to excitations of the giant, form a 
  complete, orthonormal basis that diagonalizes the 2-point function.  
\end{enumerate}
We hope that, if nothing else, this work stimulates more research on this facinating class of solutions of the type IIA superstring on $AdS_{4}\times \mathbb{CP}^{3}$.

%% file: 08-Acknow.tex
\section{Acknowledgements} \label{section - acknow}

We would like to thank Robert de Mello Koch and Nitin Rughoonauth for useful discussions and comments on the manuscript, and Alex Hamilton for collaboration on the initial stages of this project. The work of JM is supported by the National Research Foundation (NRF) of South Africa's Thuthuka and Key International Scientific Collaboration programs. DG is supported by a National Institute for Theoretical Physics (NITheP) Masters Scholarship.  AP is supported by an NRF Innovation Postdoctoral Fellowship. Any opinions, findings and conclusions or recommendations expressed in this material are those of the authors and therefore the NRF do not accept any liability with regard thereto.

%% file: A-Background.tex
\section{Type IIA string theory on $AdS_{4}\times\mathbb{CP}^{3}$} \label{appendix - background}

Herein we present a brief description of the $AdS_{4}\times\mathbb{CP}^{3}$ background, which is a solution of the type IIA 10D SUGRA equations of motion. Making use of a Hopf fibration of $S^{7}$ over $\mathbb{CP}^{3}$, this background can also be obtained by a Kaluza-Klein dimensional reduction of the $AdS_{4}\times S^{7}$ solution of 11D SUGRA \cite{Pope-et-al}.

The $AdS_{4}\times\mathbb{CP}^{3}$ metric is given by
\begin{equation} \label{metric-orig}
    ds^{2} = R^{2}\left\{ds_{AdS_{4}}^{2} + 4 \hspace{0.075cm} ds_{\mathbb{CP}^{3}}^{2}\right\},
\end{equation}
with $R$ the radius of the anti-de Sitter and complex projective spaces.  The anti-de Sitter metric, in the usual global coordinates, takes the form
\begin{equation}
ds_{AdS_{4}}^{2} = -\left(1+r^{2}\right)dt^{2} + \frac{dr^{2}}{\left(1+r^{2}\right)}
+ r^{2}\left(d\tilde{\theta}^{2}+\sin^{2}{\tilde{\theta}} \hspace{0.075cm} d\tilde{\varphi}^{2}\right).
\end{equation}

Let us make use of a slight variation of the parameterization of \cite{NT-giants} to describe the four homogenous coordinates $z^{a}$ of the complex projective space as follows:
\begin{eqnarray} \label{cp3 parameterization}
\nonumber && z^{1} = \cos{\zeta}\sin{\tfrac{\theta_{1}}{2}} \hspace{0.075cm} e^{i(y + \frac{1}{4}\psi - \frac{1}{2}\phi_{1})} \hspace{1.2cm}
z^{2} = \cos{\zeta}\cos{\tfrac{\theta_{1}}{2}} \hspace{0.075cm} e^{i(y + \frac{1}{4}\psi + \frac{1}{2}\phi_{1})} \\
&& z^{3} = \sin{\zeta}\sin{\tfrac{\theta_{2}}{2}} \hspace{0.075cm} e^{i(y - \frac{1}{4}\psi + \frac{1}{2}\phi_{2})} \hspace{1.2cm}
z^{4} = \sin{\zeta}\cos{\tfrac{\theta_{2}}{2}} \hspace{0.075cm} e^{i(y - \frac{1}{4}\psi - \frac{1}{2}\phi_{2})},
\end{eqnarray}
with radial coordinates $\zeta \in \left[0,\tfrac{\pi}{2}\right]$ and
$\theta_{i} \in \left[0,\pi\right]$, and angular coordinates
$y$, $\phi_{i} \in \left[0,2\pi\right]$ and $\psi \in \left[0,4\pi\right]$. These describe the magnitudes and phases of the homogenous coordinates respectively.  Note that the three inhomogenous coordinates $\tfrac{z^{1}}{z^{4}}$, $\tfrac{z^{2}}{z^{4}}$ and 
$\tfrac{z^{3}}{z^{4}}$ of $\mathbb{CP}^{3}$ are independent of the total phase $y$.
The Fubini-Study metric of the complex projective space can now be written as
\begin{eqnarray}
\nonumber && ds_{\mathbb{CP}^{3}}^{2} = d\zeta^{2}
+ \tfrac{1}{4}\cos^{2}{\zeta}\sin^{2}{\zeta}
\left[d\psi + \cos{\theta_{1}} \hspace{0.075cm} d\phi_{1} + \cos{\theta_{2}} \hspace{0.075cm} d\phi_{2}\right]^{2} \\
&& \hspace{1.5cm}
+ \hspace{0.075cm} \tfrac{1}{4}\cos^{2}{\zeta}\left(d\theta_{1}^{2} + \sin^{2}{\theta_{1}} \hspace{0.075cm} d\phi_{1}^{2}\right)
+ \tfrac{1}{4}\sin^{2}{\zeta}\left(d\theta_{2}^{2} + \sin^{2}{\theta_{2}} \hspace{0.075cm} d\phi_{2}^{2}\right). \label{metric-cp3-orig}
\end{eqnarray}

There is also a constant dilaton $e^{2\Phi} = \tfrac{4R^{2}}{k^{2}}$ and the following even dimensional field strengths:
\begin{eqnarray}
\nonumber && F_{2} = 2kJ = -\tfrac{1}{2}k\left\{\sin\left(2\zeta\right)d\zeta \wedge
\left[d\psi + \cos{\theta_{1}} \hspace{0.075cm} d\phi_{1} + \cos{\theta_{2}} \hspace{0.075cm} d\phi_{2}\right] \right.  \\
&& \hspace{3.25cm} \left. + \hspace{0.075cm}\cos^{2}{\zeta}\sin{\theta_{1}} \hspace{0.075cm} d\theta_{1} \wedge d\phi_{1}
- \sin^{2}{\zeta}\sin{\theta_{2}} \hspace{0.075cm} d\theta_{2} \wedge d\phi_{2} \right\} \label{F2-orig} \\
\nonumber && \\
&& F_{4} = -\tfrac{3}{2}kR^{2} \hspace{0.075cm} \textrm{vol}\left(AdS_{4}\right) = -\tfrac{3}{2}kR^{2}r^{2}\sin{\tilde{\theta}}
\hspace{0.075cm} dt \wedge dr \wedge d\tilde{\theta} \wedge d\tilde{\varphi}, \label{F4-orig}
\end{eqnarray}
with Hodge duals $F_{6} = *F_{4}$ and $F_{8} = *F_{2}$.  In particular, the 6-form field strength can be calculated to be
\begin{eqnarray}
\nonumber & F_{6} & = \tfrac{3}{2}\left(64\right)kR^{4} \hspace{0.075cm} \textrm{vol}\left(\mathbb{CP}^{3}\right) \\
&& = 3kR^{4}\cos^{3}{\zeta}\sin^{3}{\zeta}\sin{\theta_{1}}\sin{\theta_{2}}
\hspace{0.075cm} d\zeta \wedge d\theta_{1} \wedge d\theta_{2} \wedge d\psi \wedge d\phi_{1} \wedge d\phi_{2}. \label{F6-orig}
\end{eqnarray}

%% file: B-Numerics.tex
\section{Energy and momentum integrals} \label{appendix - numerics}

In this appendix, we provide some of the details of our numerical determination of the energy integral (\ref{eq:Energy}) at fixed momentum $P_{\chi}$, given by integral (\ref{eq:Momentum}), as a function of $\alpha$ (shown in Figure \ref{fig:EnergyCurves} of Section \ref{section - giant graviton}).

\subsection{Coordinate change}

The Lagrangian, momentum and energy (\ref{eq:Lagrangian}), (\ref{eq:Momentum}) and (\ref{eq:Energy}) of the D4-brane configuration are given, as functions of the size $\alpha$ and angular velocity $\dot{\chi}$, in terms of the
associated densities (\ref{eq:LagrangianDensity}), (\ref{eq:MomentumDensity}) and (\ref{eq:EnergyDensity})  in the radial $(y,z_{1})$ worldvolume space.  Let us now make the following coordinate change:
\begin{equation}
u \equiv (1+y)(1-z_{1}) \hspace{1.0cm} \textrm{and} \hspace{1.0cm} v \equiv (1-z_{1}).
\end{equation}
The Lagrangian, momentum and energy integrals then become
\begin{eqnarray}
&& L = \int_{1-\alpha}^{1+\alpha}du\int_{V(u)}^{1}dv \hspace{0.15cm} \mathcal{L}(u,v) \\
&& P_{\chi} = \int_{1-\alpha}^{1+\alpha}du\int_{V(u)}^{1}dv \hspace{0.15cm} \mathcal{P}_{\chi}(u,v) \\
&& H = \int_{1-\alpha}^{1+\alpha}du\int_{V(u)}^{1}dv \hspace{0.15cm} \mathcal{H}(u,v),
\end{eqnarray}
with
\begin{equation}
V(u) \equiv \frac{u^{2}}{2u - (1-\alpha^{2})}
\end{equation}
in terms of the new densities in the radial $(u,v)$ worldvolume space:
\begin{equation}
\mathcal{L}(u,v) = \frac{1}{v^{2}} \hspace{0.05cm} \tilde{\mathcal{L}}(u), \hspace{0.65cm} \mathcal{P}_{\chi}(u,v) = \frac{1}{v^{2}} \hspace{0.05cm} \tilde{\mathcal{P}}_{\chi}(u) 
\hspace{0.65cm} \textrm{and} \hspace{0.65cm}
\mathcal{H}(u,v) = \frac{1}{v^{2}} \hspace{0.05cm} \tilde{\mathcal{H}}(u).
\end{equation}
Here we are able to pull out an overall $\frac{1}{v^{2}}$ dependence and define
\begin{eqnarray}
\nonumber && \hspace{-0.25cm} \tilde{\mathcal{L}}(u) = \frac{N}{4} \left\{ \sqrt{2(1-\alpha^2)\,u-(1-\alpha^2) - u^2} \sqrt{2 (1-\alpha^2)\,u \dot{\chi}^2 -(1-\alpha^2)  -u^2} \right. \\
\nonumber && \hspace{2.0cm} \left. + \hspace{0.075cm} \dot{\chi} \left[ u^{2} + (1-\alpha^{2}) -2(1-\alpha^{2})u \right] \right\} \\
&& \\
\nonumber && \hspace{-0.25cm} \tilde{\mathcal{P}}_{\chi}(u) =
\frac{N}{4} \left\{ \frac{2\dot{\chi} (1-\alpha^2) \sqrt{2(1-\alpha^2)\,u-(1-\alpha^2) - u^2}   }
	   {\sqrt{2 (1-\alpha^2)\,u \dot{\chi}^2 -(1-\alpha^2)  -u^2}}  
    + \frac{1}{u} \left[ u^{2} + (1-\alpha^2)  -2(1-\alpha^2)u \right] \right\} \\
&& \\
&& \hspace{-0.25cm} \tilde{\mathcal{H}}(u) = \frac{N}{4} \hspace{0.05cm} \frac{1}{u} \left[u^{2} + (1-\alpha^{2})\right] 
    \frac{ \sqrt{2(1-\alpha^2)\,u-(1-\alpha^2) - u^2} } {\sqrt{2 (1-\alpha^2)\,u \dot{\chi}^2 -(1-\alpha^2)  -u^2}}.
\end{eqnarray}

Explicitly computing the integral over $v$ as follows:
\begin{equation}
\int_{1-\alpha}^{1+\alpha}du\int_{V(u)}^{1} \frac{dv}{v^{2}} = \frac{2u - \left[u^{2} + \left( 1-\alpha^{2} \right) \right]}{u^{2}},
\end{equation}
we can now write
\begin{eqnarray}
&& L = \int_{1-\alpha}^{1+\alpha}du \hspace{0.15cm} \bar{\mathcal{L}}_{\textrm{D4}}(u), \hspace{0.75cm} \textrm{with} \hspace{0.3cm}
\bar{\mathcal{L}}(u) = \frac{2u - \left[u^{2} + \left( 1-\alpha^{2} \right) \right]}{u^{2}} \hspace{0.15cm} \tilde{\mathcal{L}}(u)  \hspace{1.0cm}
\label{B-Lagrangian}  \\
&& P_{\chi} = \int_{1-\alpha}^{1+\alpha}du \hspace{0.15cm} \bar{\mathcal{P}}_{\chi}(u), \hspace{0.75cm} \textrm{with} \hspace{0.3cm} 
\bar{\mathcal{P}}_{\chi}(u) =  \frac{2u - \left[u^{2} + \left( 1-\alpha^{2} \right) \right]}{u^{2}} \hspace{0.15cm} \tilde{\mathcal{P}}_{\chi}(u) \label{B-Momentum} \\
&& H = \int_{1-\alpha}^{1+\alpha}du \hspace{0.15cm} \bar{\mathcal{H}}(u), \hspace{0.99cm} \textrm{with} \hspace{0.3cm}
\bar{\mathcal{H}}(u) = \frac{2u - \left[u^{2} + \left( 1-\alpha^{2} \right) \right]}{u^{2}} \hspace{0.15cm} \tilde{\mathcal{H}}(u). \label{B-Energy}
\end{eqnarray}

\subsection{Momentum integral}

The momentum integral (\ref{B-Momentum}) was calculated numerically using standard quadrature routines.  Our result is shown in the form of a surface $P_{\chi}(\alpha, \dot{\chi})$ in Figure \ref{fig:surface_P} below.
\begin{figure}[htb!]
  \setlength{\unitlength}{1cm}
  \begin{center}
    \begin{picture}(11.5,9.5)(0,0)
 	  \put(0,0){\includegraphics[scale=0.8, trim=3.4cm 9cm 3.4cm 9cm]{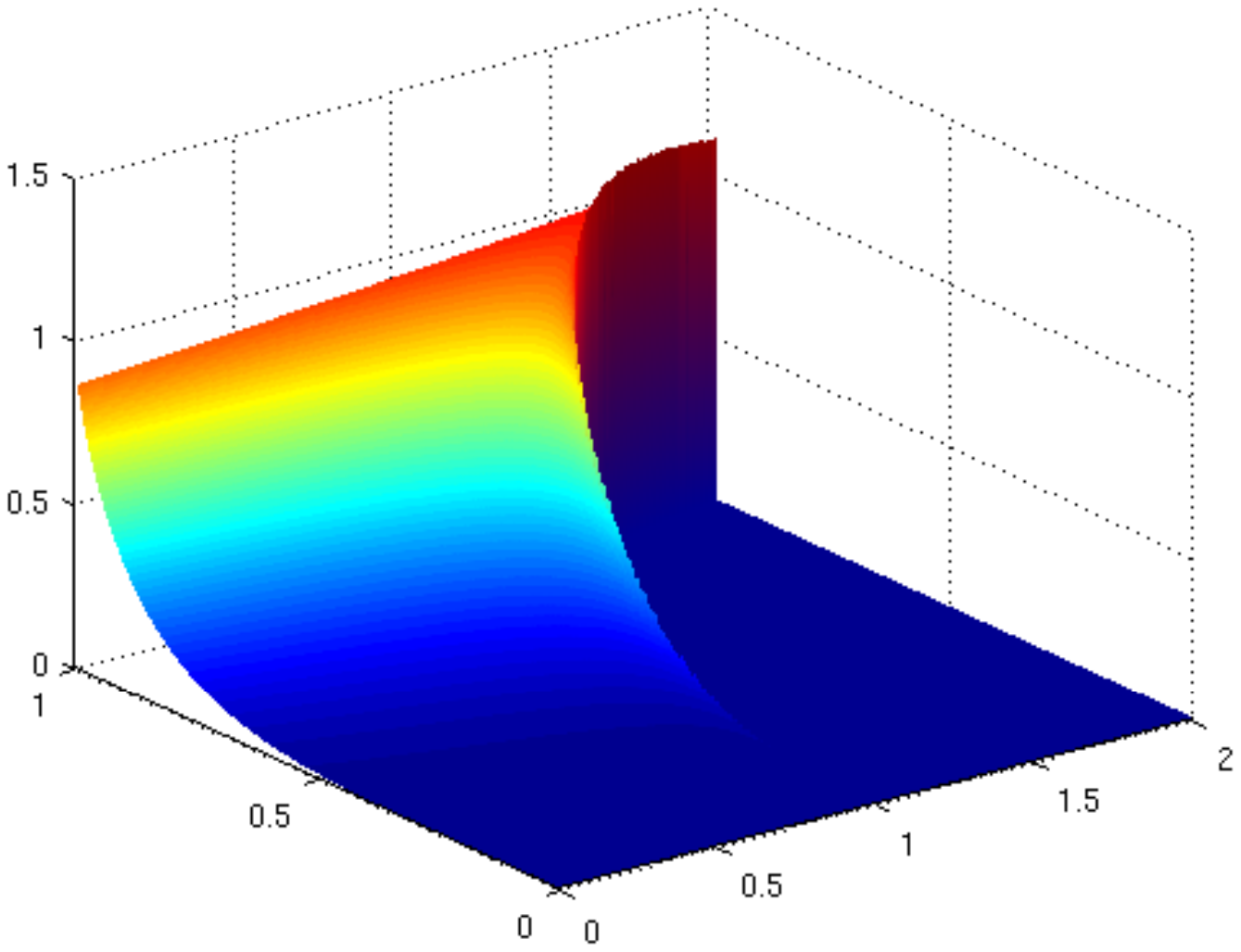}}
     \put(8.5, 0.9){$\dot\chi$}
     \put(0 ,  5){$P_\chi$}
     \put(2.2, 1.2){$\alpha$}

    \end{picture}
\end{center}
\vspace{-7mm}
 \caption{The momentum surface $P_{\chi}(\alpha, \dot\chi)$ plotted in units of the flux $N$. 
     The discontinuity curve is clearly evident.} 
 \label{fig:surface_P}
\end{figure}

The most striking feature is the presence of a singularity along the curve\footnote{Unlike the canonical sphere-giant case in $AdS_5\times S^5$, in which the singularity occurs only at $\alpha = 0$, here 
the discontinuity traces out an entire curve. This happens at angular velocities $\dot{\chi}$ 
always bigger than one (and therefore never effects the giant graviton solution).  Perhaps we can interpret this effect physically as a limiting velocity 
$(1-\alpha^{2})^{\frac{1}{2}} ~ \dot{\chi}_{\textrm{discontinuity}} = (1-\alpha^{2})^{\frac{1}{4}} \leq 1$.}
\begin{equation}
 \label{B-Discontinuity}
  \dot\chi^4 = \frac{1}{1-\alpha^2}
\end{equation}
on the $\alpha\dot{\chi}$-plane.  The existence of this singularity means that we should approach the energy integral with some caution.

\subsection{Energy integral}

\begin{figure}[htb!]
\begin{center}
 \setlength{\unitlength}{1cm}
    \begin{picture}(11.5,9.5)(0,0)
      \put(0,0){\includegraphics[scale=0.8,trim=3.4cm 9cm 3.4cm 9cm]{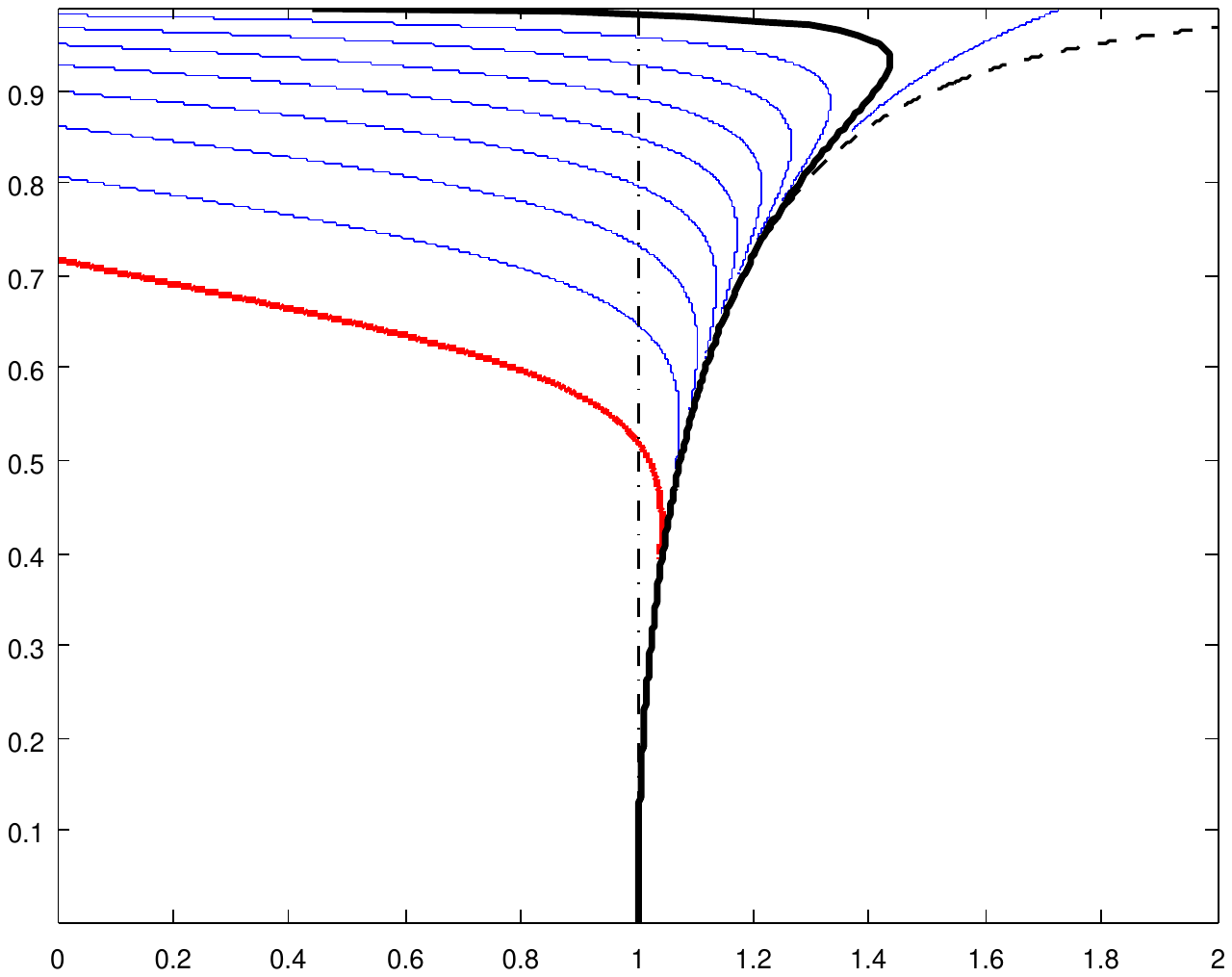}  }
      \put(6, 0.1){$\dot\chi$}
      \put(0, 5){$\alpha$}
      \put(3,   5.2){$P_{\chi} = 0.1$}
      \put(4.6, 5.3){\vector(1,0){1}}
      \put(9, 7.1){$P_{\chi} = 0.9$}
      \put(9, 7.2){\vector(-1,1){1}}
 \end{picture}
\end{center}
\vspace{-7mm}
\caption{Lines of constant momentum $P_\chi$.  The dashed curve describes the discontinuity.} 
\label{B-Contours}
\end{figure}

We would now like to calculate the energy integral (\ref{B-Energy}) at fixed momentum $P_{\chi}$ as a function of $\alpha$.  Making use of the $P_{\chi}(\alpha, \dot\chi)$ surface, it is possible to plot contours of constant 
momentum on the $\alpha\dot{\chi}$-plane (see Figure \ref{B-Contours}).

In principle, we can numerically integrate the energy (\ref{B-Energy}) along any contour $\dot{\chi}(\alpha)$ at fixed momentum $P_{\chi}$.  These contours all approach the discontinuity, however, which places
practical constraints upon how far along the contour we can perform the numerical integration.  An alternative approach to the direct integration of (\ref{B-Energy}) therefore needs to be found.

\begin{figure}[htb!]
\begin{center}
 \setlength{\unitlength}{1cm}
  \begin{picture}(14.3,11.7)
   \includegraphics[scale=1.0,trim=3.4cm 9cm 3.4cm 9cm]{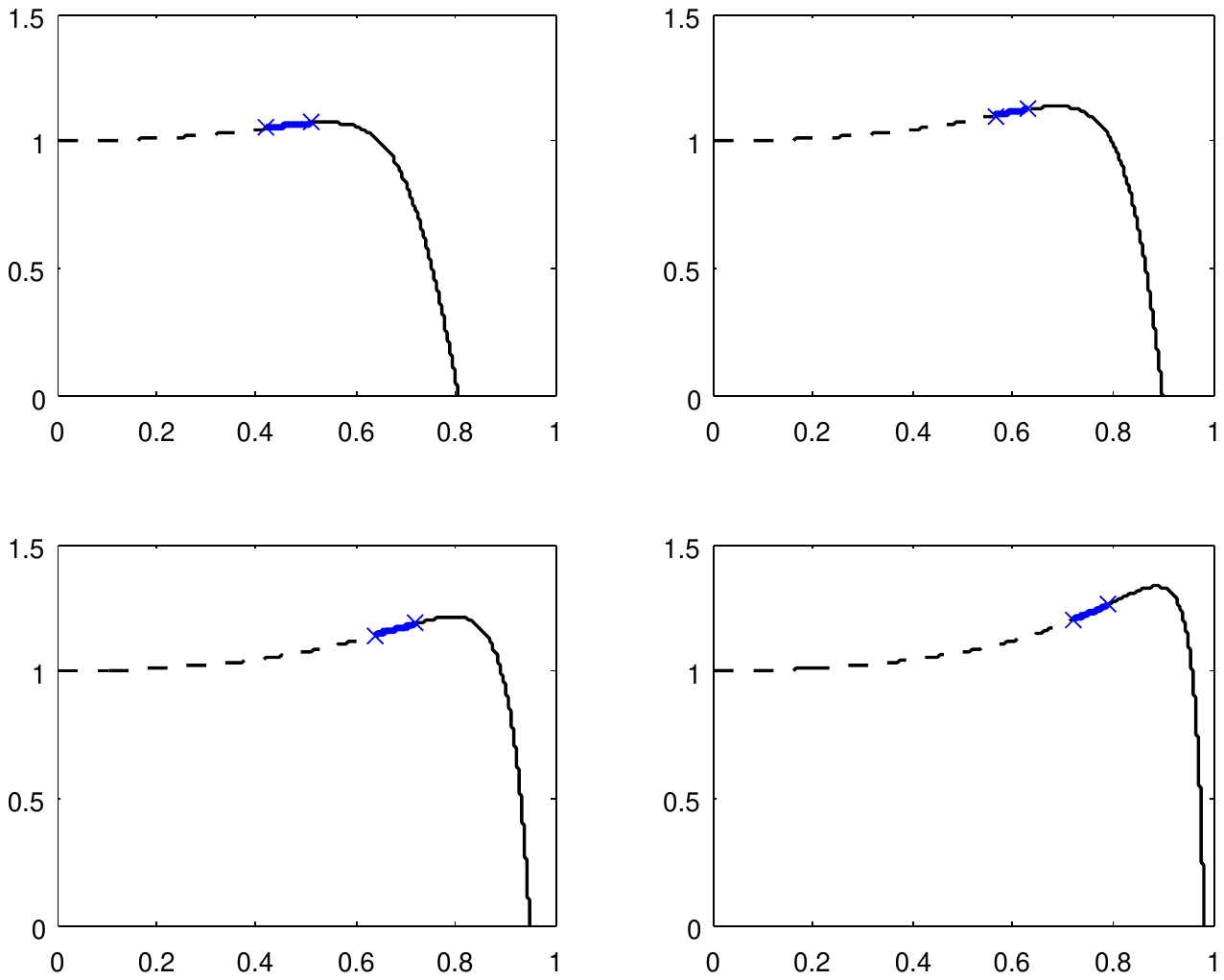}
     \put(-10.5, 6){$\alpha$}
     \put(-3.5,  6){$\alpha$}
     \put(-10.5, 0.4){$\alpha$}
     \put(-3.5, 0.4){$\alpha$}

     \put(-14, 8.9){$\dot\chi$}
     \put(-7,  8.9){$\dot\chi$}
     \put(-14, 3.3){$\dot\chi$}
     \put(-7,  3.3){$\dot\chi$}

     \put(-12.9,11.1){(a) $P_{\chi} = 0.2$}
     \put(-6,   11.1){(b) $P_{\chi} = 0.4$}
     \put(-12.9, 5.5){(c) $P_{\chi} = 0.6$}
     \put(-6,    5.5){(d) $P_{\chi} = 0.8$}
 \end{picture}
\vspace{-7mm}
\caption{The  relationship between $\dot\chi$ and $\alpha$ for fixed momentum. 
 	The curves show the results from the fixed momentum contour plot (rightmost section of the curve), together with the discontinuity curve (the leftmost dashed curve), given by $ \dot\chi^4 = \frac{1}{1-\alpha^2}.$
	The section of curve between the two $\times$'s is obtained by cubic spline interpolation.
        } \label{B-Interpolation}
\end{center}
\end{figure}

The Hamiltonian
\begin{equation}
H = \dot{\chi} P_{\chi} - L
\end{equation}
has a singularity along the same curve (\ref{B-Discontinuity}) as the momentum $P_{\chi}$.  The Lagrangian $L$, however, is devoid of any such defect.  Fixing $P_{\chi}$ and moving along this contour (in the
direction of decreasing $\alpha$), we can determine $\dot{\chi}(\alpha)$ up until a certain point, at which the contour becomes too close to the singularity to distinguish between the two and the numerics break down.
At this point, however, we can simply use the curve (\ref{B-Discontinuity}) of the discontinuity itself to obtain a good approximation for $\dot{\chi}(\alpha)$. The full contour $\dot{\chi}(\alpha)$ can then be obtained using a cubic 
spline interpolation between the numerical contour and the discontinuity curve in the vicinity of this point (the position of which depends on the particular contour in question).  We have considered $P_{\chi} = 0.2$, $0.4$, $0.6$
and $0.8$ as examples, and Figure \ref{B-Interpolation} shows the full contour $\dot{\chi}(\alpha)$, obtained using this interpolation technique, in each of these cases. Having obtained $\dot{\chi}(\alpha)$ along a fixed $P_{\chi}$
contour, there is no further hinderance to integrating the Lagrangian $L(\alpha,\dot{\chi}(\alpha))$ numerically using (\ref{B-Lagrangian}), since it is non-singular, and hence determining the energy (\ref{B-Energy}).

%% file: C-dAlembertian.tex
\section{d'Alembertian on the giant  graviton's worldvolume} \label{appendix - dAlembertian}

The metric on the worldvolume  of the giant graviton in the worldvolume coordinates $\sigma^{a} = (t,x_{1},x_{2},\varphi_{1},\varphi_{2})$ can be written as
\begin{equation}
h_{ab} = 
\begin{pmatrix}
-1 + g_{\chi\chi} & 0 & 0 & g_{\chi\varphi_{1}} & g_{\chi\varphi_{2}} \\
0 & g_{x_{1}x_{1}} & g_{x_{1}x_{2}} & 0 & 0 \\
0 & g_{x_{1}x_{2}} & g_{x_{2}x_{2}} & 0 & 0 \\
g_{\chi\varphi_{1}} & 0 & 0 & g_{\varphi_{1}\varphi_{1}} & g_{\varphi_{1}\varphi_{2}} \\
g_{\chi\varphi_{2}} & 0 & 0 & g_{\varphi_{1}\varphi_{2}} & g_{\varphi_{2}\varphi_{2}}
\end{pmatrix},
\end{equation}
in terms of the components of the angular and radial metrics of the complex projective space (evaluated at $\alpha = \alpha_{0}$).  The inverse metric $h^{ab}$ can thus be expressed in terms of cofactors as follows:
\begin{equation}
h^{ab} = 
\begin{pmatrix}
h^{tt} & 0 & 0 & h^{t \varphi_{1}} & h^{t \varphi_{2}} \\
0 & h^{x_{1}x_{1}} & h^{x_{1}x_{2}} & 0 & 0 \\
0 & h^{x_{1}x_{2}} & h^{x_{2}x_{2}} & 0 & 0 \\
h^{t \varphi_{1}} & 0 & 0 & h^{\varphi_{1}\varphi_{1}} & h^{\varphi_{1}\varphi_{2}} \\
h^{t \varphi_{2}} & 0 & 0 & h^{\varphi_{1}\varphi_{2}} & h^{\varphi_{2}\varphi_{2}}
\end{pmatrix},
\end{equation}
with temporal and angular inverse worldvolume metric components
\begin{eqnarray}
\nonumber && \hspace{-0.35cm} h^{tt} = - \frac{ \left( C_{\textrm{ang}} \right)_{11}}{\left[ \left(C_{\textrm{ang}}\right)_{11} - \det{g_{\textrm{ang}}} \right]} \hspace{1.475cm}  
h^{t \varphi_{1}} = - \frac{ \left( C_{\textrm{ang}} \right)_{12}}{\left[ \left(C_{\textrm{ang}}\right)_{11} - \det{g_{\textrm{ang}}} \right]} \\
&& \nonumber \\
\nonumber && \hspace{-0.35cm} h^{t \varphi_{2}} = - \frac{ \left( C_{\textrm{ang}} \right)_{13}}{\left[ \left(C_{\textrm{ang}}\right)_{11} - \det{g_{\textrm{ang}}} \right]} \hspace{1.225cm}
 h^{\varphi_{1} \varphi_{1}} = - \frac{\left[ \left( C_{\textrm{ang}} \right)_{22} - g_{\varphi_{2}\varphi_{2}} \right]}{\left[ \left(C_{\textrm{ang}}\right)_{11} - \det{g_{\textrm{ang}}} \right]} \\
&& \nonumber \\
&& \hspace{-0.35cm} h^{\varphi_{2} \varphi_{2}} = - \frac{\left[ \left( C_{\textrm{ang}} \right)_{33} - g_{\varphi_{1}\varphi_{1}} \right]}{\left[ \left(C_{\textrm{ang}}\right)_{11} - \det{g_{\textrm{ang}}} \right]} \hspace{1.0cm}
h^{\varphi_{1} \varphi_{2}} = - \frac{\left[ \left( C_{\textrm{ang}} \right)_{23} + g_{\varphi_{1}\varphi_{2}} \right]}{\left[ \left(C_{\textrm{ang}}\right)_{11} - \det{g_{\textrm{ang}}} \right]} \hspace{1.5cm}
\end{eqnarray}
and radial inverse worldvolume metric components
\begin{eqnarray}
&& \hspace{-0.35cm}  h^{x_{1}x_{1}} = \frac{g_{x_{2}x_{2}}}{\left(C_{\textrm{rad}}\right)_{11}} \hspace{1.2cm} 
h^{x_{2}x_{2}} =  \frac{g_{x_{1}x_{1}}}{\left(C_{\textrm{rad}}\right)_{11}} \hspace{1.2cm} 
h^{x_{1}x_{2}} = - \frac{g_{x_{1}x_{2}}}{\left(C_{\textrm{rad}}\right)_{11}}. \hspace{1.2cm}
\end{eqnarray}

 The invariant volume form on this worldvolume space is given by
\begin{eqnarray} \label{grad-squared}
& \omega & = \sqrt{-h} \hspace{0.2cm} dt \wedge dx_{1} \wedge dx_{2} \wedge d\varphi_{1} \wedge d\varphi_{2},
\end{eqnarray}
where
\begin{equation}
\sqrt{-h} = \sqrt{\left(C_{\textrm{rad}}\right)_{11}\left[ \left(C_{\textrm{ang}}\right)_{11} - \det{g_{\textrm{ang}}} \right]}.
\end{equation}

The gradient squared of an arbitrary function $f(\sigma^{a})$ can be written in the compact notation
\begin{equation} \label{dAlembertian}
\left( \p f \right)^{2} \equiv h^{ab} \left(\p_{a} f\right) \left(\p_{b} f\right)
\end{equation}
and hence the d'Alembertian operator on the worldvolume of the giant graviton takes the form
\begin{equation}
\Box \equiv \frac{1}{\sqrt{-h}} \hspace{0.075cm} \p_{a} \left( \sqrt{-h} \hspace{0.1cm} h^{ab} \hspace{0.05cm} \p_{b} \right) 
= h^{ab} \hspace{0.075cm} \p_{a} \hspace{0.025cm}  \p_{b} + \frac{1}{\sqrt{-h}} \hspace{0.075cm} \p_{a} \left( \sqrt{-h} \hspace{0.1cm} h^{ab} \right) \p_{b}.
\end{equation}